\documentclass[preprint,12pt]{elsarticle}
\usepackage{lineno}
\usepackage{amsmath}

\makeatletter
\def\ps@pprintTitle{%
 \let\@oddhead\@empty
 \let\@evenhead\@empty
 \def\@oddfoot{}%
 \let\@evenfoot\@oddfoot}
\makeatother

\usepackage{amssymb}
\usepackage{amsmath}
\usepackage{graphicx}
\usepackage{subcaption}
\usepackage[export]{adjustbox}

\usepackage{dcolumn}
\usepackage{bm}
\usepackage{xcolor,soul}
\usepackage{multirow}
\usepackage{float}
\usepackage[normalem]{ulem}
\usepackage[percent]{overpic}
\usepackage{hyperref}
\usepackage{url}

\journal{Journal of Computational Physics }

\begin{document}
\begin{frontmatter}


\author{Ilies Haouche$^a$\corref{cor1}}
\ead{ilies.haouche@univ-lille.fr}
\author{Benjamin Reichert$^a$}
\author{Micha\"el Baudoin$^{a,\,b}$}
\author{Palas Kumar Farsoiya$^{c}$\corref{cor1}}
\ead{palas.farsoiya@ch.iitr.ac.in}
\cortext[cor1]{Corresponding authors.}

\affiliation[aff1]{organization={Univ. Lille, CNRS, Centrale Lille, Univ. Polytechnique Hauts-de-France, UMR 8520, IEMN,
F59000 Lille, France},
            }
\affiliation[aff2]{organization={Institut Universitaire de France, 1 rue Descartes, 75005 Paris, France},
            }
\affiliation[aff3]{organization={Department of Chemical Engineering, Indian Institute of Technology Roorkee, Roorkee, Uttarakhand, India},
            }


\title{A hybrid Volume-of-Fluid and Phase-Field method for Direct Numerical Simulations of soluble surfactant-laden interfacial flows}



\begin{abstract}

We present a hybrid Volume-of-Fluid (VoF) and Phase-Field method for general soluble surfactant-laden interfacial flows. The scheme retains the VoF method for interface tracking and momentum solution, while a diffused Phase-Field serves as a smooth carrier for surfactant transport enabling consistent coupling between the bulk and interfacial concentration fields without the computation of surface derivatives. Adsorption and desorption kinetics are incorporated through regularized source terms localized at the interface, and the dependence of surface tension on surfactant concentration can be specified for general equation of states. The method is fully adaptive via quadtree/octree Adaptive Mesh Refinement (AMR), enabling accurate and efficient simulations in planar, axisymmetric, and three-dimensional domains with high parallel scalability.
Rigorous validation against analytical solutions for surfactant transport on deforming interfaces and for diffusion-driven adsorption in the no-flow limit confirms the accuracy and convergence of the approach. To demonstrate the coupled capabilities of the framework, we investigate the buoyancy-driven rise of a bubble in the presence of soluble surfactants, in both axisymmetric and three-dimensional configurations. By independently varying the Biot number and the Damk\"ohler number, we recover the correct asymptotic limits corresponding to clean-interface and insoluble-surfactant dynamics, and characterize the intermediate soluble regime. The resulting Marangoni stresses, induced by non-uniform interfacial surfactant concentrations, significantly reduce interfacial mobility, leading to measurable reductions in terminal velocity and pronounced modifications of the bubble trajectory. These results demonstrate the robustness of the method in capturing the complex interplay between hydrodynamics, bulk and interfacial surfactant transport, and Marangoni stresses in realistic three-dimensional geometries.

\end{abstract}

\begin{graphicalabstract}
\end{graphicalabstract}

\begin{highlights}
\item Hybrid Volume of Fluid Phase-field method for soluble surfactant interfaces.
\item Scalable simulations in 2D/axisymmetric/3D with Adaptive Mesh Refinement (AMR).
\item Benchmarking against analytical surfactant transport solutions.
\end{highlights}


\end{frontmatter}

\section{Introduction}

Soluble surfactants play a significant role in modulating interfacial dynamics in multiphase flows through their ability to adsorb to and desorb from fluid interfaces. Unlike their insoluble counterparts, soluble surfactants are governed by a coupled system of bulk convection–diffusion, interfacial transport, and kinetic exchange processes. This coupling gives rise to spatially and temporally varying surface tension fields, which in turn generate interfacial Marangoni stresses that alter the flow topology, cause or suppress instabilities, or drive new modes of motion. Such phenomena are central to the behavior of emulsions, droplet coalescence, film drainage, and interfacial instabilities in surfactant-rich environments. However, the accurate numerical simulation of these systems remains a formidable challenge due to the multiscale and multiphysics nature of the surfactant transport problem.
Direct Numerical Simulation (DNS) provides a first-principles approach for capturing surfactant-mediated interfacial flows without relying on turbulence closure models or empirical transport approximations. Yet, DNS of soluble surfactant systems entails several technical hurdles: resolving sharp interfacial gradients in surfactant concentration, capturing adsorption/desorption kinetics at moving and deformable interfaces, maintaining consistency between bulk and interfacial fluxes, and computing surface tension gradients with high fidelity. These difficulties are compounded by the stiffness of the governing equations and the need for accurate curvature and surface differential operators. 
There are many numerical studies and methods considering the Marangoni effects due to thermal gradients, e.g.\cite{abu2018conservative,seric2018direct,tripathi2018motion,mialhe2023extended}.  However, in the presence of surfactant, the Marangoni effects need to be coupled with surfactant transport equations on the interface. Transporting soluble surfactants on a moving interface has been challenging in Eulerian numerical frameworks, and is the focus of the present work.

Recent efforts have investigated how to simulate surfactant-laden flows using different numerical approaches, each offering distinct trade-offs in terms of the surfactant physics they capture, the accessible flow regimes, and the required computational framework.

\textbf{Boundary element and boundary integral methods:} Early numerical work by \cite{stone1990effects} developed a boundary integral procedure to simulate insoluble surfactant effects on drop breakup in extensional flow, restricted to the Stokes flow regime $(Re << 1)$ and incapable of handling topology changes or inertial dynamics. This approach was extended to three-dimensional Stokes flows with insoluble surfactants by \cite{li1997effect} and \cite{yon1998finite}, retaining the same fundamental constraints: the method is exact for Stokes flows but cannot be applied at finite Reynolds number, does not conserve fluid mass independently of the flow solver, and requires explicit remeshing of the interface when topology changes occur, making breakup and coalescence difficult to treat.

\textbf{Front-tracking methods:} The Front-Tracking approach was employed to study the effect of insoluble surfactants on rising bubbles by \cite{jan1991computational}. A major extension to soluble surfactants including bulk advection-diffusion and adsorption/desorption kinetics was achieved by \cite{muradoglu2008front} for two-dimensional axisymmetric flows and later extended to three-dimensional flows by \cite{muradoglu2014simulations}.  Another three-dimensional front-tracking method using the Leibniz Transport Formula was proposed in \cite{de20153d}, following related approaches in \cite{dziuk2007finite, lenz2011convergent}. While front-tracking methods are highly accurate and naturally account for adsorption/desorption kinetics through a Lagrangian interfacial grid, they carry significant implementation complexity in three dimensions and require topological treatment (explicit reconnection or cutting of the Lagrangian mesh) to handle interface reconnection and breakup. This treatment can introduce errors and is difficult to automate robustly, particularly in parallel computing environments where load balancing between the fixed Eulerian bulk grid and the dynamic interfacial mesh is non-trivial.

\textbf{Level-set methods:} Level-set approaches were developed by \cite{xu2003eulerian} to solve insoluble surfactant transport on a moving interface in the absence of flow, and later coupled to the Navier–Stokes equations by \cite{xu2006level}. While level-set methods provide smooth normal and curvature fields and handle topology changes naturally, they do not conserve fluid mass exactly requiring reinitialization of the signed-distance function, which introduces both computational overhead and additional numerical error. Mass loss in the surfactant field has been documented for this class of methods and requires post-hoc correction procedures. Level-set based methods for soluble surfactants have been demonstrated in 2D and 3D (e.g. \cite{xu2006level}), but without AMR and generally on structured uniform grids. Similar level-set frameworks were employed to investigate surfactant-laden flows \cite{piedfert2018numerical, atasi2018influence}, but those implementations were restricted to two-dimensional axisymmetric configurations.

\textbf{Hybrid Front-Tracking Level-Set methods:} A coupling of Front Tracking and Level Set was applied to study the effect of surfactants on capillary waves by \cite{ceniceros2003effects}. A similar hybrid approach was implemented in \cite{shin2018hybrid}, which employs a stationary Eulerian structured mesh for velocity and pressure alongside an unstructured Lagrangian mesh tracking the interface. While this approach handles soluble surfactants, the use of a structured Eulerian grid without adaptive mesh refinement for momentum equations limits the ability to resolve multiscale problems efficiently, and parallel scalability is constrained by the load-imbalance between the fixed Eulerian grid and the dynamically evolving unstructured interfacial mesh.

\textbf{Arbitrary Lagrangian–Eulerian (ALE) methods:} While ALE methods provide a sharp, body-fitted representation of the interface and allow accurate resolution of interfacial concentration gradients \citep{dieter2015direct, meijer2023rising}, they require continuous remeshing as the interface deforms, which becomes prohibitively complex when the mesh distorts severely.

\textbf{Diffuse-interface (Phase-Field) methods:} Diffuse-interface methods offer an attractive alternative by avoiding the explicit computation of surface differential operators and handling topology changes naturally. Methods based on the free energy approach were developed for soluble and insoluble surfactants in \cite{van2006diffuse, yun2014new}, and Cahn–Hilliard-based formulations were proposed in \cite{teigen2009diffuse} and later extended in \cite{jain2024model,jain2025model}. However, a key limitation common to Phase-Field approaches is that the method relies on the Phase-Field variable remaining at thermodynamic equilibrium across the interface; strong convection and numerical diffusion in the interface-capturing equation can cause the profile to deviate from this equilibrium, leading to concentration singularities and non-conservation of surfactant mass \cite{yamashita2024conservative}. Extending such frameworks to the fully soluble case while maintaining mass conservation and coupling with a mass-conserving, geometrically accurate VoF interface representation has remained an open problem.

\textbf{Volume of Fluid methods:} VoF methods are well suited for sharp and mass-conserving interface representation and support AMR and high parallel scalability, but surfactant transport in the strict VoF framework had long been challenging because the interface has zero thickness, making the definition and numerical computation of surface differential operators required for the surfactant transport equation geometrically cumbersome on a piecewise-linear reconstructed interface \cite{james2004surfactant}. Advances have been made in algebraic VoF method \cite{antritter2024two} to simulate surfactant laden flows, however a sharp interface representation is achieved in geometric VoF methods.   A recent sharp geometric VoF method for insoluble surfactants has been proposed by \cite{xue2025sharp}, demonstrating that this challenge can be overcome for the insoluble case. However, extending a purely sharp VoF approach to soluble surfactants, where a coupled bulk advection-diffusion equation must be solved consistently with the interfacial transport and adsorption/desorption kinetics, while avoiding artificial surface derivatives, has not been accomplished within a fully adaptive, scalable, open-source framework.

In this context, there is a need for a geometric VoF-based method for three-dimensional two-phase flows that accounts for surfactant transport, adsorption–desorption kinetics, and the resulting Marangoni stresses crucially, for surfactants that are soluble in the bulk fluid. Such a method is well suited for free-surface flows where distinct phases are separated by a sharp interface \cite{popinet2018numerical} and would allow investigating problems involving interface reconnection, such as jet break-up and breaking waves \cite{riviere2021sub,Deike2018,mostert2022high}, in the presence of realistic, industrially and biologically relevant soluble surfactants. Moreover, the use of adaptive mesh refinement is essential for resolving the multiscale gradients that arise simultaneously in the bulk concentration field, the interfacial concentration, and the flow all of which must be resolved consistently in soluble surfactant problems \cite{popinet2009accurate,mostert2022high}. Making these methods available within an open-source framework is also desirable for the community.
In the present study, we address this need by extending the geometric VoF Phase-Field framework of \cite{farsoiya2024coupled} which was limited to insoluble surfactants confined to the interface to the fully soluble case, in which surfactant exchanges dynamically between the bulk and the interface through adsorption and desorption kinetics. This extension is non-trivial: soluble surfactants introduce a coupled bulk advection–diffusion equation, a kinetic source/sink term linking bulk and interfacial concentrations, and a substantially stiffer numerical system due to the wide range of timescales governing adsorption, desorption, and hydrodynamic transport. The present approach retains the use of volume fractions for sharp, mass-conserving interface representation and momentum solution, while the Phase-Field serves exclusively as a smooth carrier for surfactant transport both at the interface and in the adjacent bulk. The Phase-Field is evolved via the Allen-Cahn-based ACDI formulation of \cite{jain2022accurate}, avoiding the computational cost of repeated signed-distance reinitialization from volume fractions \cite{limare2023hybrid}. The Marangoni force is incorporated via the continuum surface force formulation of \cite{seric2018direct}, improved by \cite{tripathi2018motion}, and the surface tension equation of state accounts for the nonlinear dependence of interfacial tension on the local surfactant concentration through a general equation of state. The entire framework is implemented within Basilisk \cite{basilisk}, an open-source code with well-documented test cases, support for adaptive mesh refinement (AMR), high parallel scalability, and applicability to a wide range of interfacial fluid mechanics problems.
Here, we illustrate the method on the canonical problem of a bubble rising in an otherwise quiescent surfactant-laden fluid. We first validate the surfactant transport module in the absence of flow against analytical solutions for both insoluble and soluble limiting cases. We then investigate the coupled problem, demonstrating the effect of soluble surfactants including the distinct roles of adsorption and desorption kinetics, quantified through the Damk\"ohler and Biot numbers on axisymmetric and three-dimensional bubble dynamics. Quantitative measurements of terminal velocity and surfactant redistribution, alongside qualitative features of the flow structure and trajectory, are provided for this complex and practically important problem.


\section{Governing equations}
We consider incompressible two-phase flows laden with soluble surfactants, in which surfactant molecules are advected and diffused in the bulk, adsorb onto or desorb from the fluid interface, and modulate the local interfacial tension. The resulting surface tension gradients generate Marangoni stresses that drive complex interfacial dynamics. The physical model is solved using the open-source framework Basilisk \cite{basilisk,popinet2009accurate}, combining a geometric Volume-of-Fluid (VoF) method for conservative interface tracking with a diffuse Phase-Field formulation for computing interfacial geometry and applying regularized surface differential operators. This hybrid approach enables consistent and accurate coupling between the interfacial surfactant dynamics and the underlying flow. This section describes the full system of governing equations for fluid motion, phase transport, surfactant evolution, and interfacial force balance.

\subsection{Volume-of-Fluid representation of the interface}
The Volume-of-Fluid (VoF) method is used to represent the distribution of two immiscible fluids 1 and 2 in a conservative manner via a scalar field
$c\big(\mathbf{x},\,t\big)\in \big[0,\,1\big]$, which denotes the volume fraction of fluid 1 in each computational cell, with $\mathbf{x}$ the spatial coordinate and $t$ the time. 
The evolution of $c\big(\mathbf{x},\,t\big)$ is governed by the incompressible advection equation:
\begin{gather}
    \frac{\partial c}{\partial t}+\big(\mathbf{u}\cdot\boldsymbol{\nabla}\big) c=0,
\end{gather}
where $\mathbf{u}$ is the fluid velocity field. The VoF method ensures exact mass conservation of each fluid and allows to capture large interfacial deformations. In our formulation, $c\big(\mathbf{x},\,t\big)$ governs the spatial distribution of material properties such as density $\rho$ and viscosity $\mu$:
\begin{gather}
    \rho\big(c\big)=\rho_1c + \rho_2\big(1-c\big),\,\,\,\,\,\,\,\,\,\,\,\,\,\,\,\,\,\,\,\,\mu\big(c\big)=\mu_1c + \mu_2\big(1-c\big)
\end{gather}
with subscripts $1$ and $2$ referring to each fluid phase. While $c\big(\mathbf{x},\,t\big)$ serves as the primary marker for the interface and ensures mass conservation, an auxiliary Phase-Field variable $\phi\big(\mathbf{x},\,t\big)$ is introduced to compute geometric quantities such as normal vectors, curvature, and surface gradients in a regularized and smooth fashion.

\subsection{Flow equations for incompressible two-phase fluids}
The two immiscible fluids are modeled using a single set of incompressible Navier–Stokes equations with spatially varying properties:
\begin{eqnarray}
& \boldsymbol{\nabla}\cdot\mathbf{u}=0, \\
&\frac{\partial \mathbf{u}}{\partial t}+\mathbf{u}\cdot \boldsymbol{\nabla}\mathbf{u} = \frac{1}{\rho}\left[ - \boldsymbol{\nabla} p + \boldsymbol{\nabla}\cdot(\mu(\boldsymbol{\nabla} \mathbf{u}+\boldsymbol{\nabla} \mathbf{u}^T))  + \sigma\kappa\mathbf{n}\delta_s + \delta_m{\boldsymbol{\nabla}_s \sigma}\right] + \mathbf{g}, \label{nseq} 
\end{eqnarray}
with $\mathbf{u}\big(\mathbf{x},\,t\big)$ the fluid velocity, $p\big(\mathbf{x},\,t\big)$ the pressure,  $\sigma$ the surface tension, $\kappa=-\boldsymbol{\nabla}\cdot\mathbf{n}$ the mean interface curvature and $\delta_s$ and $\delta_m$ are defined below.

\subsection{Surface tension equation of state}
The interfacial tension $\sigma\big(\mathbf{x},\,t\big)$ depends on the interfacial concentration $\Gamma\big(\mathbf{x},\,t\big)$. The relation between both is usually modeled through a nonlinear equation of state (EOS). A commonly used model valid at low surfactant concentrations is the Langmuir adsorption isotherm \cite{manikantan2020surfactant,tricot1997}, which reads:
\begin{gather}
    \sigma\big(\mathbf{x},\,t\big)=\sigma_0 + \mathcal{R}T\Gamma_{\infty}\ln\bigg(1-\frac{\Gamma\big(\mathbf{x},\,t\big)}{\Gamma_{\infty}}\bigg),
\end{gather}
where $\mathcal{R}$ is the ideal gas constant, $T$ is the absolute temperature, $\Gamma_{\infty}$ is the maximum interfacial surfactant concentration, and $\sigma_0$ is the surface tension for a clean interface. We can introduce the Gibbs elasticity number defined by:
\begin{gather}
    \mathrm{E}=\frac{\mathcal{R}T\Gamma_{\infty}}{\sigma_0},
\end{gather}
which is a measure of sensitivity of the surface tension to the surfactant concentration. In the low surfactant concentration limit, this can be reduced to a linear model:
\begin{gather}
    \sigma\big(\mathbf{x},\,t\big)=\sigma_0\bigg( 1 - \mathrm{E}\frac{\Gamma\big(\mathbf{x},\,t\big)}{\Gamma_{\infty}}\bigg).
\end{gather}

\subsection{Capillary and Marangoni interfacial forces}
Interfacial forces arising from surface tension are incorporated into the Navier–Stokes equations as volumetric body forces localized near the interface. Two principal contributions are considered: the isotropic capillary force due to curvature and the tangential Marangoni force due to surface tension gradients induced by surfactant inhomogeneity \citep{landau2013}. The capillary force is expressed as:
\begin{equation}
\mathbf{F}_{\text{cap}} = \sigma\big(\mathbf{x},\,t\big) \kappa \delta_s \mathbf{n},
\end{equation}
where \( \sigma\big(\mathbf{x},\,t\big) \) is the surface tension (which depends on the local surfactant concentration, see below), \( \kappa = -\boldsymbol{\nabla} \cdot \mathbf{n} \) is the mean curvature of the interface and $\delta_s = |\boldsymbol{\nabla} c|$.

The Marangoni force accounts for tangential variations of the interfacial tension and is given by:
\begin{equation}
\mathbf{F}_{\text{Ma}} = \delta_m \boldsymbol{\nabla}_s \sigma,
\end{equation}
where \( \boldsymbol{\nabla}_s = (\mathbf{I} - \mathbf{n} \otimes \mathbf{n}) \cdot \boldsymbol{\nabla} \) is the surface gradient operator projecting onto the tangential plane of the interface and $\delta_m = A_f / V_c$ i.e. the ratio of interface area and cell volume \cite{tripathi2018motion}. 

\subsection{Phase-Field method}
To transport the surfactant, we introduce the Phase-Field $\phi\big(\mathbf{x},\,t\big)$ as a partial derivative equation model ACDI (Accurate Conservative Diffuse-Interface), which is an Allen-Cahn-based second-order Phase-Field model proposed by \cite{jain2022accurate}:
\begin{gather}
    \frac{\partial \phi}{\partial t} + \boldsymbol{\nabla} \cdot \left(\mathbf{u}\phi\right) = \boldsymbol{\nabla}\cdot\bigg(\zeta\bigg\{\epsilon\boldsymbol{\nabla}\phi-\frac{1}{4}\bigg[1-\tanh^2\bigg(\frac{\psi}{2\epsilon}\bigg)\bigg]\frac{\boldsymbol{\nabla} \psi}{\|\boldsymbol{\nabla}\psi\|}\bigg\}\bigg),
\end{gather}
which smoothly varies between $0$ and $1$ across the interface. In this equation, the auxiliary signed-distance-like $\psi$ defined by:
\begin{equation}
    \psi = \epsilon \ln \left(\frac{\phi + \varepsilon}{1-\phi + \varepsilon}\right),
\end{equation}
where $\epsilon$ defines the thickness of the Phase-Field and should be less than the minimum grid spacing $\Delta x_{\text{min}}$ to ensure numerical stability and $\varepsilon$ is added to ensure the stability of $\psi$ when $\phi$ goes to $0$ or $1$ \citep{jain2022accurate}. $\zeta$ is a velocity scale parameter, which must be greater than the maximum velocity value $\zeta \ge \|\mathbf{u}_{\text{max}}\|$ to always bound $\phi\big(\mathbf{x},\,t\big)$ between $0$ and $1$ \citep{jain2022accurate}. In the present study, we used $\epsilon = 0.75 \Delta x_{\text{min}}$, $\varepsilon=10^{-6}$ and $\zeta = 1.1||\mathbf{u}_{\text{max}}\|$. This formulation ensures that $\phi\big(\mathbf{x},\,t\big)$ maintains a smoothed hyperbolic tangent profile centered at the interface and used for surfactant transport in the computational domain. The Phase-Field $\phi$ is coupled to volume fraction $c$ through reinitialization at a suitable frequency using the signed distance function $\chi$,
\begin{gather}
    \phi = \frac{1}{2}\left( 1 - \tanh\left\{\frac{\chi}{2\epsilon}\right\}\right), \label{redistance}
\end{gather}
The signed distance function $\chi$ is reconstructed from the volume fraction $c$ by iteratively solving the Hamilton–Jacobi equation,
\begin{gather}
\frac{\partial \chi}{\partial \tau} + \text{sign}(\chi_0)\left(|\nabla \chi| - 1\right) = 0, \
\chi_0 = \chi(x,0),
\end{gather}
where $\tau$ is a pseudo-time and $\chi_0$ represents the initial level-set field \cite{min2007second,min2010reinitializing,russo2000remark}.

\subsection{Surfactant transport equations}
The surfactant is modeled using two scalar fields: the bulk concentration $F\big(\mathbf{x},\,t\big)$ and the interfacial concentration $\Gamma\big(\mathbf{x},\,t\big)$. The numerical framework to handle surfactant transport using the Phase-Field is proposed in \cite{jain2025model}. The interfacial transport equation reads:
\begin{gather}
    \frac{\partial \Gamma}{\partial t}+\boldsymbol{\nabla}_s\cdot \big(\Gamma\mathbf{u}_s\big)+\Gamma\big(\boldsymbol{\nabla}_s\cdot\mathbf{n}\big)\big(\mathbf{u}\cdot\mathbf{n}\big)=D_f\nabla_s^2\Gamma + J \label{Eqst}.
\end{gather}
with $J = r_a F_s ( \Gamma_{\infty}-\Gamma )-r_d\Gamma$, $\Gamma_\infty$ the saturated surface concentration of surfactant, $F_s$ the bulk concentration of surfactant at the interface calculated numerically according to $F_s = F / \phi$, $r_a F_s$ the rate of adsorption of surfactant from the bulk to the surface, and $r_d$ the rate of desorption of surfactant from the surface into the bulk. Note that to avoid division by zero, a small constant is added numerically when computing $1/\phi$ to estimate $F_s$.

To avoid the numerical computation of $\boldsymbol{\nabla}_s$, the surface concentration $\Gamma$ is turned into a volumetric concentration $f\big(\mathbf{x},\,t\big)$ located at the interface \cite{jain2024model,jain2025model}. In this framework, equation (\ref{Eqst}) becomes:
\begin{equation}
    \frac{\partial f}{\partial t} + \boldsymbol{\nabla} \cdot \left(f \mathbf{u}\right) = \boldsymbol{\nabla}\cdot \left(D_f\boldsymbol{\nabla} f -D_f\frac{2(0.5-\phi)}{\epsilon}\mathbf{n} f\right) + j
\end{equation}
with $f=\Gamma \delta_\phi$, $\delta_\phi=|\boldsymbol{\nabla}\phi|$ the surface delta function estimated numerically according to the formula $\delta_\phi=\phi\big(1-\phi\big)/\epsilon$, and $j$ the adsorption-desorption source term  accounting for the exchange between the interface and the bulk \citep{martinez2020}:
\begin{gather}
    j=r_a F_s\big(f_{\infty}-f \big)-r_d f.
\end{gather}
The transport of the surfactant in the bulk is given by:
\begin{equation}
    \frac{\partial F}{\partial t} + \boldsymbol{\nabla} \cdot \left(F \mathbf{u}\right) = \boldsymbol{\nabla}\cdot \left(D_F\boldsymbol{\nabla} F -D_F\frac{1-\phi}{\epsilon}\mathbf{n} F\right) - j,
    \label{eqbulk}
\end{equation}
In the bulk equation, it is localized at the interface  since the field $f\big(\mathbf{x},t\big)$ is nonzero only there as shown in Fig.~\ref{fig:interface-profile}. The field $\phi\big(\mathbf{x},\,t\big)$ is the Phase-Field (details in next section), $\epsilon$ is the interface-thickness-scale parameter, $f\big(\mathbf{x},t\big)$ is the interfacial surfactant concentration and $F\big(\mathbf{x},t\big)$ is the bulk surfactant concentration. There is a source/sink terms due to species adsorption into and desorption out of the interface. $f_{\infty}$ is the saturated surface concentration of surfactant, $D_f$ and $D_F$ are the interfacial and bulk diffusivity of the surfactant. Finally, the normal vector $\mathbf{n}$ is calculated by $\mathbf{n}=\boldsymbol{\nabla} \phi/|\boldsymbol{\nabla}\phi |$.

\subsection{Illustration of the diffuse interface structure}

\begin{figure}[h!]
    \centering
    \includegraphics[width=0.75\textwidth]{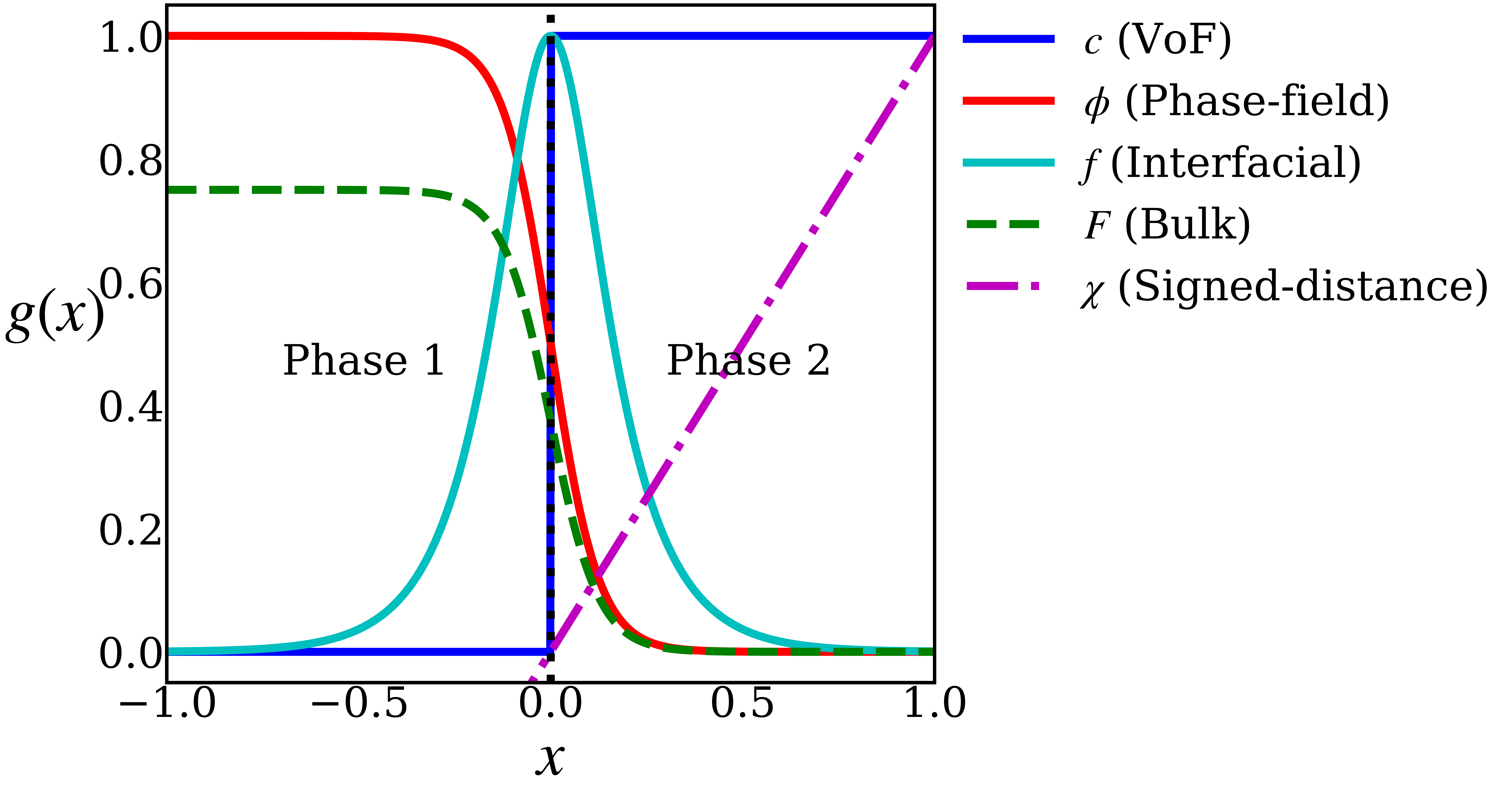}
    \caption{One-dimensional profile across a flat interface comparing the classical volume-of-fluid (VoF) step function, the Phase-Field variable \(\phi\), the interfacial surfactant concentration \(f\), the bulk concentration \(F\) and the signed-distance function $\chi$.}
    \label{fig:interface-profile}
\end{figure}

To better understand the regularized representation of the interface and the localization of surfactant concentrations in the diffuse interface framework, we illustrate in Fig.~\ref{fig:interface-profile} a one-dimensional profile across a flat interface separating two immiscible fluids. 

The Phase-Field variable \(\phi\big(\mathbf{x},\,t\big)\), which smoothly transitions from 0 in phase 1 to 1 in phase 2, defines the interface region as the transition layer of finite thickness \(\epsilon\). The interfacial surfactant concentration \(f\big(\mathbf{x},\,t\big)\), which replaces the classical sharp-interface variable \(\Gamma\), is localized within this diffuse layer and defined by \(f = \Gamma\,\delta_\epsilon\), where \(\delta_\epsilon = \phi(1-\phi)/\epsilon\) serves as a regularized delta function centered at the interface. The bulk concentration \(F\big(\mathbf{x},\,t\big)\) exhibits a depletion layer near the interface due to adsorption. As surfactant molecules are transferred from the bulk to the interface, the bulk concentration decreases in the vicinity of the interface and gradually increases away from it, recovering its far-field value in the bulk.  For comparison, we also plot the corresponding VoF field, which exhibits a sharp jump at the interface.

This figure highlights how the diffuse-interface formulation provides a natural and smooth framework to localize interfacial quantities without explicitly tracking a sharp boundary.

\section{Numerical methods}
In this section, we describe the numerical methods used to solve the coupled Phase-Field and soluble surfactant transport equations introduced in Section 2. We focus on the detailed numerical treatment of (i) the evolution of the diffuse interface through a regularized Phase-Field formulation and (ii) the transport and exchange dynamics of soluble surfactants in both the bulk and the interfacial regions.

The system thus consists of three primary scalar fields: the diffuse interface field \(\phi(\mathbf{x}, t)\), the bulk surfactant concentration \(F(\mathbf{x}, t)\), and the interfacial surfactant concentration \(f(\mathbf{x}, t)\). These fields are evolved in time on the same computational mesh using conservative, semi-implicit time integration schemes detailed below.

The interface is represented through the Phase-Field \(\phi\big(\mathbf{x},\,t\big)\), which varies smoothly from 0 to 1 across a diffuse layer of thickness \(\epsilon\). This auxiliary field is used to regularize interfacial quantities and to define geometrical operators required in the surfactant transport equations. Its evolution is governed by a transport–regularization equation that maintains a localized diffuse profile and preserves the consistency of the interfacial layer during the simulation. In the present framework, $\phi\big(\mathbf{x},\,t\big)$ is not introduced as an independent sharp-interface replacement, but as an auxiliary diffuse representation coupled to the interface-capturing strategy used in the overall formulation. This construction follows the general Phase-Field-based regularization strategy adopted for surfactant transport in \cite{jain2025model}, while remaining consistent with the coupled VoF–Phase-Field methodology employed in the present work.

The surfactant transport equations are discretized in a conservative finite-volume form and solved separately in the bulk and at the interface. The equations include advection, anisotropic diffusion aligned with the interface geometry, and source/sink terms modeling adsorption and desorption. The coupling between the bulk and interfacial fields is mediated through regularized delta functions derived from the Phase-Field gradient, which localize interfacial interactions over a finite support. Special care is taken in the discretization of the adsorption–desorption term to ensure numerical stability, conservation of total surfactant mass, and correct asymptotic behavior in the thin-interface limit (see Appendix~C for details).

Spatial discretization is performed on a dynamically refined quadtree/octree grid \citep{popinet2003}, where resolution is concentrated near the interface to resolve steep gradients in \(\phi\big(\mathbf{x},\,t\big)\), \(f\big(\mathbf{x},\,t\big)\), and \(F\big(\mathbf{x},\,t\big)\) and spatial derivatives (gradients, divergences, Laplacians) are evaluated using second-order accurate  finite-volume consistent approximations. Time integration is handled via operator splitting, with explicit treatment of advection and either semi-implicit treatment of diffusion, depending on the stiffness of the problem. For the Phase-Field reinitialization equation, we introduce an artificial diffusion velocity that stabilizes the profile and avoids drift or smearing over long times.

\begin{figure}[h!]
    \centering
    \includegraphics[width=0.52\linewidth]{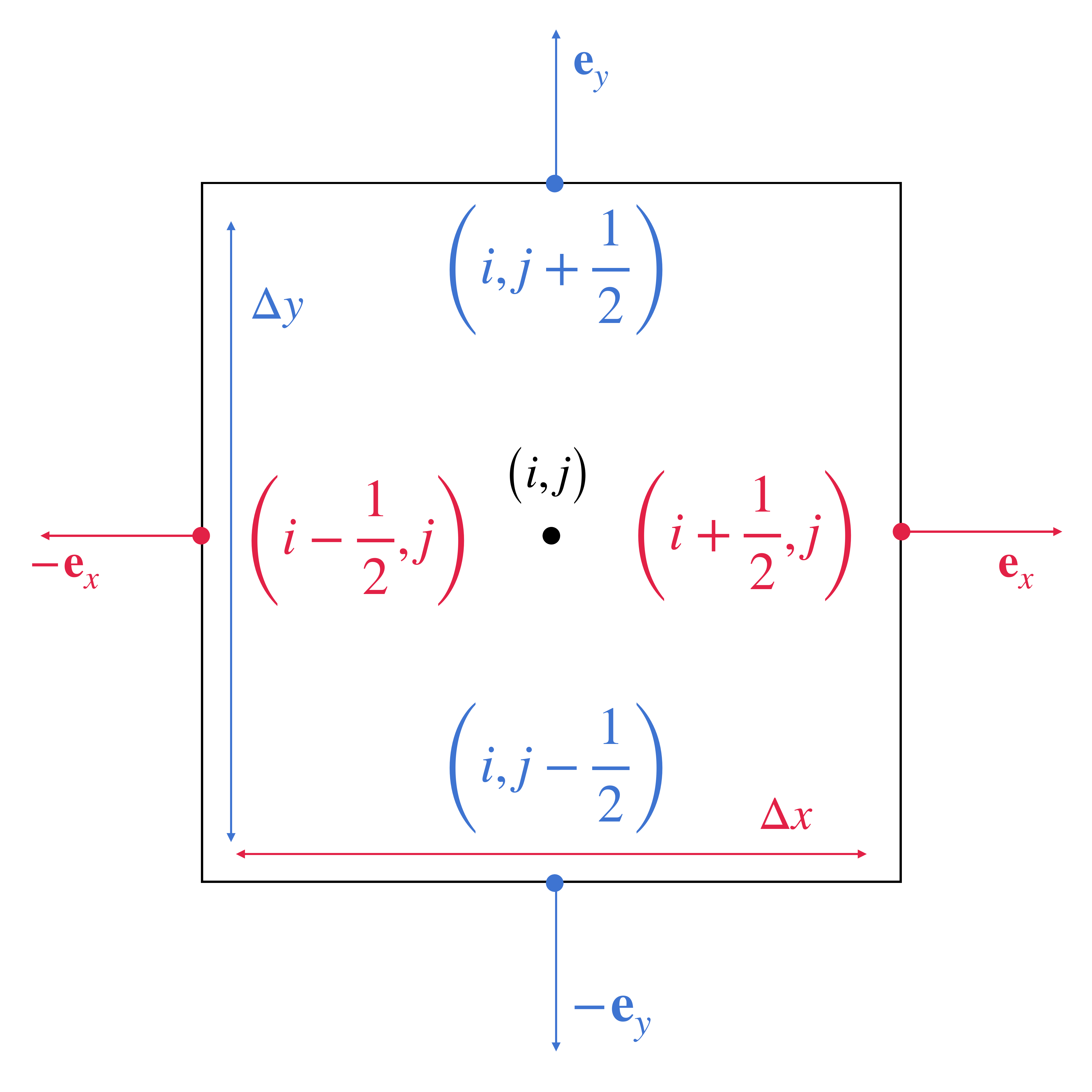}
    \caption{
Schematic of a two-dimensional finite-volume control cell. The scalar quantity \(\xi_{i,j}\) is stored at the cell center, while fluxes are evaluated at the cell faces located at \((i \pm \tfrac{1}{2}, j)\) and \((i, j \pm \tfrac{1}{2})\). The unit normal directions \(\mathbf{e}_x\) and \(\mathbf{e}_y\), as well as the grid spacings \(\Delta x\) and \(\Delta y\), define the orientation and size of the control volume used for flux integration.}
    \label{fig:staggeredgrid}
\end{figure}

In the following subsections, we provide detailed descriptions of the discretization and numerical treatment of the Phase-Field transport and regularization equation, the interfacial and bulk surfactant transport equations, the computation of regularized delta functions and interfacial geometry, the numerical handling of the adsorption/desorption source term, and the adaptive mesh refinement strategy used to concentrate resolution near the interface.

To unify the presentation, we note that the governing equations for \(\phi\big(\mathbf{x},\,t\big)\), \(F\big(\mathbf{x},\,t\big)\), and \(f\big(\mathbf{x},\,t\big)\) share a common structure: they can all be written as advection–diffusion equations with an additional anisotropic (anti-diffusive) drift term and a source term. We therefore introduce a generalized scalar transport equation of the form:
\begin{gather}
    \frac{\partial \xi}{\partial t} + \boldsymbol{\nabla} \cdot (\xi \mathbf{u}) = \boldsymbol{\nabla} \cdot \left( \boldsymbol{\mathcal{D} }\boldsymbol{\nabla} \xi + \mathbf{B} \xi \right) + q,
\end{gather}
where $\boldsymbol{\mathcal{D}}=D\mathbf{I}$ is the diffusivity matrix and $\mathbf{I}$ is the identity matrix. Here, \(\xi(\mathbf{x}, t)\) represents a generic scalar field, corresponding to \(\phi\big(\mathbf{x},\,t\big)\), \(F\big(\mathbf{x},\,t\big)\), or \(f\big(\mathbf{x},\,t\big)\), depending on the context; $\boldsymbol{\nabla}\cdot\big(\mathbf{B}\xi\big)$ is a anti-diffusive term and \(q(\mathbf{x}, t)\) is a source or sink term. This formulation allows us to present a unified discretization strategy for all three fields, while highlighting the specific differences in the subsequent subsections.

\subsection{Time discretization}
We consider here the temporal discretization of the generalized scalar transport equation introduced above:
\begin{equation}
    \frac{\partial \xi}{\partial t} + \boldsymbol{\nabla} \cdot (\xi \mathbf{u}) = \boldsymbol{\nabla} \cdot \left( \boldsymbol{\mathcal{D}} \boldsymbol{\nabla} \xi + \mathbf{B} \xi \right) + q,
\end{equation}
where \(\xi\big(\mathbf{x},\,t\big)\) denotes a generic scalar field representing the Phase-Field \(\phi\big(\mathbf{x},\,t\big)\), bulk surfactant concentration \(F\big(\mathbf{x},\,t\big)\), or interfacial surfactant concentration \(f\big(\mathbf{x},\,t\big)\).

A first-order operator splitting strategy is adopted to separately advance each physical process—advection, diffusion (including anti-diffusion term), and source terms—over a time step \(\Delta t\). This approach provides flexibility in handling different numerical schemes tailored to the stability and accuracy requirements of each operator.

For a given time interval \([t^n, t^{n+1} = t^n + \Delta t]\), the scalar field is updated in two substeps as follows:
\begin{enumerate}
    \item \textbf{Advection step:}
    \begin{equation}
        \frac{\xi^* - \xi^n}{\Delta t} + \boldsymbol{\nabla} \cdot (\xi^n \mathbf{u}^n) = 0
    \end{equation}
    The advection equation is solved explicitly using a geometric unsplit Bell–Collela–Glaz \citep{bcg1989} scheme with directionally split fluxes \cite{popinet2003,popinet2009}. This conservative method ensures boundedness and preserves the sharpness of scalar gradients.

    \item \textbf{Diffusion, anti-diffusion, and source step:}
    \begin{equation}
        \frac{\xi^{n+1} - \xi^*}{\Delta t} = \boldsymbol{\nabla} \cdot \left( \boldsymbol{\mathcal{D}} \boldsymbol{\nabla} \xi^{n+1} + \mathbf{B} \xi^{n+1} \right) + q^{n+1}
    \end{equation}
    This step includes both diffusion and anti-diffusive terms. It is treated implicitly using Euler implicit method in time and a finite-volume method and solved using the modified multigrid Poisson solver in space. Source term is also integrated implicitly. For the surfactant fields, this includes the adsorption/desorption exchange terms localized at the interface, while for the Phase-Field equation, the source term is set to zero.
\end{enumerate}

This splitting approach ensures modularity, ease of implementation, and straightforward incorporation of source terms localized near the interface. Although the scheme is formally first-order accurate in time, it is found to be sufficient for the present study, given the dominant role of advection and the use of fine adaptive meshes. Higher-order time integration methods could be introduced in future work to improve temporal accuracy, especially for long-time simulations or regimes with stiff adsorption–desorption kinetics.

We focus here on the second substep of the time-splitting scheme, which includes both diffusion and anti-diffusive terms, along with localized source terms. This stage is solved implicitly to enhance numerical stability and allow for the treatment of stiff source dynamics.

The numerical resolution relies on solving an elliptic problem inspired by the method proposed by \citep{popinet2003, popinet2009}, where we extend the standard formulation to account for surfactant-specific physics. In particular, we introduce anisotropic anti-diffusive contributions, localized source terms (e.g., adsorption/desorption), and interface-based localization through regularized delta functions. These additions required modifications to the elliptic solver and to the surfactant module to ensure consistent coupling with the Phase-Field representation and interfacial geometry.

The general form of the modified elliptic equation takes the following form:
\begin{equation}
    \mathcal{L}(a) = \boldsymbol{\nabla} \cdot \left( \alpha \boldsymbol{\nabla} a \right) + \lambda a = b.
\end{equation}
To account for the anti-diffusive term, we generalize the elliptic operator to the following form:
\begin{equation}
    \mathcal{L}(a) = \boldsymbol{\nabla} \cdot \left( \alpha \boldsymbol{\nabla} a + \boldsymbol{\beta} a \right) + \lambda a = b.
\end{equation}
This allows us to recast the second substep equation:
\begin{equation}
    \frac{\xi^{n+1} - \xi^*}{\Delta t} = \boldsymbol{\nabla} \cdot \left( \boldsymbol{\mathcal{D}} \boldsymbol{\nabla} \xi^{n+1} + \mathbf{B} \xi^{n+1} \right) + q^{n+1}
\end{equation}
into the equivalent Poisson-like formulation:
\begin{equation}
    - \boldsymbol{\nabla} \cdot \left( \boldsymbol{\mathcal{D}} \boldsymbol{\nabla} \xi^{n+1} + \mathbf{B} \xi^{n+1} \right) + \frac{\xi^{n+1}}{\Delta t} = \frac{\xi^*}{\Delta t} + q^{n+1}.
\end{equation}
By comparison with the generalized Poisson equation, we identify the mapping:
\begin{gather}
    a = \xi^{n+1},\,\,\,\alpha = -\boldsymbol{\mathcal{D}},\,\,\,\boldsymbol{\beta} = -\mathbf{B},\,\,\,\lambda = \frac{1}{\Delta t},\text{ and}\,\,b = \frac{\xi^*}{\Delta t} + q^{n+1}
\end{gather}

Note that if the source term \(s^{n+1}\) depends explicitly on \(\xi^{n+1}\), it can be absorbed into the coefficient \(\lambda\), resulting in a modified reaction term. This is particularly relevant for stiff adsorption–desorption kinetics in the surfactant transport equations.

\subsection{Finite-volume discretization}

The governing transport equations are discretized using a finite-volume approach on a Cartesian mesh. The spatial discretization relies on a staggered grid as shown in Fig.~\ref{fig:staggeredgrid}, where scalar quantities such as the interfacial and bulk concentrations are stored at cell centers, while fluxes and velocity components are defined at cell faces. This configuration ensures a consistent evaluation of advective and diffusive fluxes across cell interfaces and improves numerical stability, particularly in convection-dominated regimes.
We now detail the finite-volume discretization of the equation \citep{flmbook} introduced earlier:
\begin{equation}
    \int_K \frac{\partial \xi}{\partial t} \, dV
    = \int_K \boldsymbol{\nabla} \cdot \left( \boldsymbol{\mathcal{D}} \boldsymbol{\nabla} \xi + \mathbf{B}\, \xi \right) dV
    + \int_K q \, dV,
\end{equation}
where \(\boldsymbol{\mathcal{D}} = D \mathbf{I}\).

Applying the divergence theorem yields
\begin{equation}
    \frac{d}{dt} (\xi_{i,j} V)
    = \oint_{\partial K} \left( D \boldsymbol{\nabla} \xi + \mathbf{B}\, \xi \right) \cdot \mathbf{n} \, dS
    + q_{i,j} V,
\end{equation}
with \(V = \Delta x \Delta y\). The fluxes are evaluated as the sum over the four cell faces:
\begin{gather}
    \sum_{\text{faces}} (D \boldsymbol{\nabla} \xi)\cdot \mathbf{n} S
    = \Delta y \left[ \left(D \frac{\partial \xi}{\partial x}\right)_{i+\frac{1}{2},j}
    - \left(D \frac{\partial \xi}{\partial x}\right)_{i-\frac{1}{2},j} \right] \nonumber\\
    + \Delta x \left[ \left(D \frac{\partial \xi}{\partial y}\right)_{i,j+\frac{1}{2}}
    - \left(D \frac{\partial \xi}{\partial y}\right)_{i,j-\frac{1}{2}} \right],
\end{gather}
\begin{gather}
    \sum_{\text{faces}} (\mathbf{B}\xi)\cdot \mathbf{n} S
    = \Delta y \left[ (B_x \xi)_{i+\frac{1}{2},j} - (B_x \xi)_{i-\frac{1}{2},j} \right] \nonumber\\
    + \Delta x \left[ (B_y \xi)_{i,j+\frac{1}{2}} - (B_y \xi)_{i,j-\frac{1}{2}} \right].
\end{gather}

Time integration is performed using an implicit Euler scheme:
\begin{gather}
    \frac{\xi_{i,j}^{n+1} - \xi_{i,j}^n}{\Delta t}
    = \frac{1}{\Delta x} \left[ \left(D \frac{\partial \xi^{n+1}}{\partial x}\right)_{i+\frac{1}{2},j}
    - \left(D \frac{\partial \xi^{n+1}}{\partial x}\right)_{i-\frac{1}{2},j} \right] \nonumber\\
    + \frac{1}{\Delta y} \left[ \left(D \frac{\partial \xi^{n+1}}{\partial y}\right)_{i,j+\frac{1}{2}}
    - \left(D \frac{\partial \xi^{n+1}}{\partial y}\right)_{i,j-\frac{1}{2}} \right] \nonumber\\
    + \frac{1}{\Delta x} \left[ (B_x \xi^{n+1})_{i+\frac{1}{2},j} - (B_x \xi^{n+1})_{i-\frac{1}{2},j} \right] \nonumber\\
    + \frac{1}{\Delta y} \left[ (B_y \xi^{n+1})_{i,j+\frac{1}{2}} - (B_y \xi^{n+1})_{i,j-\frac{1}{2}} \right]
    + q^{n+1}_{i,j}.
\end{gather}
Diffusive fluxes are approximated using centered differences, ensuring second-order accuracy. Convective fluxes are discretized using a first-order upwind scheme to ensure numerical stability in advection-dominated regimes. The anti-diffusive flux associated with \(\mathbf{B}\) is treated as a convective contribution, with \(\mathbf{B}\) acting as an effective velocity field. The upwind scheme is therefore applied based on the sign of the face-centered components of \(\mathbf{B}\).
\begin{gather}
    \frac{\partial \xi}{\partial x}\Big|_{i+\frac{1}{2},j}
    \approx \frac{\xi_{i+1,j}^{n+1} - \xi_{i,j}^{n+1}}{\Delta x}, \quad
    \xi_{i+\frac{1}{2},j} =
\begin{cases}
\xi_{i,j} & \text{if } \big(B_x\big)_{i+\frac{1}{2},j} > 0, \\
\xi_{i+1,j} & \text{if } \big(B_x\big)_{i+\frac{1}{2},j} \leq 0.
\end{cases},
\end{gather}
with:
\begin{gather}
    \big(B_x\big)_{i+\frac{1}{2},j}\approx \frac{\big(B_x\big)_{i+1,j}+\big(B_x\big)_{i,j}}{2}
\end{gather}
A similar discretization is applied in the $y$-direction, including centered approximations for diffusive fluxes and upwind evaluation of convective fluxes.

The extension of the present formulation to three-dimensional and two-dimensional axisymmetric configurations are given in Appendix~A and Appendix-B respectively. The finite-volume discretization described above follows standard formulations for advection–diffusion equations \citep{flmbook}, and is reported in full detail to allow independent reproduction of the method. Consistent implementation within the Basilisk framework enables coupled bulk–interface surfactant transport on adaptive quadtree/octree meshes, with a unified treatment of soluble surfactants in planar, axisymmetric, and fully three-dimensional configurations. Combined with adaptive mesh refinement, this allows interfacial concentration gradients and Marangoni stresses to be accurately resolved across a wide range of scales within a single computational framework.

\subsection{Courant-Friedrichs-Lewy criterion}

To ensure the numerical stability of the explicit advection and diffusion schemes, we enforce a Courant–Friedrichs–Lewy (CFL) condition based on the most restrictive physical process, following \citep{jain2022accurate}. In the case of soluble surfactant transport, the scalar field \(\xi\big(\mathbf{x},\,t\big)\) is governed by an advection–diffusion equation augmented with an anti-diffusive term. As such, the effective velocity driving the transport is composed of both the convective and anti-diffusive contributions. We define the effective velocity magnitude as a summation of the convective part and the anti-diffusive part:
\begin{equation}
    u_{\text{eff}} = |\mathbf{u}|_{\max} + \frac{D}{\epsilon},
\end{equation}
where \(|\mathbf{u}|_{\max}\) is the maximum velocity magnitude in the domain, \(D\) is the diffusivity, and \(\epsilon\) is the width of the interfacial layer (used in the regularization of the delta function). This term accounts for the strongest gradients introduced by the anti-diffusion normal to the interface.

The time step \(\Delta t\) is then chosen according to the minimum of two criteria:
\begin{equation}
    \Delta t \leq \min(\Delta t_{\text{conv}}, \Delta t_{\text{diff}}),
\end{equation}
with
\begin{align}
    \Delta t_{\text{conv}} &= \frac{\Delta x}{u_{\text{eff}}}, \\
    \Delta t_{\text{diff}} &= \frac{\Delta x^2}{2N_dD},
\end{align}
where $\Delta x$ is the local grid size and $N_d$ is the dimension.

Combining the two restrictions leads to a final conservative estimate for the global space step:
\begin{equation}
    \Delta x \leq \frac{2D}{|\mathbf{u}|_{\max} + \dfrac{D}{\epsilon}}.
\end{equation}

This condition guarantees both stability of the explicit advection step and boundedness of the diffusive and anti-diffusive contributions. In practice, we apply an additional safety factor to the computed \(\Delta t\) to further prevent instabilities in regions of high velocity or sharp interfacial curvature.
Note that in the special case where the regularization thickness \(\epsilon\) is chosen equal to the mesh size, i.e., \(\epsilon = \Delta x\), the CFL condition simplifies to a constraint directly on the mesh size:
\begin{equation}
    \Delta x \leq \frac{D}{|\mathbf{u}|_{\text{max}}} \quad \Leftrightarrow \quad Pe_c \leq 1
\end{equation}
where \(Pe_c = \dfrac{\Delta x |\mathbf{u}|_{\text{max}}}{D}\) is the local cell Péclet number. This condition ensures that advection does not dominate over diffusion at the grid scale, a necessary requirement for numerical stability and accuracy when solving advection-diffusion equations with strong interface gradients.

\section{Validation of surfactant transport without flow}
Before applying the numerical method to complex interfacial flows, it is essential to assess its accuracy, consistency, and robustness through a series of validation studies. These validations aim to verify the correct implementation of the surfactant transport model, including the interfacial and bulk equations, the adsorption/desorption kinetics, the Phase-Field geometry, and the numerical discretization strategy.

The tests are designed to investigate various aspects of the model, both in static and dynamic settings and in different spatial dimensions. In all cases, the fluid velocity and volume fraction fields may be prescribed to isolate the surfactant dynamics and allow a direct comparison with known or expected reference behaviors. The Phase-Field variable is initialized accordingly and evolved consistently with the numerical scheme presented earlier.

These validation studies provide quantitative and qualitative evidence that the implemented model accurately captures the underlying physical mechanisms, conserves surfactant mass, and behaves as expected under adaptative mesh refinement. They also serve to establish confidence in the numerical approach before extending it to more challenging and physically rich simulations. In particular, the first two test cases aim to verify that the formulation consistently recovers the classical insoluble surfactant behavior when adsorption and desorption are switched off. Note that all the code used for the simulations of this section is available on \citep{haouche_sandbox} for reproduction of results.

\subsection{Expanding sphere in three dimensions}
We first consider the expansion of a spherical interface covered with a uniform initial distribution of insoluble surfactant in a three-dimensional domain as a canonical test case \citep{stone1990}. A sphere of diameter \( d_s \) is initialized at the center of a cubic domain, and a radial velocity field is imposed as:
\begin{gather}
\mathbf{u}(r, \theta, \phi) = K r\, \mathbf{e}_r,
\end{gather}
where \( K \) is a constant, \( r \) is the radial distance from the center, \( \mathbf{e}_r \) is the radial unit vector. To model insoluble surfactants, the adsorption and desorption terms are turned off (leading to $J=0$). Considering the uniform distribution of surfactant and the radial and isotropic velocity field, the diffusion and advection terms cancel and equation (\ref{Eqst}) simplifies into:
\begin{equation}
    \frac{d \Gamma}{d t} + 2K \Gamma = 0.
\end{equation}
This equation admits the analytical solution:
\begin{equation}
\Gamma(t) = \Gamma_0 \, e^{-2 K t}, \\
\end{equation}
reflecting the surface area growth of the sphere due to isotropic expansion, with $\Gamma_0 = \Gamma(0)$. In dimensionless form, this equation becomes:
\begin{equation}
\Tilde{\Gamma}\big(\tilde{t}\big) = \frac{\Gamma\big(t\big)}{\Gamma_0} = e^{-2\Tilde{t}},
\end{equation}
with $\tilde{t} = Kt$.

We numerically reproduce this test case using three uniform grid resolutions: \( N = 32 \), \( 64 \), and \( 128 \), and set the initial interfacial concentration to \( \Gamma_0 = 1 \). The surface Péclet number, defined as \( \mathrm{Pe}_f = K d_s^2 / D_f \), is set to \( 16 \) in all cases.

\begin{figure}[h!]
  \centering
  \begin{subfigure}[t]{0.32\textwidth}
    \centering
    \includegraphics[width=\textwidth]{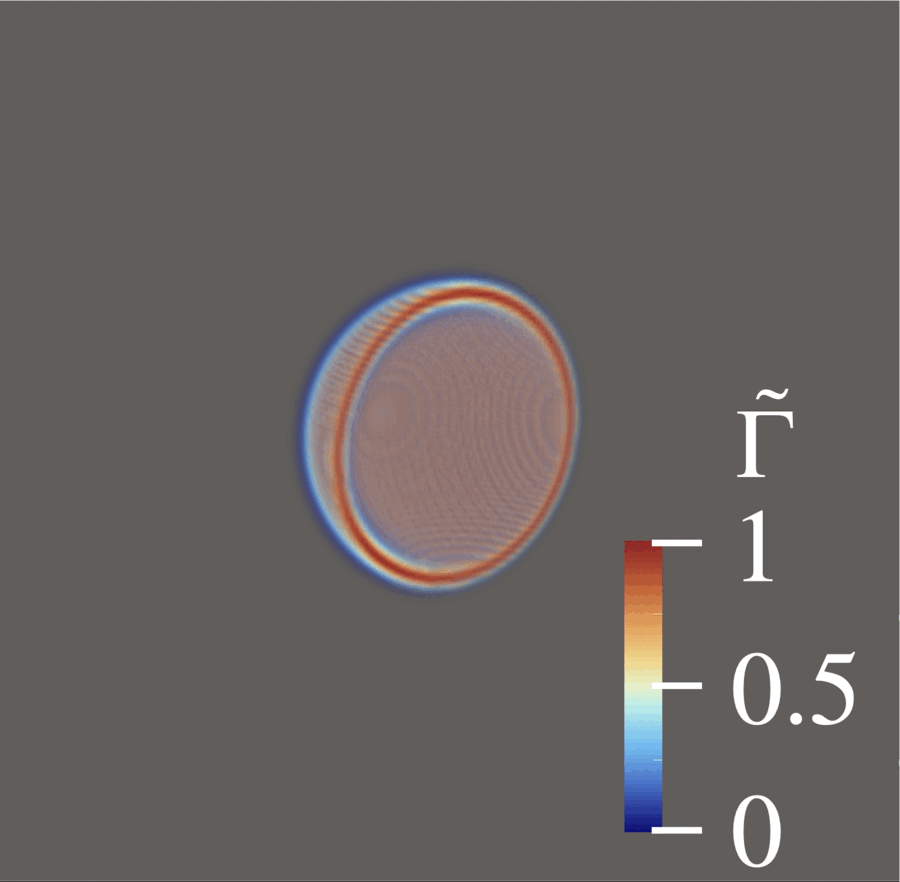}
    \caption{$\tilde{t}=0$}
    \label{fig:sub1}
  \end{subfigure}
  \hfill
  \begin{subfigure}[t]{0.32\textwidth}
    \centering
    \includegraphics[width=\textwidth]{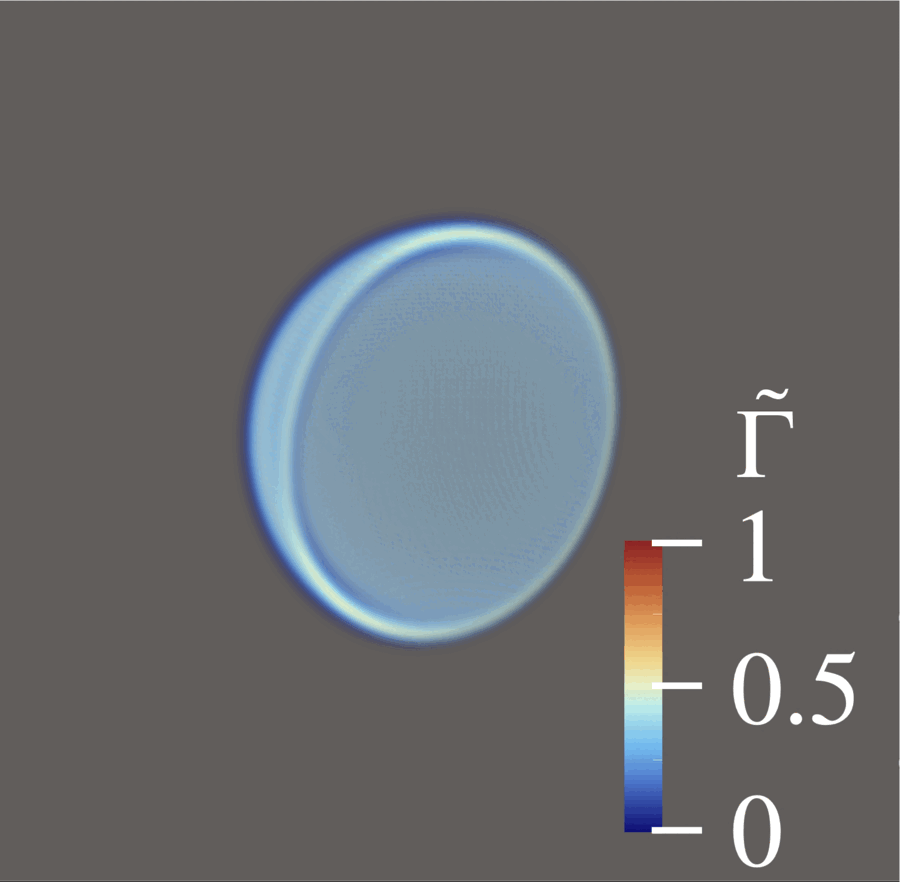}
    \caption{$\tilde{t}=0.35$}
    \label{fig:sub2}
  \end{subfigure}
  \hfill
  \begin{subfigure}[t]{0.32\textwidth}
    \centering
    \includegraphics[width=\textwidth]{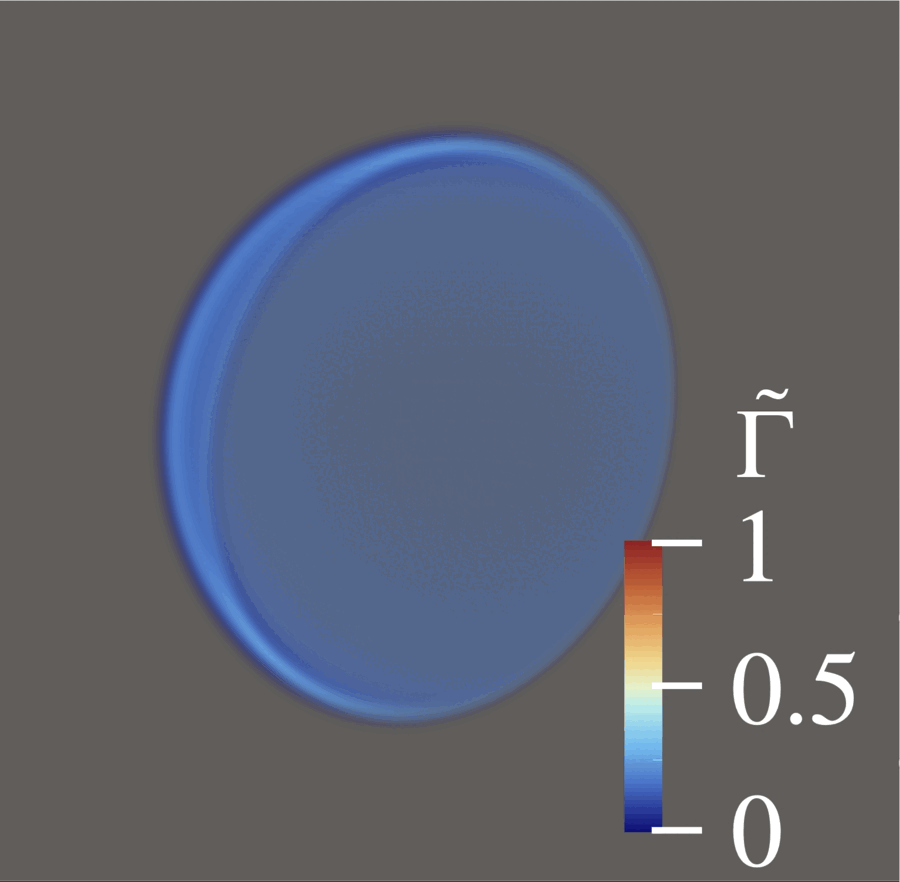}
    \caption{$\tilde{t}=0.6$}
    \label{fig:sub3}
  \end{subfigure}
  \caption{Snapshots of the dimensionless interfacial surfactant concentration \(\tilde{\Gamma}\) on an expanding sphere at (a) the initial time \( \tilde{t}=Kt = 0\), (b) an intermediate time \( \tilde{t}= 0.35\), and (c) the final simulation time \( \tilde{t}= 0.6\), for the highest resolution \(N = 128\). The imposed radial flow causes the sphere to expand uniformly, resulting in a decrease of surfactant concentration due to interface stretching. The color scale represents the magnitude of \(\tilde{\Gamma}\big(\tilde{t}\big)\).
}
  \label{fig:expansionsnap}
\end{figure}

\begin{figure}[h!]
    \centering
            \begin{overpic}[width=1\linewidth]{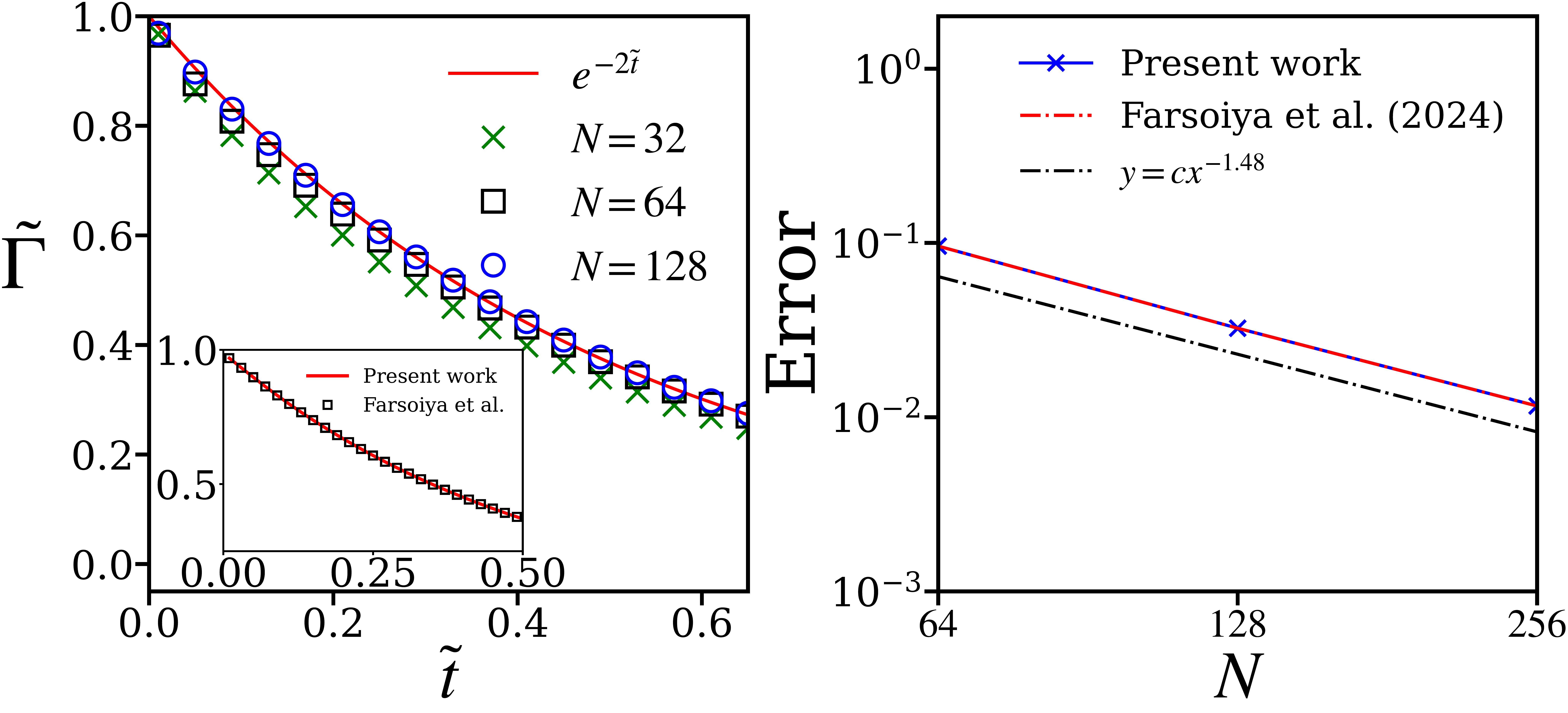}
        \put(1,44.5){\scriptsize (a)}
        \put(51,44.5){\scriptsize (b)}
     \end{overpic}
    \caption{(a) Time evolution of the interfacial surfactant concentration \(\Tilde{\Gamma}(\Tilde{t})\) for three different uniform grid resolutions: \(N = 32\), \(64\), and \(128\), compared with the analytical solution \(\tilde{\Gamma}\big(\tilde{t}\big) = e^{-2\tilde{t}}\) (solid red line). As the resolution increases, the numerical results converge toward the analytical solution, demonstrating the consistency and accuracy of the method. (b) Corresponding relative error \(|\tilde{\Gamma}_{\text{num}} - \tilde{\Gamma}_{\text{ana}}|/\tilde{\Gamma}_{\text{ana}}\) as a function of grid resolution, with $\tilde{\Gamma}_{\text{num}}$ the surfactant concentration obtained numerically and $\tilde{\Gamma}_{\text{ana}}$ the analytical solution.}
    \label{fig:expansion}
\end{figure}

Fig.~\ref{fig:expansionsnap}(a-c) shows the initial, intermediate, and final distributions of the interfacial dimensionless surfactant concentration $\Tilde{\Gamma}\big(\tilde{t}\big)$ on the expanding sphere for the highest resolution \( N = 128 \). Fig.~\ref{fig:expansion}(a) compares the temporal evolution of \( \Tilde{\Gamma}(\Tilde{t}) \) with the analytical solution for all three resolutions. Fig.~\ref{fig:expansion}(b) displays the corresponding relative error, confirming convergence toward the theoretical prediction as the resolution increases.

\subsection{Validation of the advection–diffusion equation on a rotating interface}

To verify the accuracy of the advection--diffusion solver for insoluble surfactants on moving interfaces, we consider a canonical test case in which a circular bubble of diameter $d_b$, initially covered by a non-uniform interfacial distribution of surfactant $\tilde{\Gamma}(\tilde{\theta},0) = 2 + \sin\tilde{\theta}$, is placed in a solid-body rotational flow \citep{jain2025model}:
\[
\mathbf{u}\bigg(r=\frac{d_b}{2},\theta\bigg) = \frac{d_b \omega}{2}\,\mathbf{e}_\theta,
\]
which induces a rigid-body rotation of the interface at a constant angular velocity \(\omega\).

Adsorption and desorption are disabled, so that only interfacial transport is active leading to the following simplified version of equation (\ref{Eqst}):
\begin{equation}
\frac{\partial \tilde{\Gamma}}{\partial \tilde{t}}
+ \frac{\partial \tilde{\Gamma}}{\partial \tilde{\theta}}
=
\frac{4}{\mathrm{Pe}_f} \frac{\partial^2 \tilde{\Gamma}}{\partial \tilde{\theta}^2}
\end{equation}
This test case enables direct comparison with an analytical solution \citep{teigen2009diffuse}:
\[
\tilde{\Gamma}(\tilde{\theta},\,\tilde{t}) = 2 + \sin\big(\tilde{\theta} - \tilde{t}\big) e^{-4 \tilde{ t} / \mathrm{Pe}_f},
\]
where the surface Péclet number is defined as \(\mathrm{Pe}_f = \omega d_b^2 / D_f\), with \(D_f\) the interfacial diffusivity and $\tilde{t}$ defined as $\tilde{t}=t\omega/2\pi$. 

\begin{figure}[h!]
    \centering
        \begin{overpic}[width=1\linewidth]{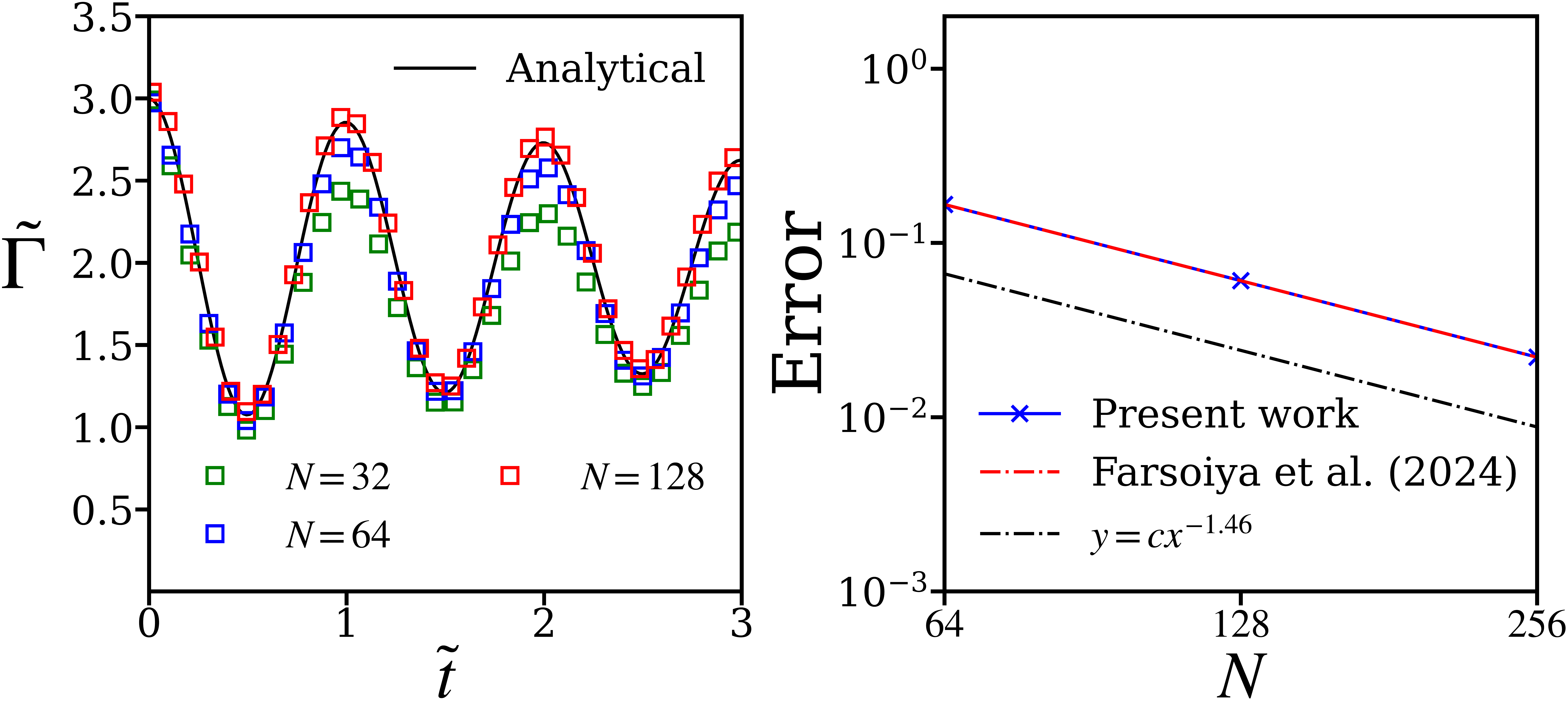}
        \put(8,47){\scriptsize (a)}
        \put(60,47){\scriptsize (b)}
    \end{overpic}
    \caption{(a) Temporal evolution of the interfacial surfactant concentration \( \tilde{\Gamma}(\tilde{t})\) at three uniform grid resolutions (\(N = 32, 64, 128\)), compared with the analytical solution \( \tilde{\Gamma}(\tilde{\theta},\tilde{t}) = 2 + \sin(\tilde{\theta} + \tilde{t}) e^{-4\tilde{t}/\mathrm{Pe}_f} \). As the resolution increases, numerical results converge toward the analytical prediction, demonstrating the accuracy and consistency of the interfacial transport solver. (b) Relative \ error \( |\tilde{\Gamma}_{\text{num}} - \tilde{\Gamma}_{\text{ana}}|/\tilde{\Gamma}_{\text{ana}} \) as a function of the grid resolution.}
    \label{fig:rotation}
\end{figure}

Numerically, a circular bubble of diameter \(d_b\) is initialized at the center of a square 2D domain of size \(\big[L_x \times L_y\big] = [2 d_b \times 2 d_b ] \), with a smooth Phase-Field profile. Simulations are performed for \(Pe_f = 160\), using several uniform grid resolutions. Fig.~\ref{fig:rotation}(a) shows the time evolution of the surfactant profile and confirms the excellent agreement with the analytical solution at different times. Relative errors and convergence rates are also computed Fig.~\ref{fig:rotation}(b), showing close to second-order accuracy with respect to grid refinement.

\begin{figure}[h!]
  \centering
  \begin{subfigure}[t]{0.32\textwidth}
    \centering
    \includegraphics[width=\textwidth]{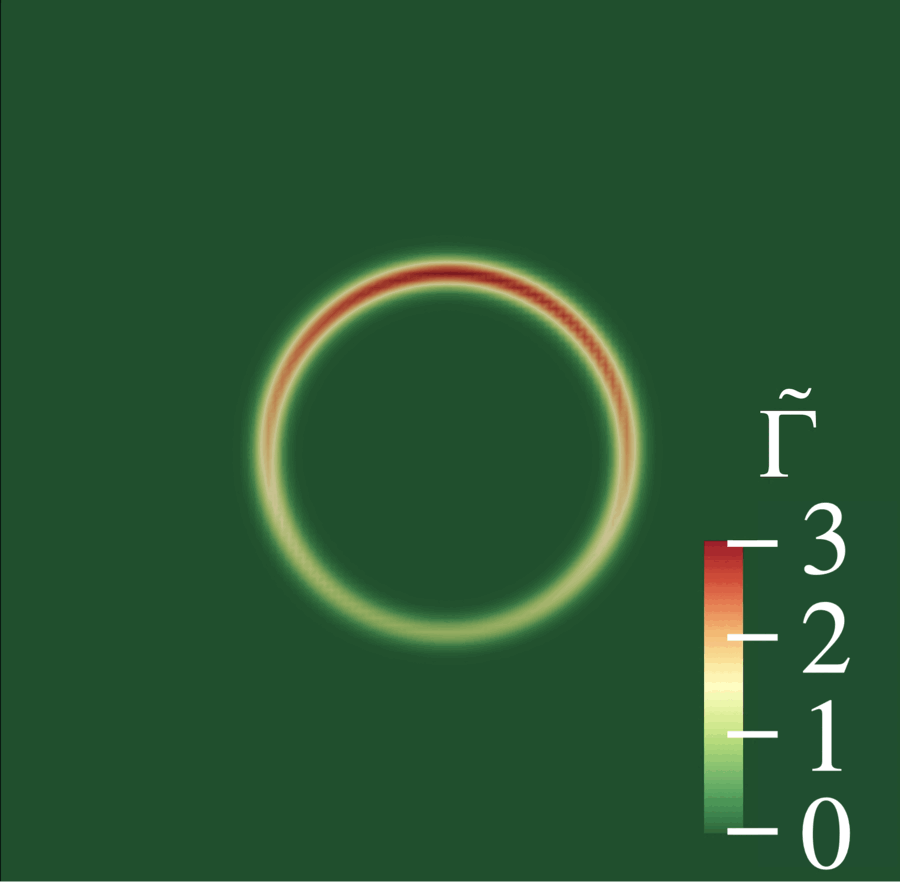}
    \caption{$\tilde{t}=0$}
    \label{fig:sub1}
  \end{subfigure}
  \hfill
  \begin{subfigure}[t]{0.32\textwidth}
    \centering
    \includegraphics[width=\textwidth]{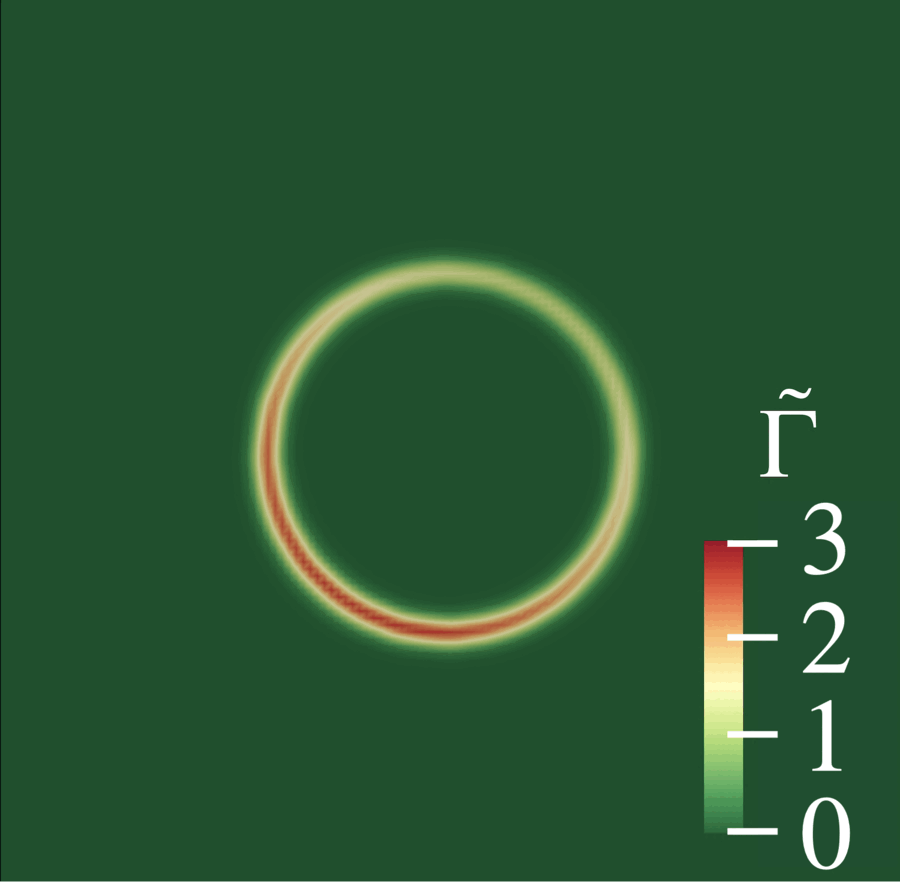}
    \caption{$\tilde{t}=0.35$}
    \label{fig:sub2}
  \end{subfigure}
  \hfill
  \begin{subfigure}[t]{0.32\textwidth}
    \centering
    \includegraphics[width=\textwidth]{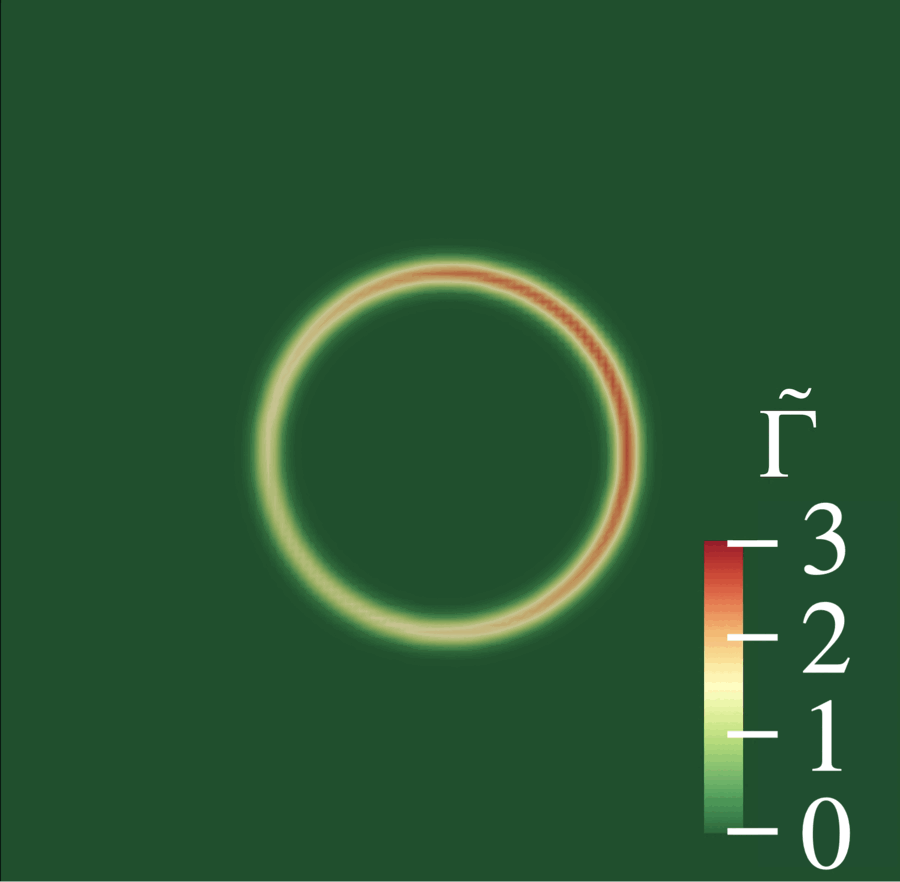}
    \caption{$\tilde{t}=0.6$}
    \label{fig:sub3}
  \end{subfigure}
  \caption{Snapshots of the dimensionless interfacial surfactant concentration \(\tilde{\Gamma} \) on an rotating circle at (a) initial time \( \tilde{t} = 0\), (b) intermediate time \( \tilde{t} = 0.35\), and (c) final simulation time \( \tilde{t} = 0.6\) for the highest resolution \(N = 128\). The color scale represents the magnitude of \(\tilde{\Gamma}\), with initially non-uniform distribution evolving according to the analytical prediction \(\tilde{\Gamma}(\tilde{\theta},\tilde{t}) = 2 + \sin\big(\tilde{\theta} + \tilde{t}\big) e^{-4 \tilde{ t} / \mathrm{Pe}_f}\). 
}
  \label{fig:rotationsnap}
\end{figure}
Together with the expanding sphere test, this canonical example confirms the accuracy of the interfacial transport solver in the absence of adsorption and desorption, thereby validating the insoluble surfactant limit of our model. These benchmarks demonstrate that disabling the kinetic source terms reliably recovers classical insoluble dynamics, ensuring that the soluble formulation remains consistent and robust across limiting regimes.

Having established that our model faithfully reproduces the behavior of insoluble surfactants, we now turn to the fully coupled soluble case. In this regime, both interfacial and bulk concentrations evolve simultaneously through advection, diffusion, and kinetic exchange. The resulting dynamics are significantly richer, as adsorption and desorption processes redistribute surfactant between the interface and the surrounding fluid, thereby modifying Marangoni stresses and the overall hydrodynamic response. The following test cases are designed to assess the accuracy of the coupled solver, explore the role of kinetic parameters in recovering clean and insoluble limits, and demonstrate the ability of the method to capture the full complexity of soluble surfactant transport in deforming multiphase flows.

\subsection{Adsorption--desorption on a flat interface}

We now consider the transport of a soluble surfactant in the vicinity of a flat, stationary interface in 2D. This configuration is used to validate the bulk--interface coupling terms in a simplified setting, without hydrodynamic motion or interface deformation. All lengths are scaled by a characteristic length \(L\), and tilded variables denote dimensionless quantities.

The computational domain is a rectangular box of size \(\big[-4L,4L\big]\times\big[-4L,4L\big] \), containing a flat interface initially located at \(\tilde{y}=0\), where $(\tilde{x},\tilde{y})$ denote the horizontal and vertical cartesian coordinates respectively. No flow is imposed, i.e. \(\mathbf{u}=\mathbf{0}\), so that the surfactant dynamics are governed solely by bulk diffusion and interfacial exchange. Since the interface remains flat and motionless, the problem is effectively one-dimensional in the direction normal to the interface.

The purpose of this test is to assess the numerical treatment of the source term under three different kinetic laws: (i) pure adsorption solely dependent on the bulk concentration evaluated at the interface, $F_s$, (ii) Langmuir-type adsorption and (iii) pure desorption.

In all cases, the governing equations reduce to a diffusion equation in the bulk coupled to an ordinary differential equation for the interfacial concentration. The bulk concentration $F\big(y,t\big)$ satisfies:
\begin{equation}
    \frac{\partial F}{\partial t}
    =
    D_F \frac{\partial^2 F}{\partial y^2}
    - j,
\end{equation}
while the interfacial concentration evolves according to:
\begin{equation}
    \frac{d\Gamma}{dt} = J.
\end{equation}

The initial and boundary conditions depend on the kinetic regime considered. 

\subsubsection{Constant adsorption flux}

We consider the canonical adsorption model \( J = r_a F_s \Gamma_\infty \) introduced in \citep{muradoglu2008front, teigen2009diffuse}. In contrast to the original studies, which were conducted in spherical \citep{muradoglu2008front} and cylindrical \citep{teigen2009diffuse} geometries, we examine here a two-dimensional configuration with a flat interface. In this case, Eqs. \eqref{Eqst} and \eqref{eqbulk}  reduce to the following dimensionless form:
\begin{align}
    \frac{\partial \tilde{F}}{\partial \tilde{t}} &= \frac{\partial^2 \tilde{F}}{\partial \tilde{y}^2} - \mathrm{Da} \tilde{F}_s  \tilde{\delta}_\phi, \\
    \frac{d \tilde{\Gamma}}{d \tilde{t}} &= \tilde{F}_s,
\end{align}
using the scalings:
\begin{gather}
    \tilde{y} = \frac{y}{L} = 
    \frac{F_\infty}{\Gamma_\infty} y, \quad
    \tilde{t} = \frac{D_F F_\infty^2}{\Gamma_\infty^2} t, \quad   
    \tilde{F} = \frac{F}{F_\infty}, \quad
    \tilde{\Gamma} = \frac{\Gamma}{\Gamma_\infty},
\end{gather}
and with $\mathrm{Da} = r_a \Gamma_\infty^2 / D_F F_\infty$ the Damköhler number comparing the adsorption kinetic and bulk diffusion and $F_\infty$ the bulk concentration far from the interface.

The bulk is initialized with a uniform concentration \( \tilde{F}(0) = 1 \), while the interface is initially clean, \( \tilde{\Gamma}(0) = 0 \). Analytical solutions can be derived and read:
\begin{align}
\tilde{F}(\tilde{y}, \tilde{t}) &=
\operatorname{erf}\left(\frac{\tilde{y} }{2\sqrt{\tilde{t}}}\right)
+ \exp \bigg( \mathrm{Da}(\tilde{y}+\mathrm{Da}\tilde{t}) \bigg)
\operatorname{erfc}\left(\frac{\tilde{y}}{2\sqrt{\tilde{t}}} + \mathrm{Da} \sqrt{\tilde{t}}\right), \\
\tilde{\Gamma}(\tilde{t}) &=
\frac{1}{\mathrm{Da}} \left[ \exp \big(\mathrm{Da}^2 \tilde{t} \big) \operatorname{erfc}\left(\mathrm{Da} \sqrt{ \tilde{t}}\right)
- 1 + \frac{2 \mathrm{Da} \sqrt{\tilde{t}}}{\sqrt{\pi}} \right].
\end{align}

\begin{figure}
    \centering
    \includegraphics[width=0.5\linewidth]{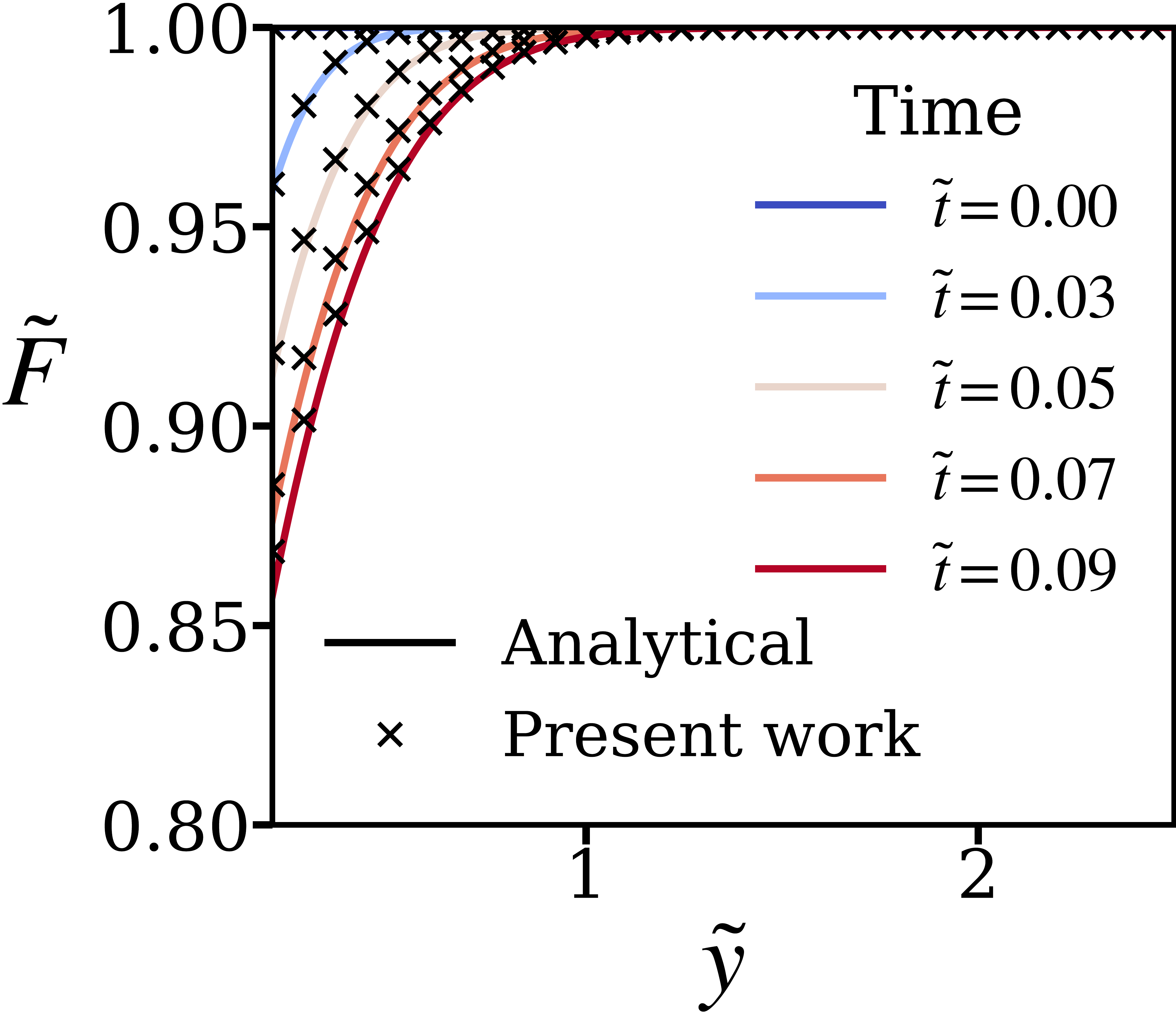}
    \caption{Spatial profiles of the dimensionless bulk surfactant concentration \( \tilde{F}(\tilde{y}, \tilde{t}) \) at different dimensionless times $\tilde{t}$ compared with the analytical solution.}
    \label{fig:cbevolution}
\end{figure}

\begin{figure}[h!]
    \centering
    \begin{overpic}[width=1\linewidth]{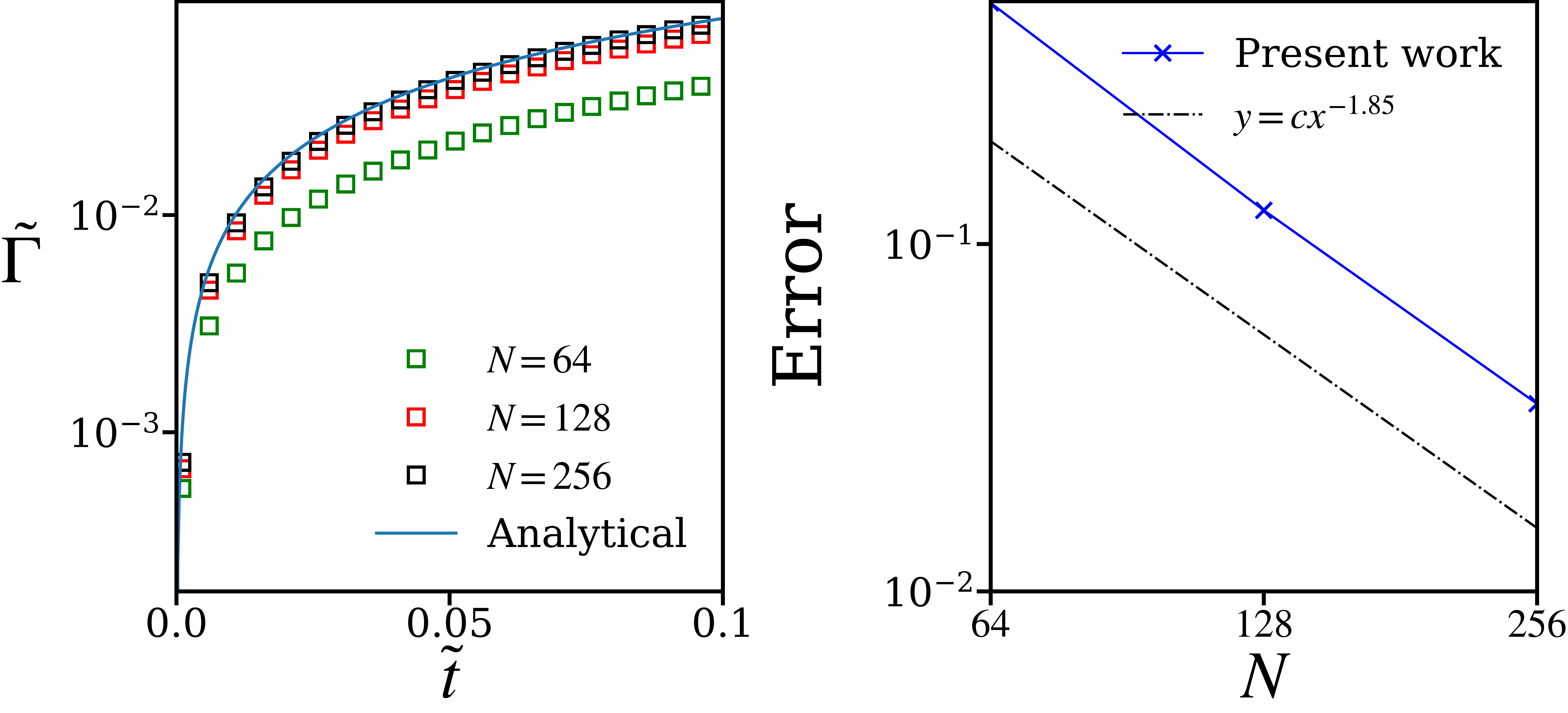}
        \put(7,45){\scriptsize (a)}
        \put(58,45){\scriptsize (b)}
    \end{overpic}
    \caption{Validation of surfactant adsorption on a stationary flat interface in 2D. 
(a) Temporal evolution of the dimensionless interfacial surfactant $\tilde{\Gamma}$ concentration for three different uniform resolutions $N$, showing convergence toward the analytical solution (blue solid line). 
(b) Relative error between numerical and analytical solutions, confirming the convergence behavior with mesh refinement.}
    \label{fig:gammaadsorption}
\end{figure}

The spatial profiles of the bulk concentration at different times are simulated for $\mathrm{Da} = 1$ (see Fig.~\ref{fig:cbevolution}) and compared to the analytical solution. The numerical simulation converges toward the analytical solution when the mesh is refined. The temporal evolution of the interfacial concentration and the corresponding relative error are presented in Fig.~\ref{fig:gammaadsorption}. The results also exhibit convergence behavior with mesh refinement.

\begin{figure}[h!]
  \centering
  \begin{subfigure}[t]{0.42\textwidth}
    \centering
    \includegraphics[width=\textwidth]{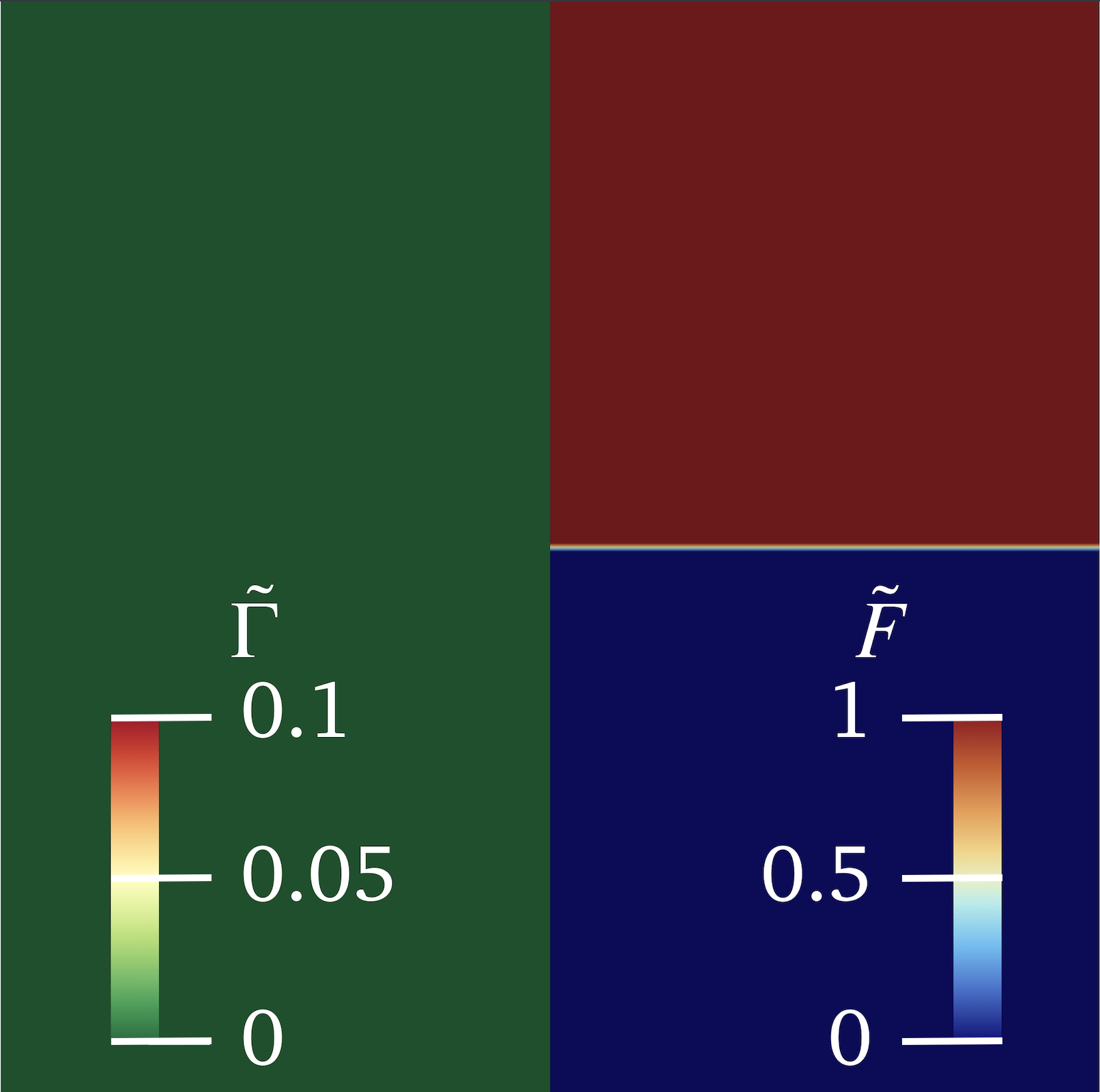}
    \caption{\( \tilde{t} = 0 \)}
  \end{subfigure}
  \hspace{0.04\textwidth}
  \begin{subfigure}[t]{0.42\textwidth}
    \centering
    \includegraphics[width=\textwidth]{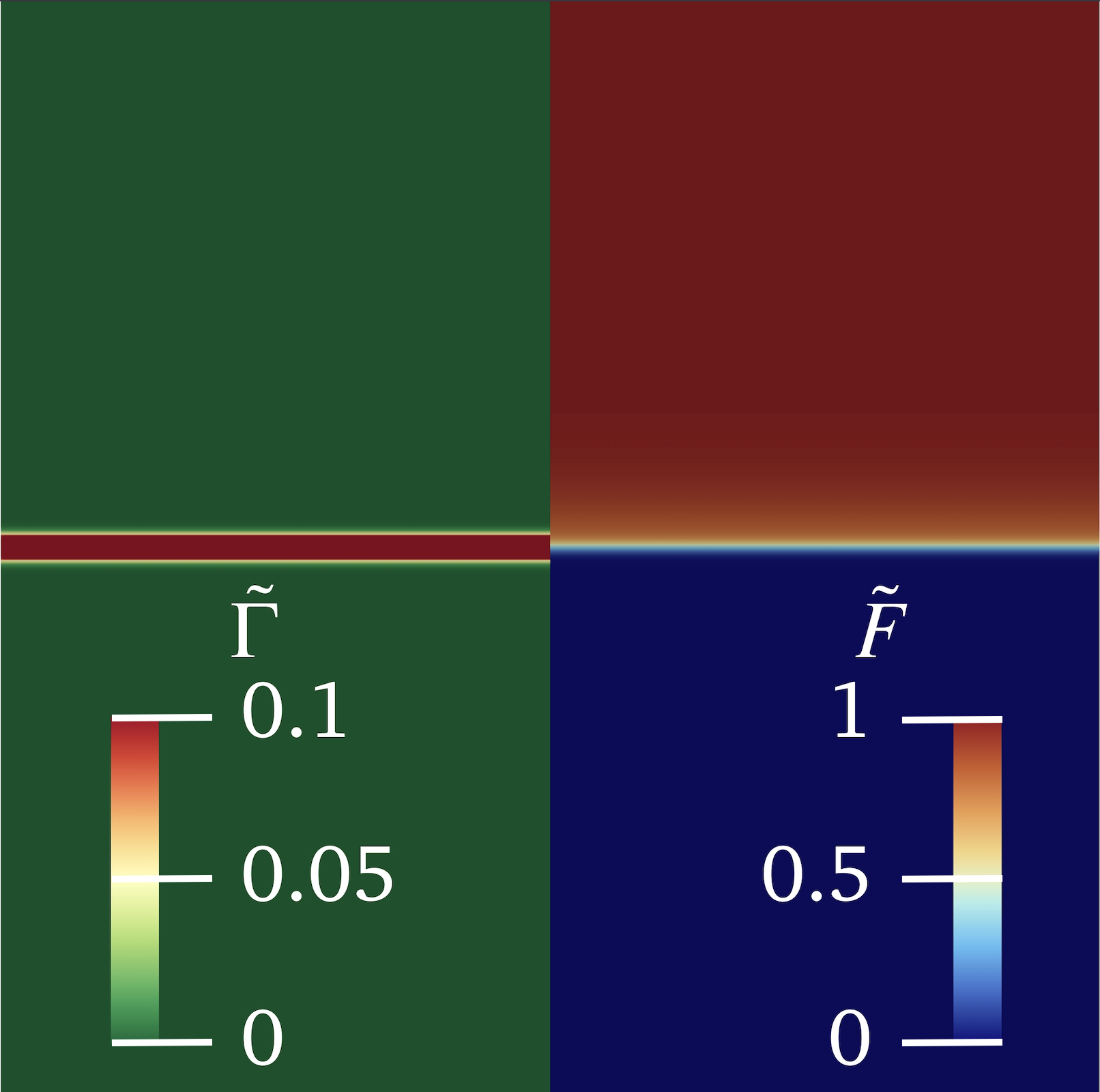}
    \caption{\( \tilde{t} = 0.1 \)}
  \end{subfigure}

  \caption{Snapshots of the interfacial (left) and bulk (right) dimensionless surfactant concentrations during the adsorption process. 
(a) Initial distributions corresponding to a clean interface and a uniform bulk concentration. 
(b) Final distributions ($\tilde{t} = 0.1$) showing surfactant accumulation along the interface and the formation of a depletion region in the bulk near the interface due to adsorption.}
  \label{fig:adsorptionsnap}
\end{figure}

Finally, snapshots of the interfacial and bulk concentrations are displayed in Fig.~\ref{fig:adsorptionsnap}, illustrating the progressive accumulation of surfactant along the interface and the formation of a depletion region in the bulk near the interface.

\subsubsection{Pure desorption}

We now consider the pure desorption regime from an initially loaded interface at interfacial concentration $\Gamma_0$, for which the source term reduces to \( J = -r_d \Gamma \). In this configuration, surfactant is released from the interface into the bulk, in the absence of adsorption.

Using the following scalings:
\begin{gather}
    \tilde{y} = \frac{F_*}{\Gamma_0}y, \quad
    \tilde{t} = \frac{D_FF_*^2}{\Gamma_0^2} t, \quad
    \tilde{F} = \frac{F}{F_*}, \quad
    \tilde{\Gamma} = \frac{\Gamma}{\Gamma_0},
\end{gather}
with $F_*$ a reference bulk concentration, the governing equations become:
\begin{align}
    \frac{\partial \tilde{F}}{\partial \tilde{t}} &= \frac{\partial^2 \tilde{F}}{\partial \tilde{y}^2} + \mathrm{Bi}\tilde{\Gamma} \tilde{\delta}_\phi, \\
    \frac{d \tilde{\Gamma}}{d \tilde{t}} &= -\mathrm{Bi}\tilde{\Gamma}.
\end{align}
with $\mathrm{Bi}=r_d\Gamma_0^2/D_FF_*^2$ the Biot number comparing the desorption and bulk diffusion. Initially, we assume (i) no surfactant in the bulk, $\tilde{F}\big(0\big)=0$ and (ii) an interface loaded with a uniform concentration of surfactant, $\tilde{\Gamma}\big(0\big)=1$. The analytical solutions is given by:
\begin{align}
\tilde{F}(\tilde{y}, \tilde{t}) &= \frac{\mathrm{Bi}}{\sqrt{\pi}} \, e^{-\mathrm{Bi}\tilde{t}} 
\int_0^{\tilde{t}} 
\frac{\exp\left(\mathrm{Bi}\tilde{\tau} - \frac{\tilde{y}^2}{4\tilde{\tau}}\right)}
{\sqrt{\tilde{\tau}}}
\, d\tilde{\tau}, \\
\tilde{\Gamma}(\tilde{t}) &= e^{-\mathrm{Bi}\tilde{t}}.
\end{align}
The numerical simulations are performed for $\mathrm{Bi}=1$. As time evolves, surfactant is released from the interface and diffuses into the bulk, leading to a progressive decrease of the interfacial concentration and a corresponding enrichment of the bulk phase.

\begin{figure}[h!]
    \centering
    \begin{overpic}[width=1\linewidth]{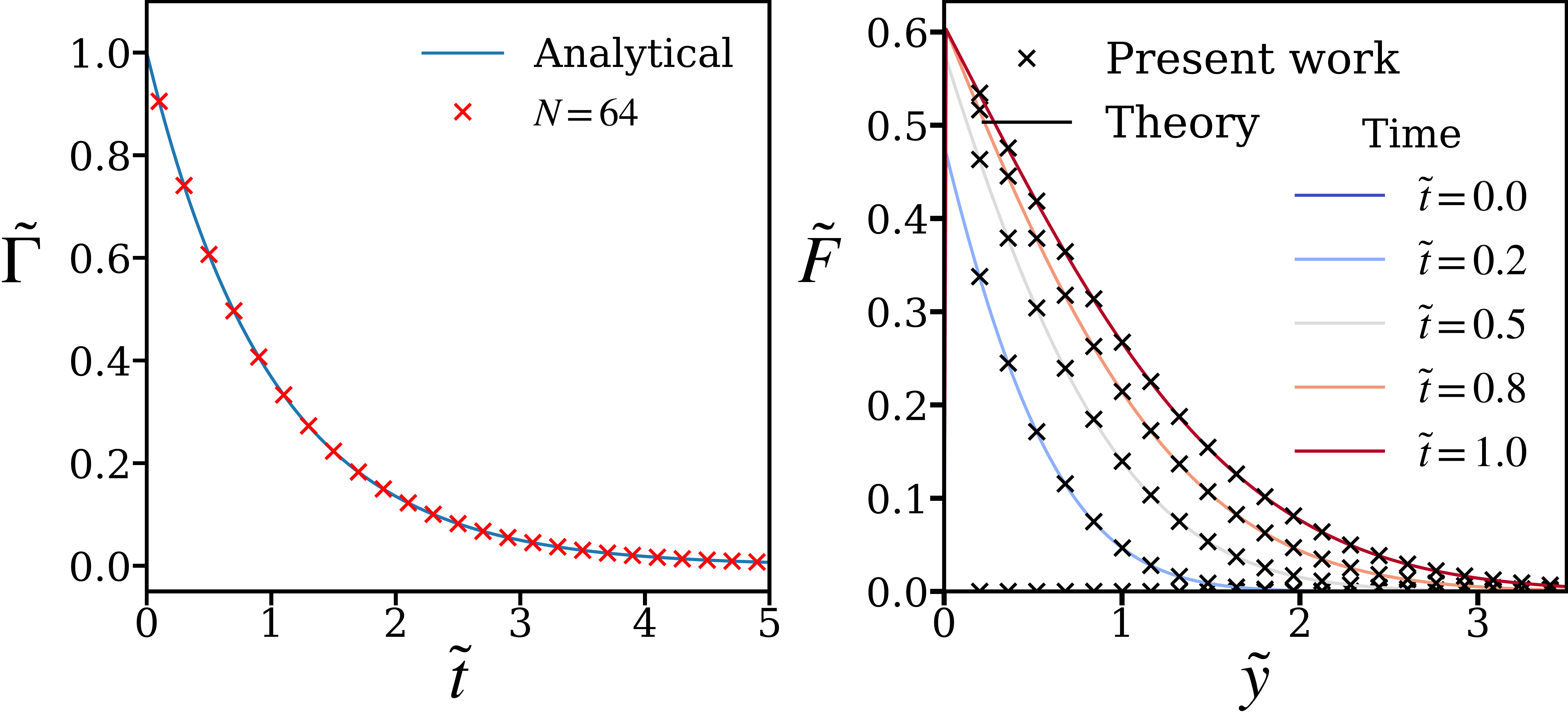}
        \put(7,46.5){\scriptsize (a)}
        \put(58,46.5){\scriptsize (b)}
    \end{overpic}
    \caption{Validation of surfactant desorption on a stationary flat interface in 2D. 
(a) Temporal evolution of the interfacial surfactant concentration \( \tilde{\Gamma}(\tilde{t}) \), showing convergence toward the analytical solution. 
(b) Spatial profiles of the bulk concentration \( \tilde{F}(\tilde{y}, \tilde{t}) \) at different time instants, illustrating the diffusion of surfactant released from the interface into the bulk.}
    \label{fig:desorption}
\end{figure}

\begin{figure}[h!]
  \centering
  \begin{subfigure}[t]{0.42\textwidth}
    \centering
    \includegraphics[width=\textwidth]{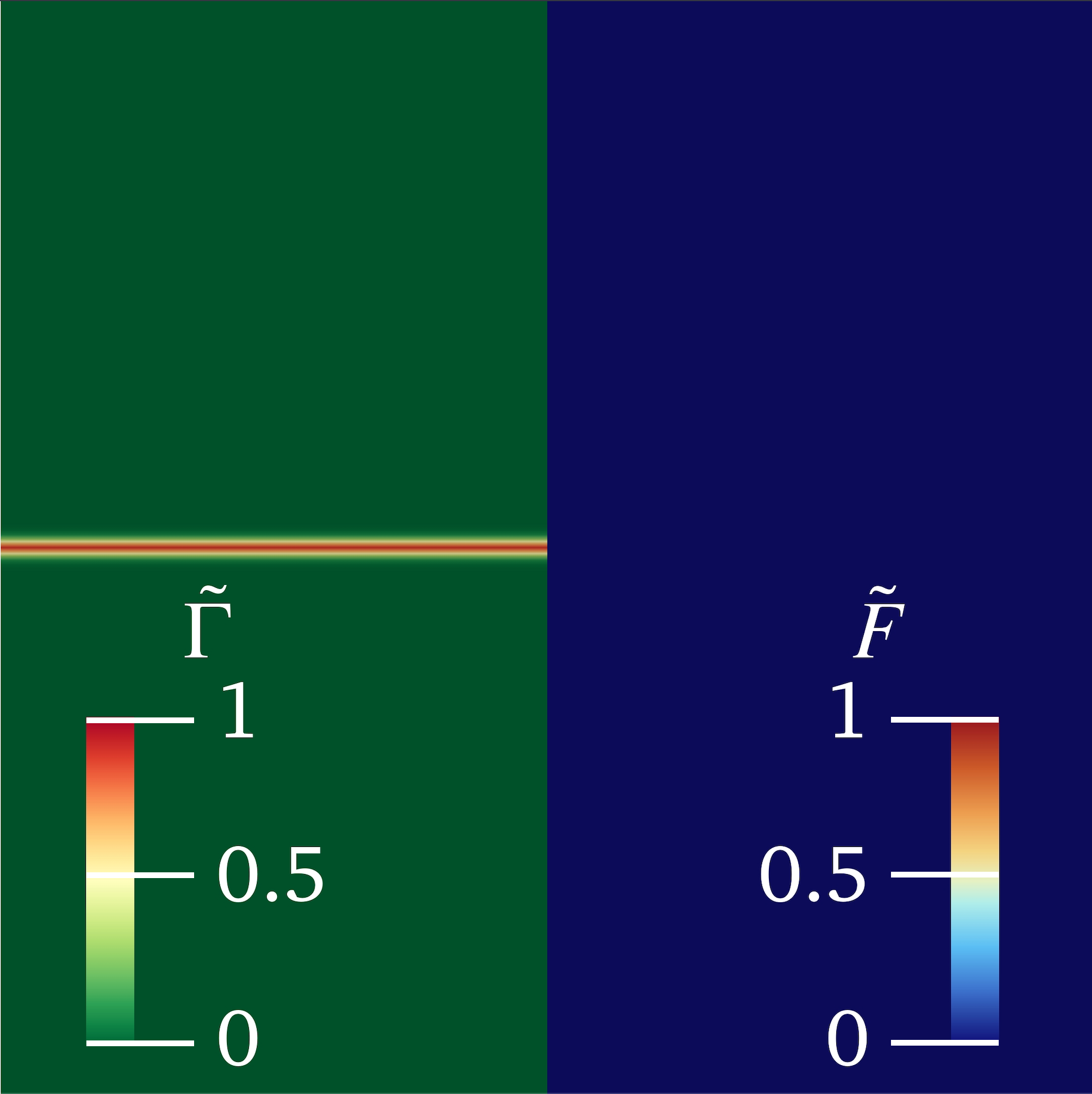}
    \caption{\( \tilde{t} = 0 \)}
  \end{subfigure}
  \hspace{0.04\textwidth}
  \begin{subfigure}[t]{0.42\textwidth}
    \centering
    \includegraphics[width=\textwidth]{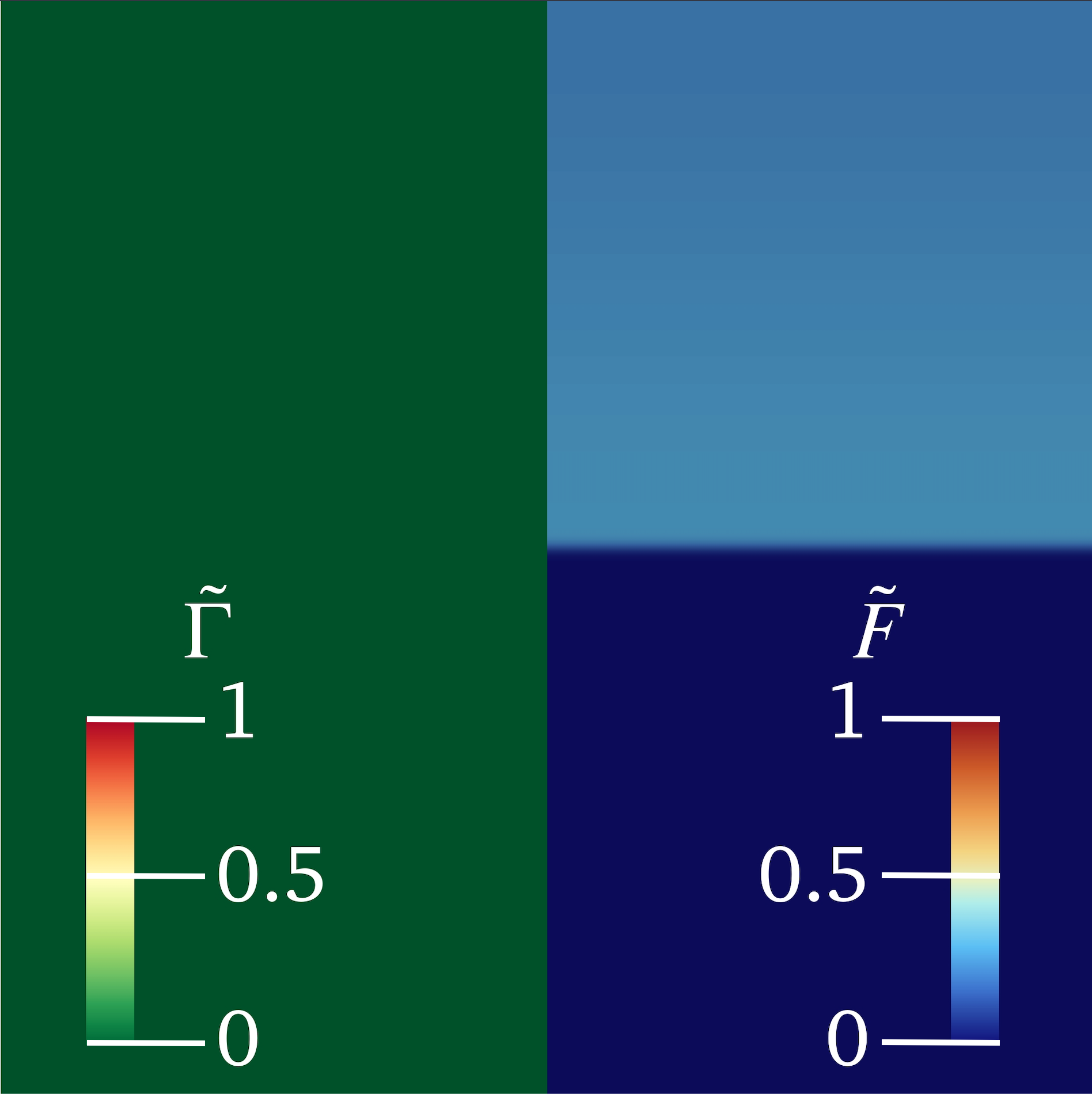}
    \caption{\( \tilde{t} = 0.1 \)}
  \end{subfigure}

  \caption{Snapshots of the interfacial and bulk surfactant concentrations during the desorption process. 
(a) Initial distributions corresponding to a uniformly covered interface and a clean bulk. 
(b) Final distributions ($\tilde{t} = 0.1$) showing the depletion of surfactant at the interface and the corresponding enrichment of the bulk near the interface due to desorption.}
  \label{fig:desorptionsnap}
\end{figure}

The temporal evolution of the interfacial concentration and the spatial distribution of the bulk concentration are shown in Fig.~\ref{fig:desorption}. The results demonstrate excellent agreement with the analytical solution. Finally, snapshots of the interfacial and bulk concentrations are displayed in Fig.~\ref{fig:desorptionsnap}, illustrating the progressive depletion of surfactant at the interface and the corresponding enrichment of the bulk in the vicinity of the interface due to desorption.

\subsubsection{Langmuir adsorption--desorption kinetics}

We now consider the full Langmuir adsorption--desorption model, which accounts for both saturation effects at the interface and reversible exchange with the bulk $J = r_a F_s (\Gamma_\infty - \Gamma) - r_d \Gamma$.

This formulation describes adsorption onto a finite number of interfacial sites with a maximum packing concentration \(\Gamma_\infty\), together with desorption back into the bulk. The competition between adsorption and desorption leads to a nonlinear evolution of the interfacial concentration and, at equilibrium, recovers the classical Langmuir isotherm.

Using the scalings
\begin{gather}
     \tilde{y} = y \frac{\Gamma_\infty}{F_\infty}, \quad
     \tilde{t} = t \frac{D_F F_\infty^2}{\Gamma_\infty^2}, \quad
     \tilde{F} = \frac{F}{F_\infty}, \quad
     \tilde{\Gamma} = \frac{\Gamma}{\Gamma_\infty},
\end{gather}
the governing equations become
\begin{align}
    \frac{\partial \tilde{F}}{\partial \tilde{t}} &= \frac{\partial^2 \tilde{F}}{\partial \tilde{y}^2}
    - \left[ \mathrm{Da} \, \tilde{F}_s (1 - \tilde{\Gamma}) - \mathrm{Bi}\, \tilde{\Gamma} \right]\tilde{\delta}_\phi, \\
    \frac{d \tilde{\Gamma}}{d \tilde{t}} &= \mathrm{Da} \, \tilde{F}_s (1 - \tilde{\Gamma}) - \mathrm{Bi}\, \tilde{\Gamma},
\end{align}
where:
\begin{equation}
    \mathrm{Da} = \frac{r_a \Gamma_\infty^2}{D_F F_\infty},
    \qquad
    \mathrm{Bi} = \frac{r_d \Gamma_\infty^2}{D_F F_\infty^2}.
\end{equation}
Together, these parameters control the balance between interfacial accumulation, bulk depletion, and reversible mass transfer.

At equilibrium, assuming a nearly constant bulk concentration at the interface ($F_s \approx F_\infty$), adsorption and desorption balance, yielding
\begin{equation}
    \Gamma_{eq} = \frac{r_a F_\infty}{r_a F_\infty + r_d}\,\Gamma_\infty,
\end{equation}
or, in dimensionless form:
\begin{equation}
    \tilde{\Gamma}_{eq} = \frac{\mathrm{Da}}{\mathrm{Da} + \mathrm{Bi}}.
\end{equation}

\begin{figure}[h!]
    \centering
    \begin{overpic}[width=1\linewidth]{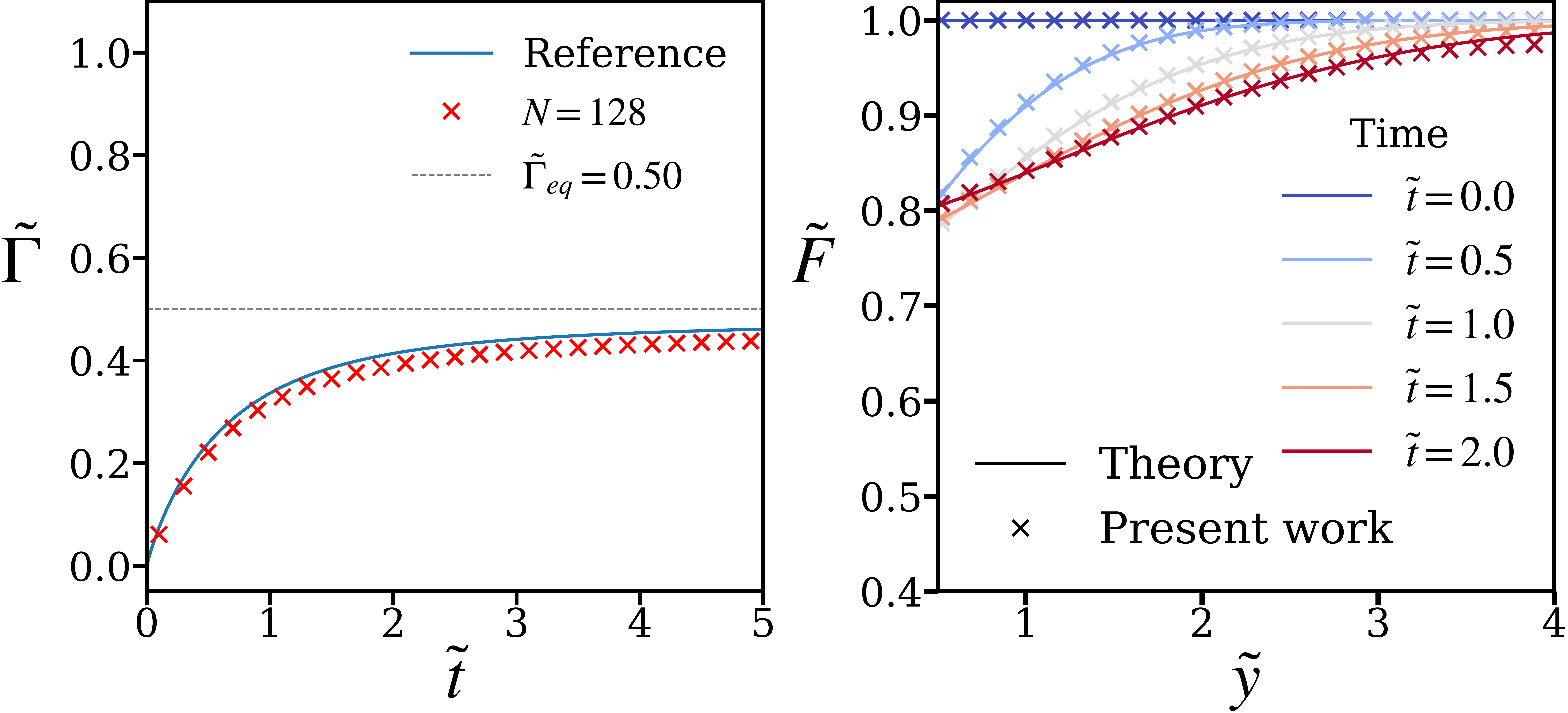}
        \put(7,46.5){\scriptsize (a)}
        \put(58,46.5){\scriptsize (b)}
    \end{overpic}
    \caption{Validation of Langmuir adsorption--desorption kinetics on a stationary flat interface in 2D. 
(a) Temporal evolution of the interfacial surfactant concentration \( \tilde{\Gamma}(\tilde{t}) \), showing convergence toward the reference solution and toward equilibrium.
(b) Spatial profiles of the bulk concentration \( \tilde{F}(\tilde{y}, \tilde{t}) \) at different dimensionless times $\tilde{t}$, illustrating the coupled adsorption--desorption dynamics and bulk depletion near the interface.}
    \label{fig:lang}
\end{figure}

The temporal evolution of the interfacial concentration and the spatial distribution of the bulk concentration obtained with the code are shown in Fig.~\ref{fig:lang}. These numerical results are compared against a reference solution obtained from a one-dimensional discretization of the reduced problem using second-order finite differences in space and an implicit backward differentiation formula (BDF) time integrator. These comparisons demonstrate the convergence toward the equilibrium interfacial concentration and the consistency of the coupled bulk--interface solver in the absence of flow.

This configuration provides a unified framework encompassing the limiting cases previously considered. In particular, the constant adsorption flux regime is recovered for \(\tilde{\Gamma} \ll 1\), while the pure desorption regime is obtained in the limit \(\mathrm{Da} \to 0\). As such, this test case enables a comprehensive validation of the source-term implementation across adsorption-, desorption-, and saturation-dominated regimes.

Overall, this test confirms the ability of the numerical method to capture nonlinear interfacial saturation effects, bulk depletion, and reversible mass transfer between the interface and the bulk, thereby validating the coupled formulation over a wide range of kinetic regimes.

Having validated the individual and coupled contributions of adsorption, desorption, and diffusion, we now turn to fully coupled simulations involving both surfactant transport and hydrodynamic motion.

\section{Effect of insoluble/soluble surfactants on the terminal rise velocity of an axisymmetric bubble}

In this test case, we investigate the buoyancy-driven motion of a deformable bubble rising in a quiescent liquid under gravity, within a two-dimensional axisymmetric domain. The dynamics of the bubble are influenced by the presence of either insoluble or soluble surfactants, which induce Marangoni stresses through local surface tension gradients. The objective is to quantify the impact of surfactant properties on the terminal rise velocity of the bubble and to explore the limiting regimes corresponding to the clean or surface-incompressible interface ($\boldsymbol{\nabla}_s\cdot \mathbf{u}_s = 0$).

The system is initialized with a spherical bubble of diameter \( d_b \), centered at the origin. The bubble is initially at rest, and gravity acts downward. The fluid motion is governed by the incompressible Navier–Stokes equations, coupled either with the advection–diffusion equation for the interfacial surfactant concentration \( \Gamma(\mathbf{x},t) \) in the insoluble case, or with coupled equations for both bulk and interfacial concentrations \( F(\mathbf{x},t) \) and \( \Gamma(\mathbf{x},t) \), including source terms accounting for adsorption and desorption processes, in the soluble case.

The surface tension depends nonlinearly on the local interfacial concentration via the following equation of state:
\begin{equation}
    \sigma\big(\mathbf{x},\,t\big) = \sigma_0 \left(1 + \mathrm{E} \ln\left[1- \frac{\Gamma\big(\mathbf{x}, \,t\big)}{\Gamma_\infty}\right]\right),
\end{equation}
where \( \sigma_0 \) denotes the clean surface tension, \( \Gamma_\infty \) the saturation interfacial concentration, and \( \mathrm{E} \) the Gibbs elasticity number, which quantifies the sensitivity of surface tension to interfacial concentration variations.

The problem is non-dimensionalized using the bubble diameter \( d_b \) as the characteristic length, \( \sqrt{g d_b} \) as the velocity scale, and \( \sqrt{d_b/g} \) as the time scale. The main dimensionless parameters governing the system are:
\begin{gather}
    \mathrm{E} =\frac{\mathcal{R}T\Gamma_\infty}{\sigma_0},\quad 
    \rho_r = \frac{\rho_l}{\rho_g},\quad 
    \mu_r = \frac{\mu_l}{\mu_g},\quad 
    \mathrm{K} = \frac{r_a F_{0}}{r_d},\quad 
    \mathrm{Bo} = \frac{\rho_l g d_b^2}{\sigma_0}, \nonumber\\
    \mathrm{Pe}_f = \frac{d_b \sqrt{g d_b}}{D_f},\quad 
    \mathrm{Pe}_F = \frac{d_b \sqrt{g d_b}}{D_F},\quad 
    \mathrm{Bi} = \frac{r_d d_b}{\sqrt{g d_b}},\quad 
    \mathrm{Ga} = \frac{\rho_l d_b \sqrt{g d_b}}{\mu_l},
\end{gather}
where the indices $l$ and $g$ denote the liquid and gas phase respectively, $F_0$ the initial uniform bulk concentration of surfactant, $\mathrm{Bo}$ the Bond number, $\mathrm{Pe}_f$ the surface Péclet number, $\mathrm{Pe}_F$ the bulk Péclet number, $\mathrm{Bi}$ the Biot number and $\mathrm{Ga}$ the Galilei number.

Using these dimensionless groups, we define the Marangoni number and the interfacial Damköhler number as:
\begin{equation}
    \mathrm{Ma} = \frac{\mathrm{Ga}  \mathrm{E}}{\mathrm{Bo}}, \,\,\,\,\,\mathrm{Da} = \mathrm{K} \mathrm{Bi},
\end{equation}
which can be expressed as a function of the physical parameters as:
\begin{equation}
    \mathrm{Ma} = \frac{\mathcal{R} T \Gamma_\infty}{\mu_l \sqrt{g d_b}}.\,\,\,\mathrm{Da} = \frac{r_a d_b F_{0}}{\sqrt{g d_b}}
\end{equation}

The various dimensionless groups introduced above compare different physical mechanisms: \( \mathrm{Ma} \) quantifies the strength of Marangoni stresses generated by surface tension gradients relative to viscous effects. The Bond number \( \mathrm{Bo} \) compares gravitational forces to surface tension forces. The Galilei number \( \mathrm{Ga} \) characterizes the ratio of buoyancy-driven inertial forces to viscous forces. The Péclet numbers \( \mathrm{Pe}_F \) and \( \mathrm{Pe}_f \) measure the relative importance of convective transport to diffusive transport in the bulk and at the interface, respectively. The Biot number \( \mathrm{Bi} \) quantifies the rate of surfactant removal from the interface via desorption relative to the hydrodynamic timescale of the system  and controls the rate at which interfacial surfactant is removed to the surrounding fluid. The interfacial Damköhler number \( \mathrm{Da} \) quantifies the rate of surfactant adsorption from the bulk to the interface relative to the hydrodynamic timescale. Large values of the Damköhler number correspond to fast adsorption kinetics compared to the hydrodynamic timescale.

For the simulations presented in this work, we consider the following non-dimensional parameters: density ratio \(\rho_r = 1000\), viscosity ratio \(\mu_r = 100\), Bond number \(\mathrm{Bo} = 0.5\), interfacial Péclet number \(\mathrm{Pe}_f = 100\), bulk Péclet number \(\mathrm{Pe}_F = 10\), and Galilei number \(\mathrm{Ga} = 100\).

These non-dimensional numbers provide a unified framework to describe the coupled influence of surfactant transport, hydrodynamics, and Marangoni effects on the bubble rise dynamics.

\subsection{Insoluble surfactants}

We first consider the case of insoluble surfactants. The bubble is initialized with a prescribed interfacial concentration $\tilde{\Gamma} = \Gamma / \Gamma_\infty = 0.1$, while the bulk contains no surfactant. The transport of $\Gamma$ is governed by the conservative formulation detailed in Section~2.  

\begin{figure}[h!]
    \centering
    \begin{overpic}[width=1\linewidth]{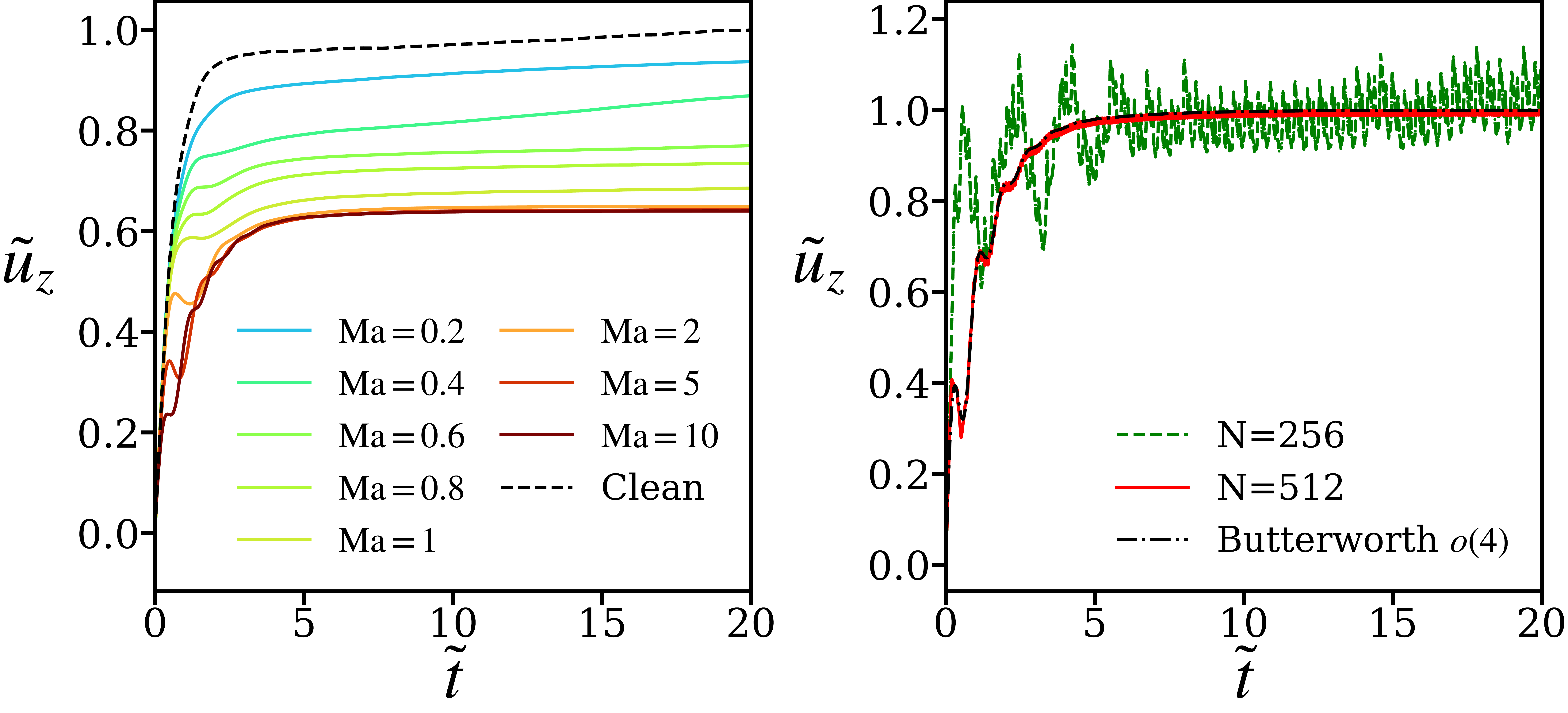}
        \put(7,46.5){\scriptsize (a)}
        \put(58,46.5){\scriptsize (b)}
    \end{overpic}

  \caption{(a) Time evolution of the vertical rise velocity of the bubble for clean and insoluble surfactant-laden cases. The presence of surfactants induces Marangoni stresses that oppose the flow, reducing the terminal velocity.
  (b) Influence of mesh resolution and signal filtering on the vertical rise velocity of the bubble. Increasing the mesh resolution reduces numerical noise but becomes computationally expensive due to the coupling between flow and interfacial transport. A fourth-order zero-phase Butterworth filter is therefore used to smooth the velocity signal while retaining physical trends.}
  \label{fig:velocity_insoluble}
\end{figure}

Accurately capturing the time evolution of the bubble’s rise velocity in the presence of surfactants requires high spatial resolution, especially near the interface where gradients of surfactant concentration and surface tension are steep. In the insoluble case, the accumulation of surfactant at the rear stagnation point leads to strong Marangoni stresses, which in turn introduce localized velocity gradients. These sharp gradients can lead to numerical noise in the computed velocity, particularly when using moderate grid resolutions. As illustrated in Fig.~\ref{fig:velocity_insoluble}(a), increasing the resolution from \( N = 256 \) to \( N = 512 \) indeed reduces the level of noise in the raw signal. However, further refinement becomes computationally expensive, especially when solving the coupled system involving bulk and interfacial transport. The use of a filtering approach therefore offers an efficient compromise, enabling reliable extraction of the bubble's rise velocity without the need for excessive computational cost. To reduce spurious high-frequency oscillations in the velocity signal, we therefore apply a low-pass filter to the temporal evolution of the bubble’s vertical velocity. Specifically, we use a 4th-order Butterworth filter, defined as:
\begin{equation}
    \hat{u}_f(\omega) = \frac{\hat{u}(\omega)}{\sqrt{1 + \left(\frac{\omega}{\omega_c}\right)^{2n}}},
\end{equation}
where \( \hat{u}(\omega) \) is the Fourier transform of the original signal, \( \omega_c \) is the cutoff frequency, and \( n = 4 \) is the filter order. This filter preserves the low-frequency content of the signal (including the terminal velocity), while attenuating high-frequency noise due to discretization errors.

Fig.~\ref{fig:velocity_insoluble}(b) shows the time evolution of the vertical rise velocity for both the clean and insoluble cases. As expected, the presence of surfactants reduces the terminal velocity due to Marangoni stresses that oppose the flow and inhibit interfacial mobility. This result shows that the model is able to account for both the surfactant transport and its coupling with the hydrodynamics.

\subsection{Soluble surfactants}

We now turn to the case of soluble surfactants. The transport dynamics are governed by coupled advection–diffusion equations for the bulk and interfacial dimensionless concentrations \( \tilde{F} \)  and \( \tilde{\Gamma} \), along with source terms accounting for adsorption and desorption, as detailed in Section~2.

To isolate the effect of each interfacial exchange mechanism, we consider two distinct initial configurations, depending on whether desorption or adsorption dominates:

\begin{itemize}
    \item In the case of \textbf{pure desorption} (adsorption neglected, \( \mathrm{Da} = 0 \)), the bulk contains no surfactant initially, \( \tilde{F}(\tilde{t}=0) = 0 \), and the interface is initialized with a finite concentration \( \tilde{\Gamma}(\tilde{t}=0) = 0.1 \).

    \item In the case of \textbf{pure adsorption} (desorption neglected, \( \mathrm{Bi} = 0 \)), the bulk is initially filled with a uniform non-dimensional concentration \( \tilde{F}(0) = 1 \), while the interface is clean, i.e., \( \tilde{\Gamma}(0) = 0 \).

\end{itemize}

To investigate how the system transitions toward the classical limiting behaviors, namely the clean or surface-incompressible interface cases, we perform a series of simulations by varying the relevant non-dimensional parameters: the Biot number \( \mathrm{Bi} \) for desorption and the Damköhler number \( \mathrm{Da} \) for adsorption. These parameters control the strength of interfacial kinetics relative to the hydrodynamic timescale and allow us to systematically explore the effect of each transfer process on the bubble dynamics.

\subsubsection{Pure desorption: convergence toward the clean case}

\begin{figure}[h!]
    \centering

    \begin{subfigure}{0.32\textwidth}
        \centering
        \includegraphics[width=\linewidth]{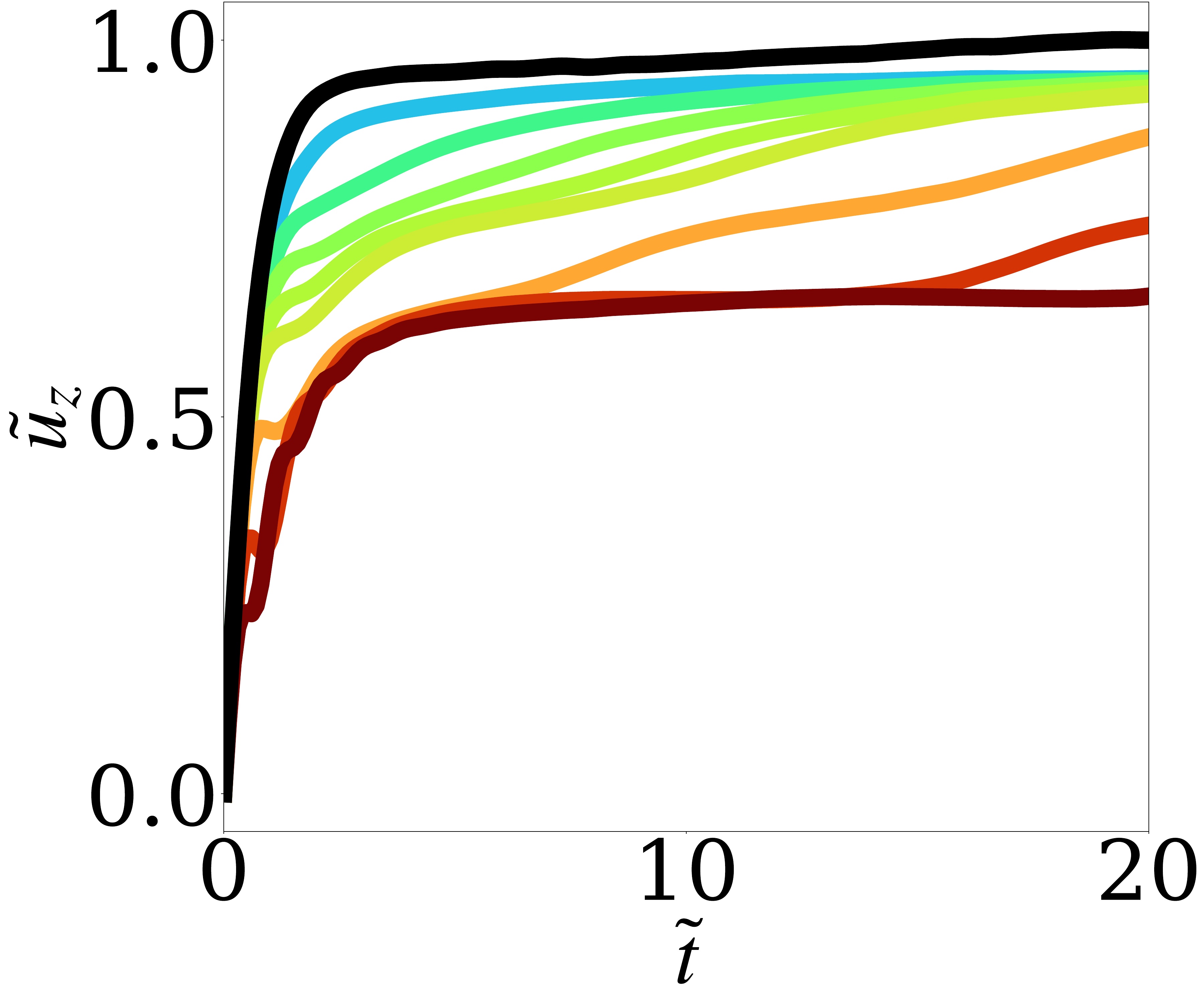}
        \caption{$\mathrm{Bi}=0.1$}
    \end{subfigure}
    \hfill
    \begin{subfigure}{0.32\textwidth}
        \centering
        \includegraphics[width=\linewidth]{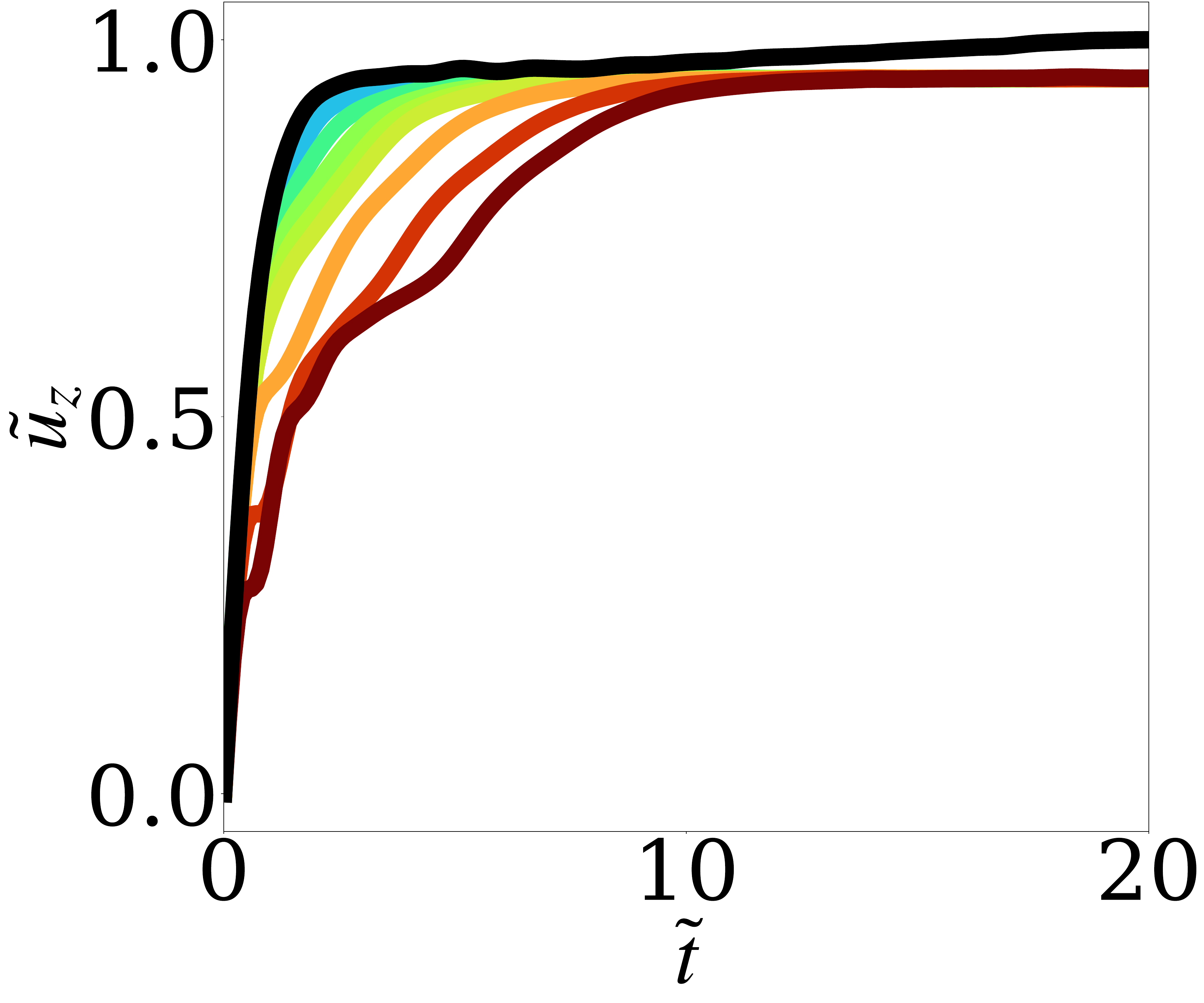}
        \caption{$\mathrm{Bi}=0.5$}
    \end{subfigure}
    \hfill
    \begin{subfigure}{0.32\textwidth}
        \centering
        \includegraphics[width=\linewidth]{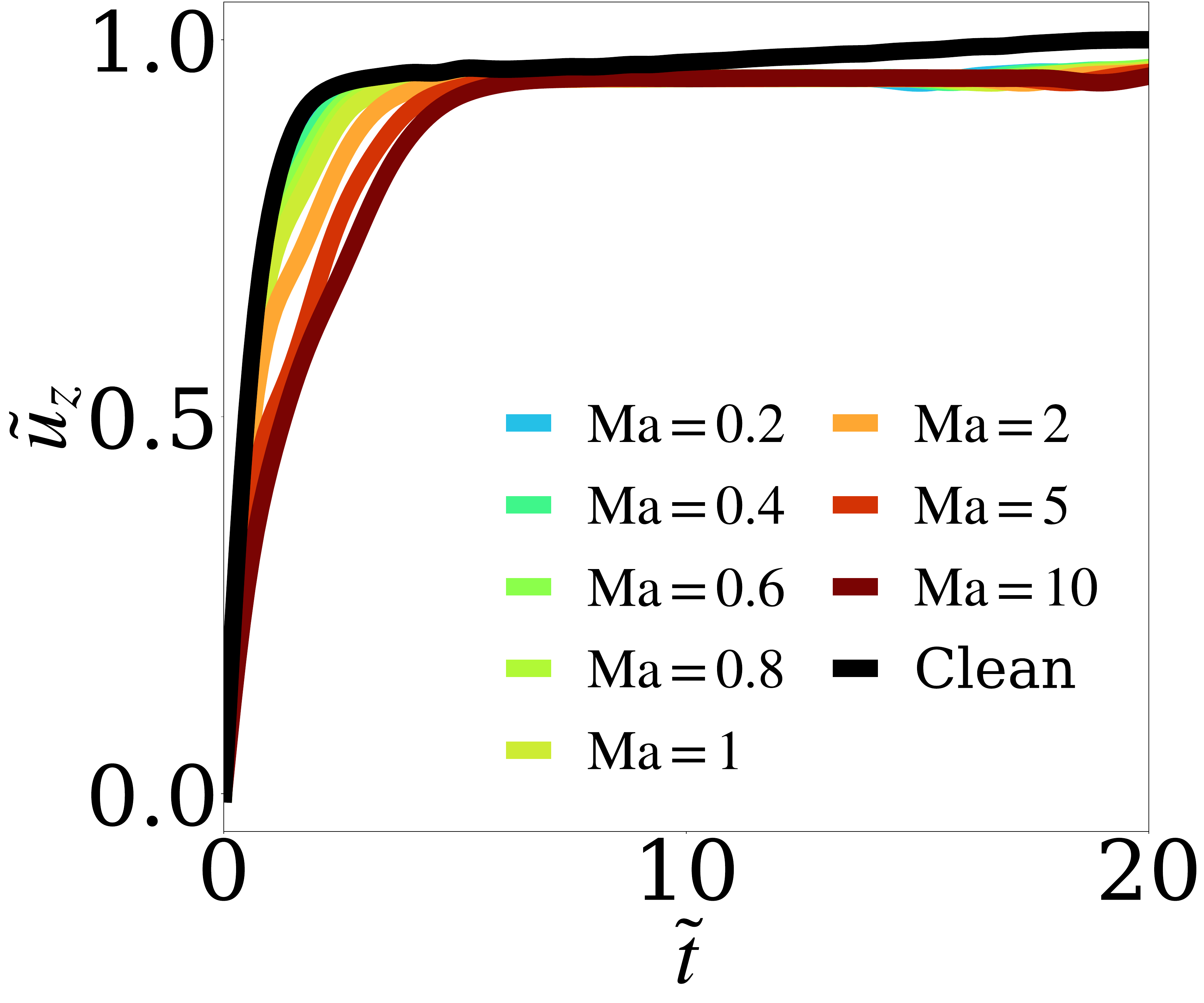}
        \caption{$\mathrm{Bi}=1$}
    \end{subfigure}

    \caption{Time evolution of the bubble's vertical rise velocity in the pure desorption regime for increasing Biot numbers: \( \mathrm{Bi} = 0.1 \) (a), \( \mathrm{Bi} = 0.5 \) (b), and \( \mathrm{Bi} = 1 \) (c). As \( \mathrm{Bi} \) increases, desorption becomes more dominant, progressively depleting the interface and reducing Marangoni stresses. The system thus approaches the clean-interface limit, resulting in higher terminal rise velocities.}
    \label{fig:desorption_velocity}
\end{figure}

In this configuration, we consider a pure desorption regime (adsorption neglected), and vary the Biot number \( \mathrm{Bi} \). Fig.~\ref{fig:desorption_velocity} shows the time evolution of the vertical rise velocity for increasing desorption Biot numbers at a fixed time \(\tilde{t} = 20\). As the Biot number increases, desorption becomes more efficient and surfactant is progressively removed from the interface. Consequently, Marangoni stresses weaken and the terminal rise velocity increases, approaching the clean-interface limit. 

\begin{figure}[H]
  \centering

  \begin{subfigure}[t]{0.3\textwidth}
    \centering
    \includegraphics[width=\textwidth]{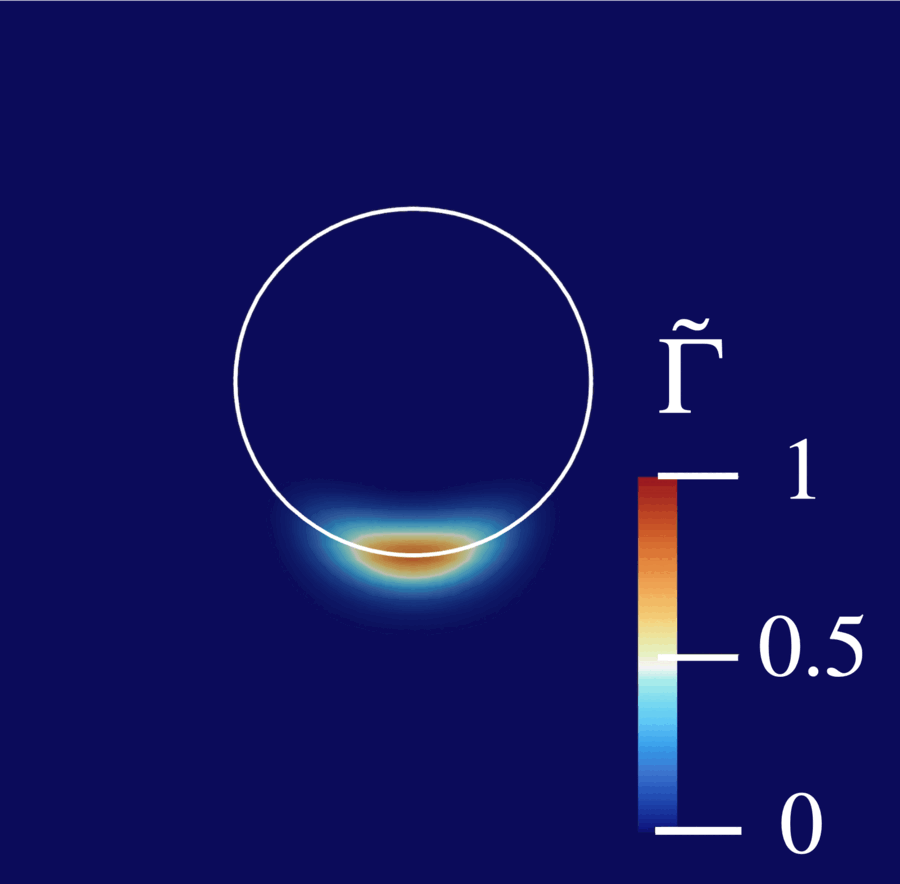}
    \caption{$\mathrm{Bi}=0.5,\,\mathrm{Ma}=0.2$}
    \label{fig:sub2}
  \end{subfigure}
  \hfill
  \begin{subfigure}[t]{0.3\textwidth}
    \centering
    \includegraphics[width=\textwidth]{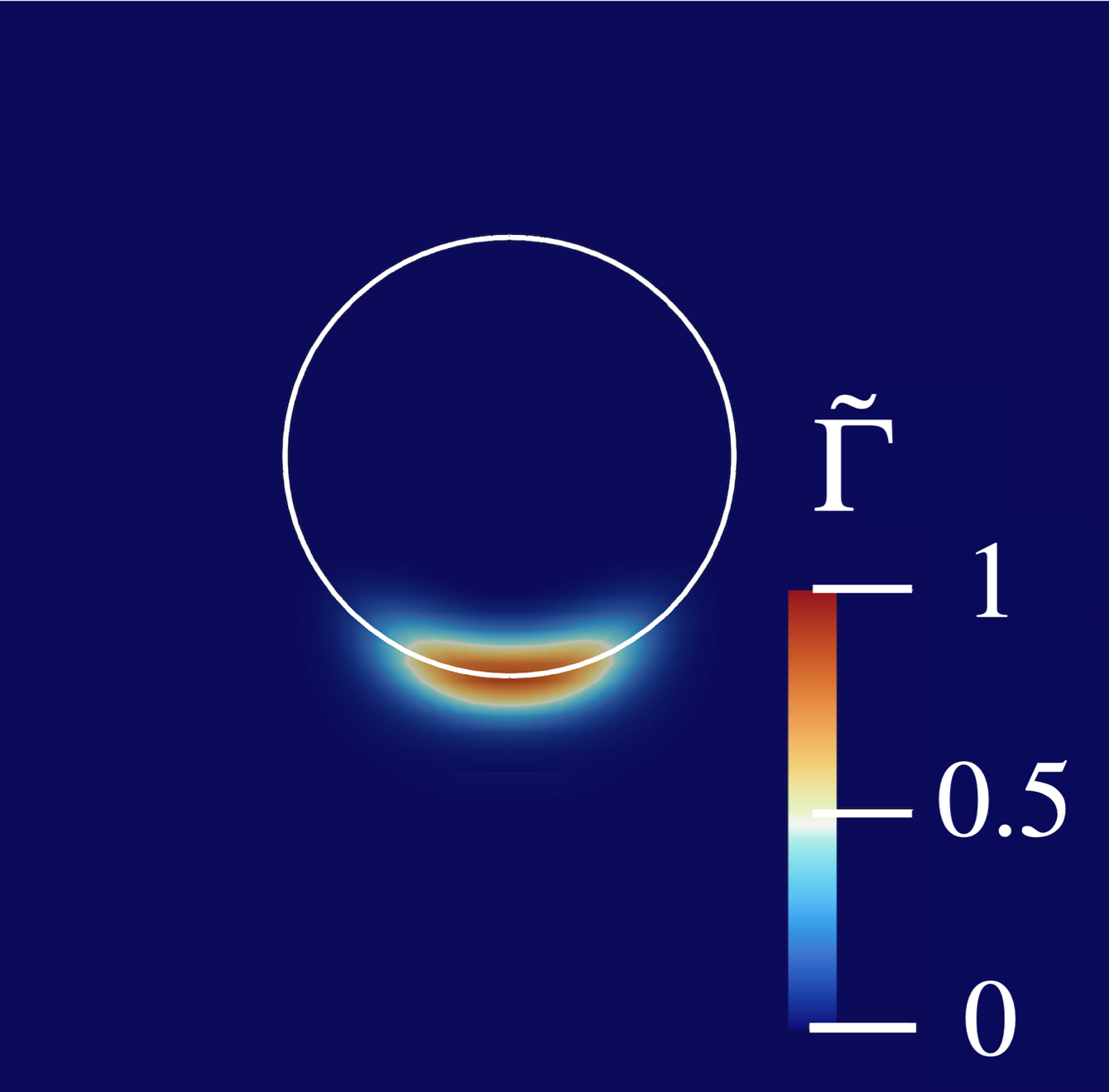}
    \caption{$\mathrm{Bi}=0.5,\,\mathrm{Ma}=1$}
    \label{fig:sub3}
  \end{subfigure}
  \hfill
  \begin{subfigure}[t]{0.3\textwidth}
    \centering
    \includegraphics[width=\textwidth]{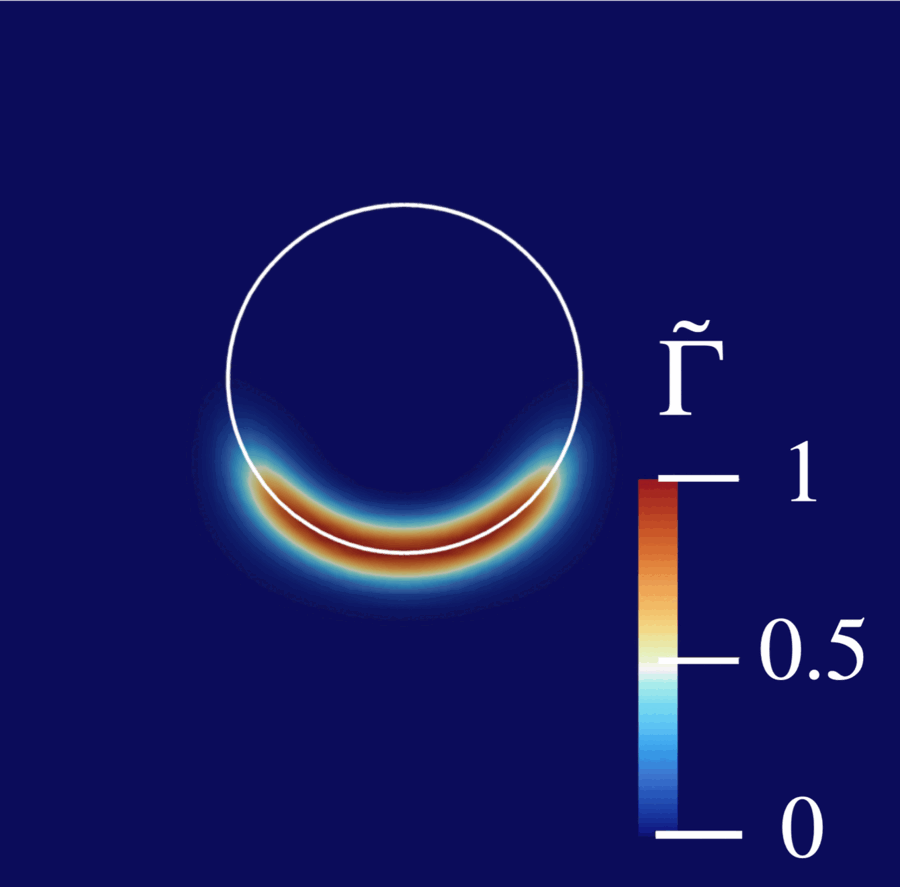}
    \caption{$\mathrm{Bi}=0.5,\,\mathrm{Ma}=10$}
    \label{fig:sub3}
  \end{subfigure}

  \vspace{0.5cm} 

  \begin{subfigure}[t]{0.3\textwidth}
    \centering
    \includegraphics[width=\textwidth]{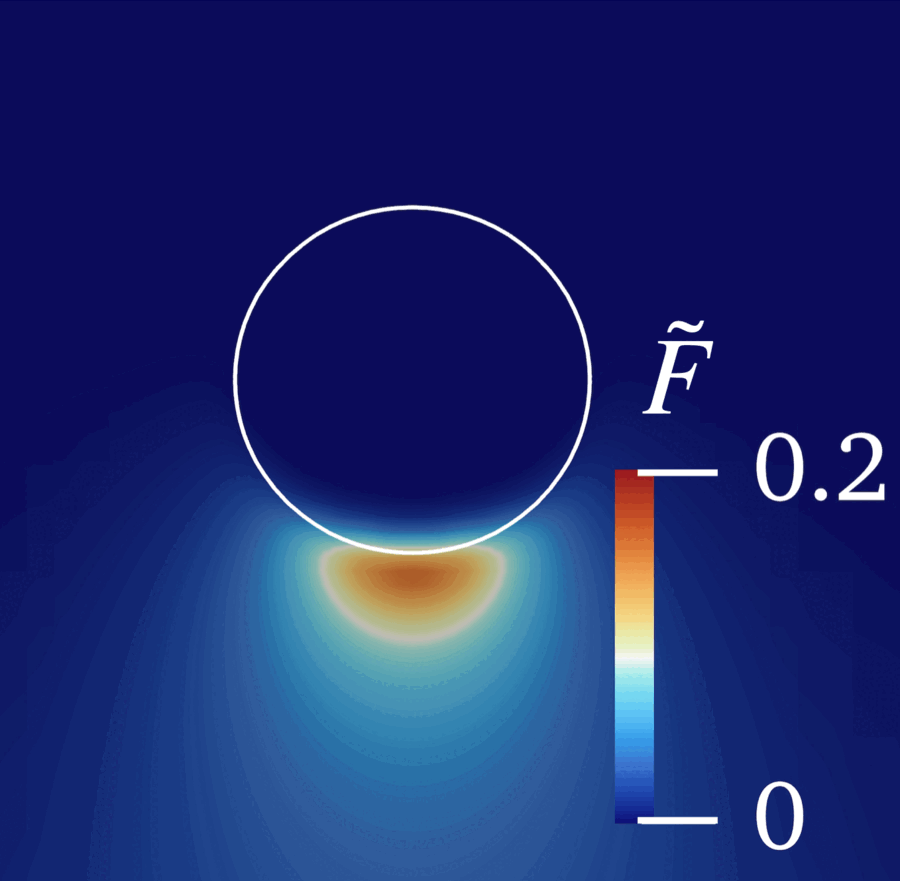}
    \caption{$\mathrm{Bi}=0.5,\,\mathrm{Ma}=0.2$}
    \label{fig:sub5}
  \end{subfigure}
  \hfill
  \begin{subfigure}[t]{0.3\textwidth}
    \centering
    \includegraphics[width=\textwidth]{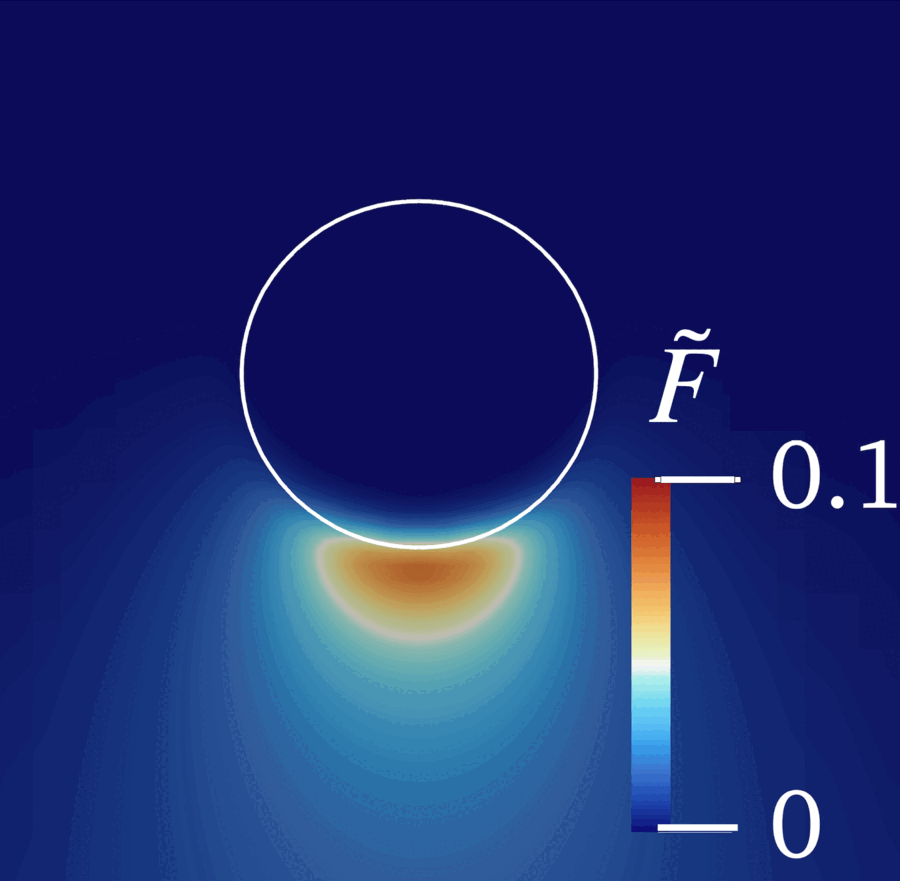}
    \caption{$\mathrm{Bi}=0.5,\,\mathrm{Ma}=1$}
    \label{fig:sub6}
  \end{subfigure}
  \hfill
  \begin{subfigure}[t]{0.3\textwidth}
    \centering
    \includegraphics[width=\textwidth]{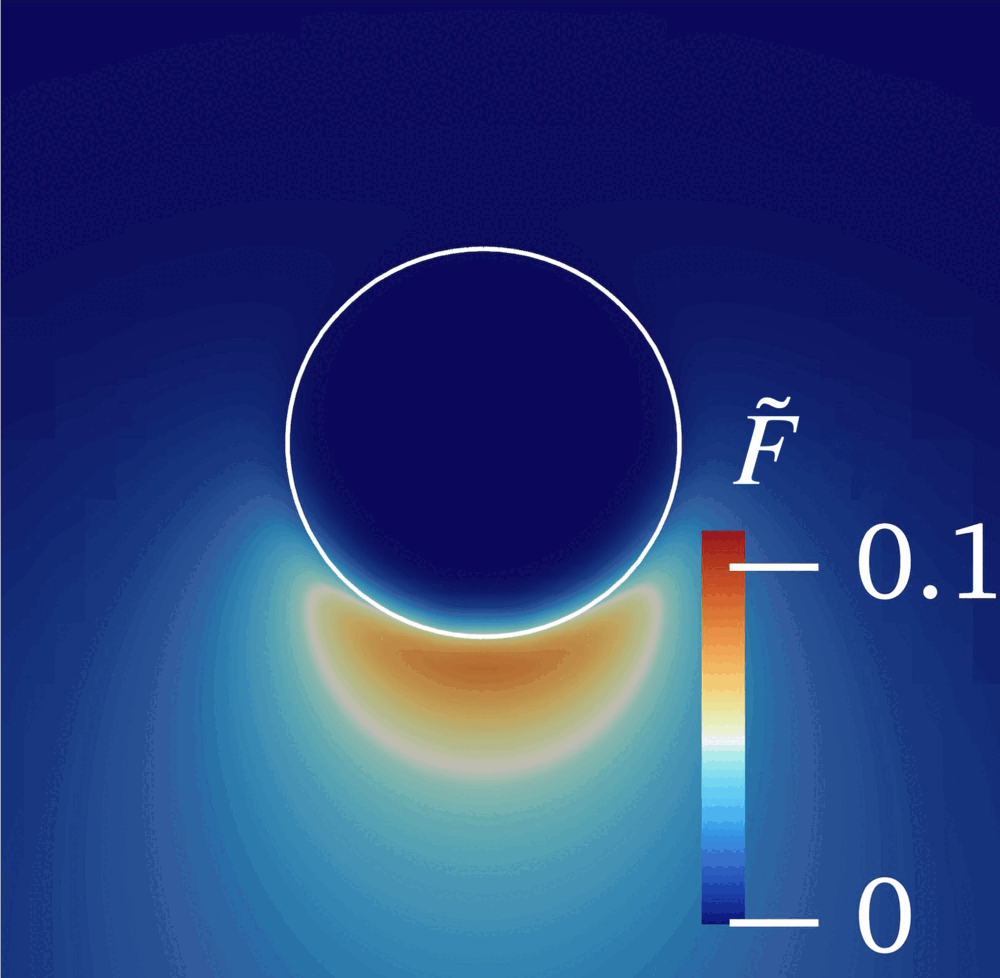}
    \caption{$\mathrm{Bi}=0.5,\,\mathrm{Ma}=10$}
    \label{fig:sub3}
  \end{subfigure}

  \vspace{0.5cm} 

  \begin{subfigure}[t]{0.3\textwidth}
    \centering
    \includegraphics[width=\textwidth]{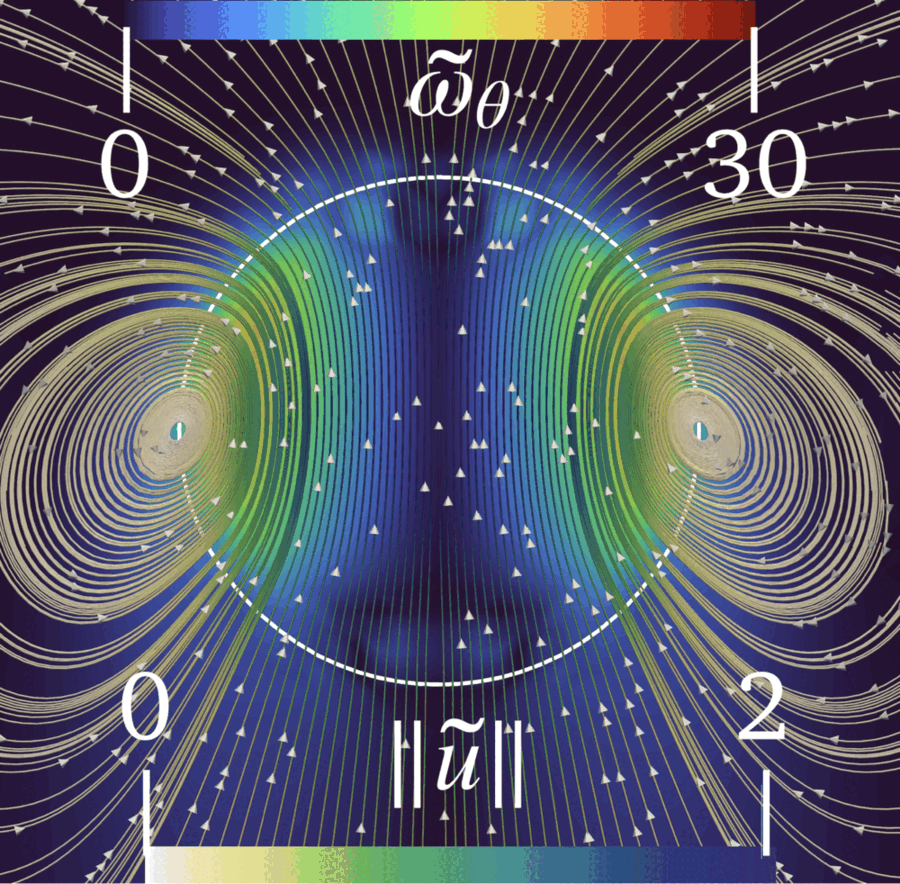}
    \caption{$\mathrm{Bi}=0.5,\,\mathrm{Ma}=0.2$}
    \label{fig:sub5}
  \end{subfigure}
  \hfill
  \begin{subfigure}[t]{0.3\textwidth}
    \centering
    \includegraphics[width=\textwidth]{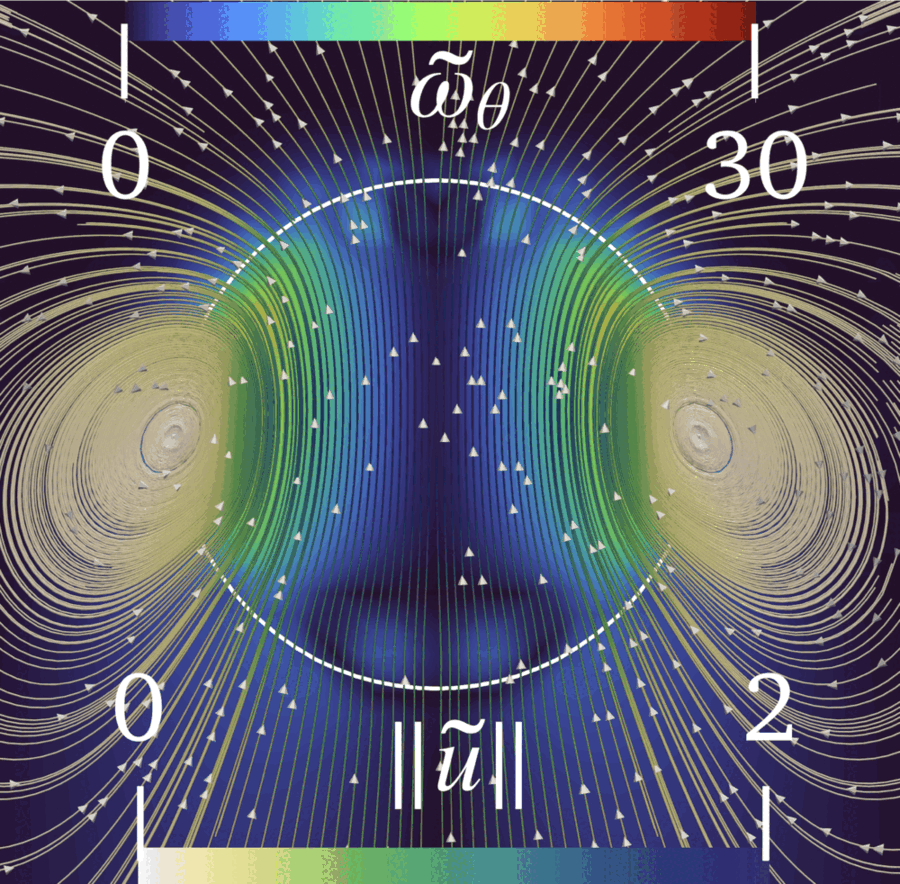}
    \caption{$\mathrm{Bi}=0.5,\,\mathrm{Ma}=1$}
    \label{fig:sub6}
  \end{subfigure}
  \hfill
  \begin{subfigure}[t]{0.3\textwidth}
    \centering
    \includegraphics[width=\textwidth]{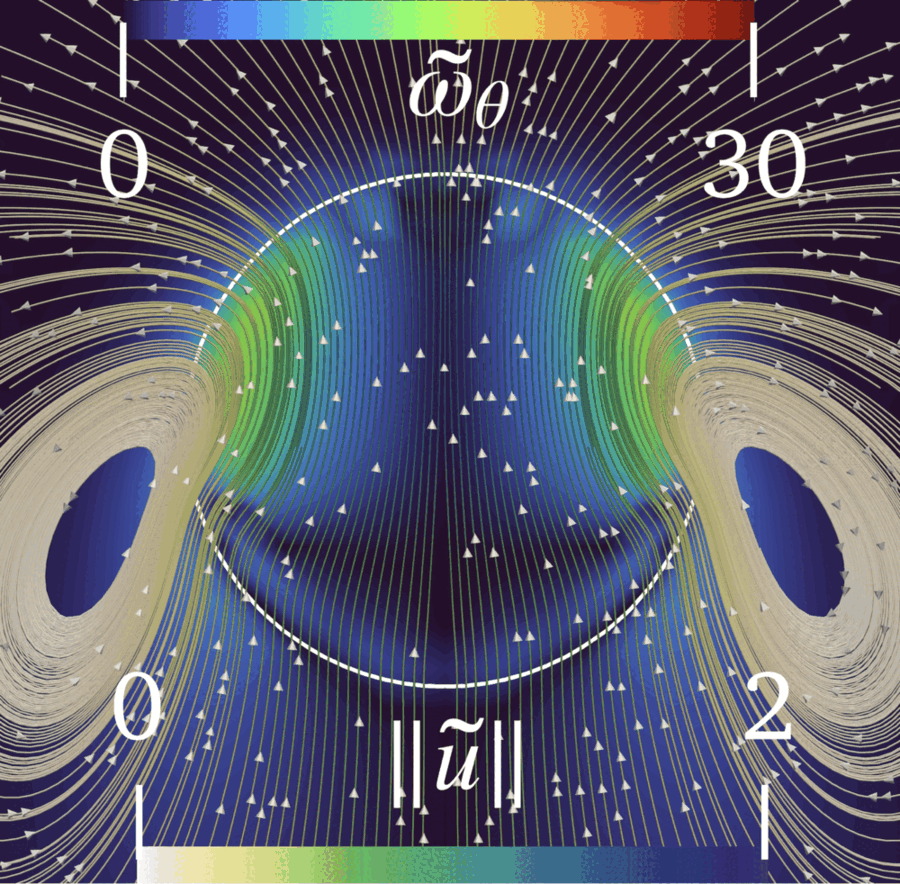}
    \caption{$\mathrm{Bi}=0.5,\,\mathrm{Ma}=10$}
    \label{fig:sub3}
  \end{subfigure}

  \caption{
Snapshots of the dimensionless interfacial surfactant concentration \(\tilde{\Gamma}\) (top row), bulk concentration \(\tilde{F}\) (middle row), and dimensionless vorticity field $\tilde{\omega}_\theta$ with streamlines (bottom row) for increasing Marangoni numbers \(\mathrm{Ma} = 0.2\), \(1\), and \(10\) at fixed Biot number \(\mathrm{Bi} = 0.5\) at time $\tilde{t}=10$. As \(\mathrm{Ma}\) increases, the stagnation cap enlarges and surfactant accumulates near the rear of the bubble. Desorption releases surfactant from the interface into the surrounding liquid, producing a local enrichment of the bulk concentration in the near-interface region before it is redistributed by diffusion and advection. The vorticity field shows a progressive weakening of the internal recirculation within the bubble at large \(\mathrm{Ma}\), consistent with the damping effect of Marangoni stresses on the flow.
}

  \label{fig:2x3matrix_des}
\end{figure}

Additional insight into the flow structure and surfactant distribution is provided in Fig.~\ref{fig:2x3matrix_des}, which presents snapshots of the interfacial concentration, bulk concentration, and vorticity field for different Marangoni numbers at a fixed desorption rate. As \(\mathrm{Ma}\) increases, surfactant accumulates in a pronounced stagnation cap near the rear of the bubble, while desorption transfers surfactant into the surrounding bulk. This redistribution modifies the hydrodynamic structure of the flow, reducing the strength of internal recirculation inside the bubble due to enhanced Marangoni stresses along the interface.

\subsubsection{Pure adsorption: approach to the surface-incompressible interface}

Next, we examine the pure adsorption regime (desorption neglected), and vary the interfacial Damköhler number \( \mathrm{Da} \). In this case, surfactant irreversibly adsorbs onto the interface. For sufficiently large \( \mathrm{Da} \), the interface rapidly accumulates surfactant approaching the surface-incompressible interface behaviour.

\begin{figure}[h!]
    \centering

    \begin{subfigure}{0.32\textwidth}
        \centering
        \includegraphics[width=\linewidth]{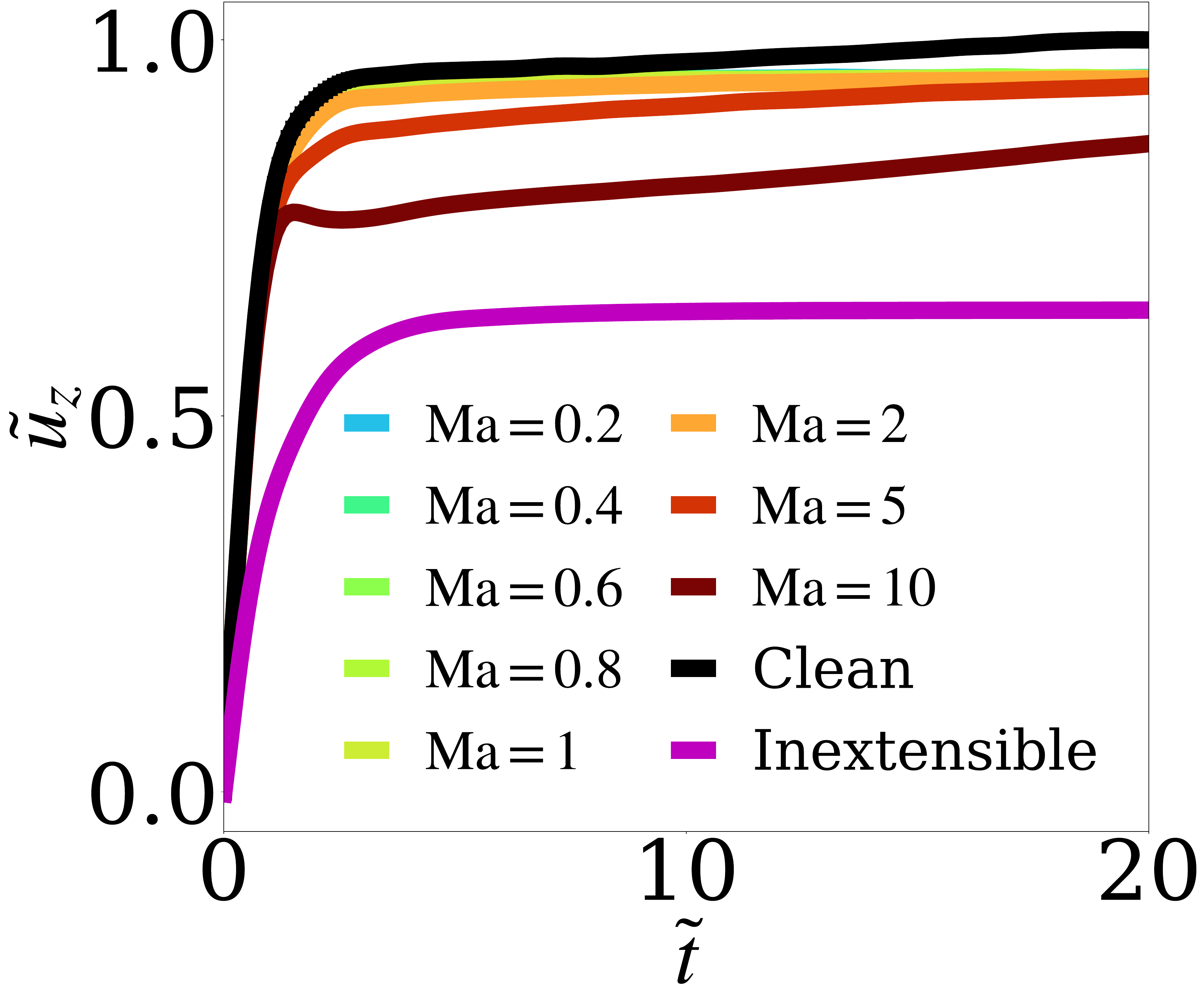}
        \caption{$\mathrm{Da}=0.1$}
    \end{subfigure}
    \hfill
    \begin{subfigure}{0.32\textwidth}
        \centering
        \includegraphics[width=\linewidth]{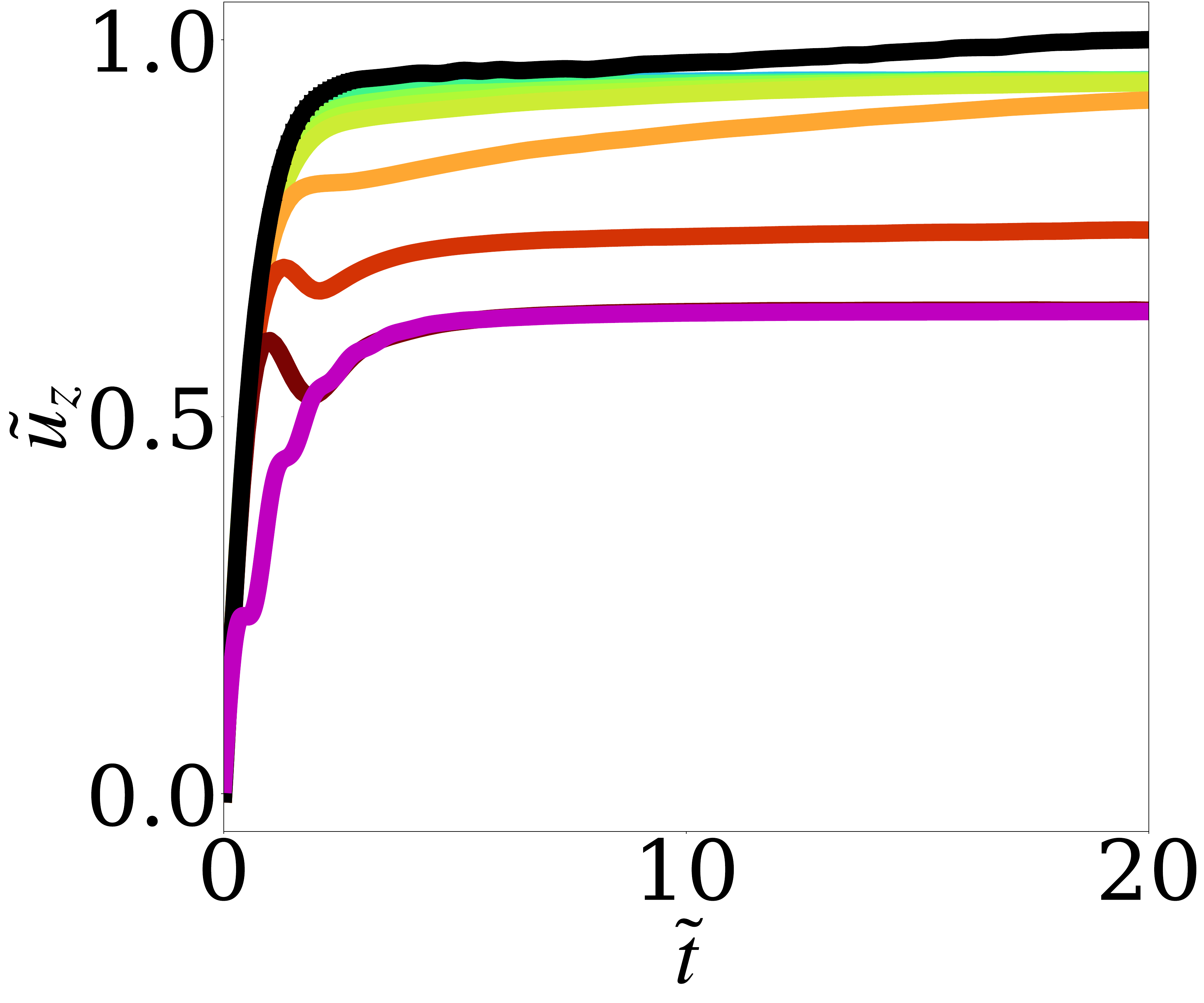}
        \caption{$\mathrm{Da}=0.5$}
    \end{subfigure}
    \hfill
    \begin{subfigure}{0.32\textwidth}
        \centering
        \includegraphics[width=\linewidth]{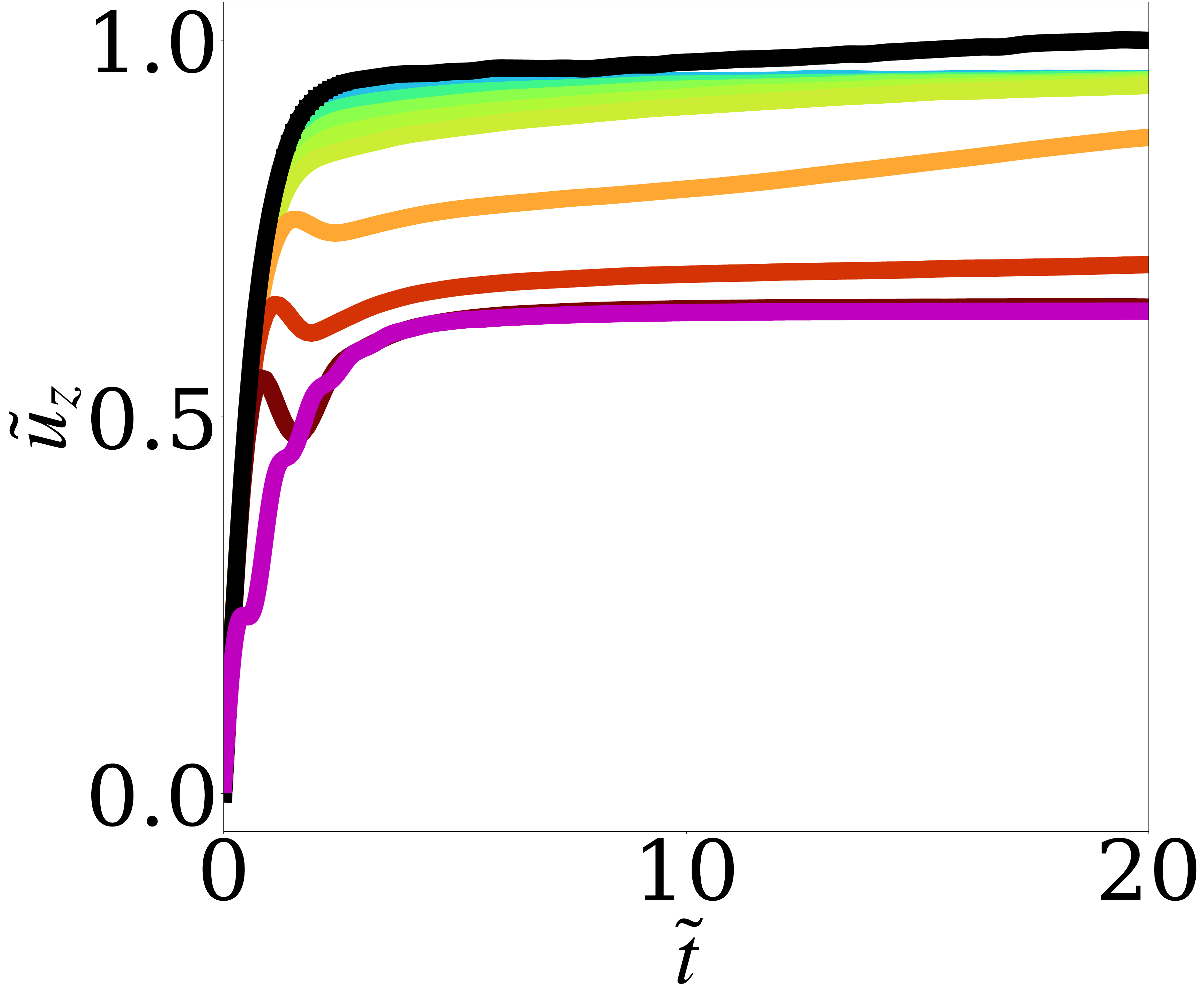}
        \caption{$\mathrm{Da}=1$}
    \end{subfigure}

    \caption{Time evolution of the bubble's vertical rise velocity in the pure adsorption regime for increasing interfacial Damköhler numbers: \( \mathrm{Da} = 0.1 \) (a), \( \mathrm{Da} = 0.5 \) (b), and \( \mathrm{Da} = 1 \) (c). As \( \mathrm{Da} \) increases, adsorption becomes more efficient, leading to stronger Marangoni stresses that slow down the bubble, causing its terminal velocity to approach the limit corresponding to an insoluble surfactant.}
    \label{fig:adsorption_velocity}
\end{figure}

Fig.~\ref{fig:adsorption_velocity} displays the time evolution of the bubble rising velocity for increasing Damköhler numbers. As \(\mathrm{Da}\) increases, adsorption becomes more efficient and surfactant accumulates more rapidly at the interface. This accumulation strengthens Marangoni stresses along the interface, which oppose the flow and progressively reduce the terminal rise velocity, driving the system toward the surface-incompressible interface limit.

\begin{figure}[h!]
  \centering

  \begin{subfigure}[t]{0.3\textwidth}
    \centering
    \includegraphics[width=\textwidth]{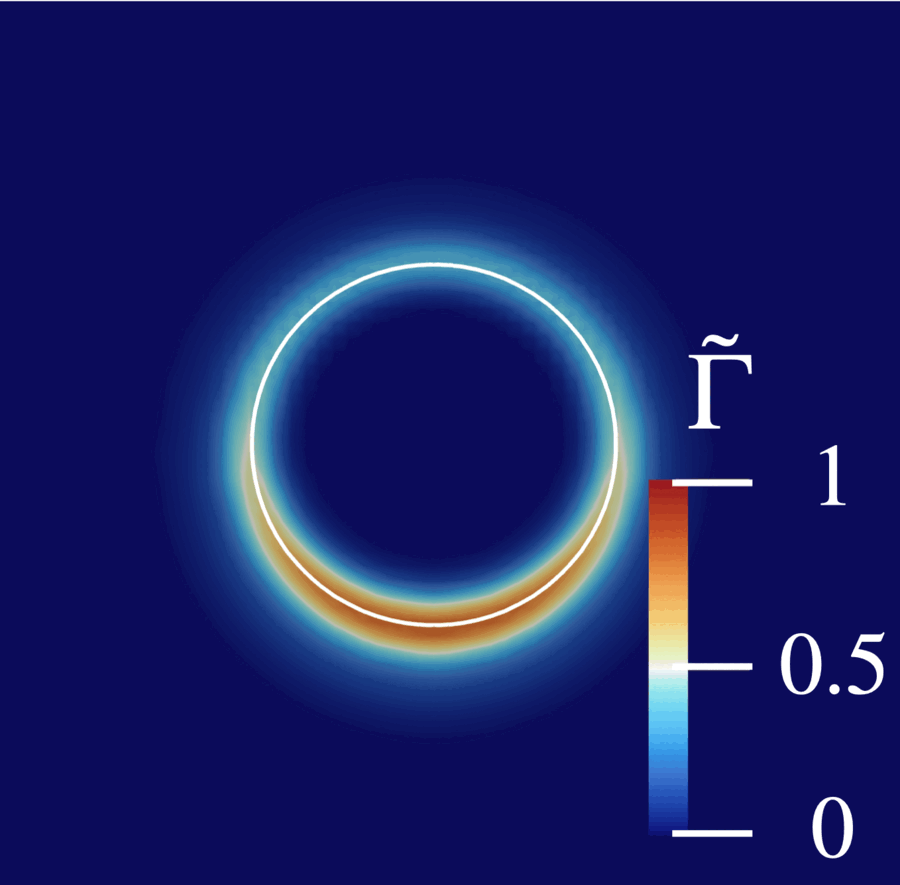}
    \caption{$\mathrm{Da}=0.5,\,\mathrm{Ma}=0.2$}
    \label{fig:sub2}
  \end{subfigure}
  \hfill
  \begin{subfigure}[t]{0.3\textwidth}
    \centering
    \includegraphics[width=\textwidth]{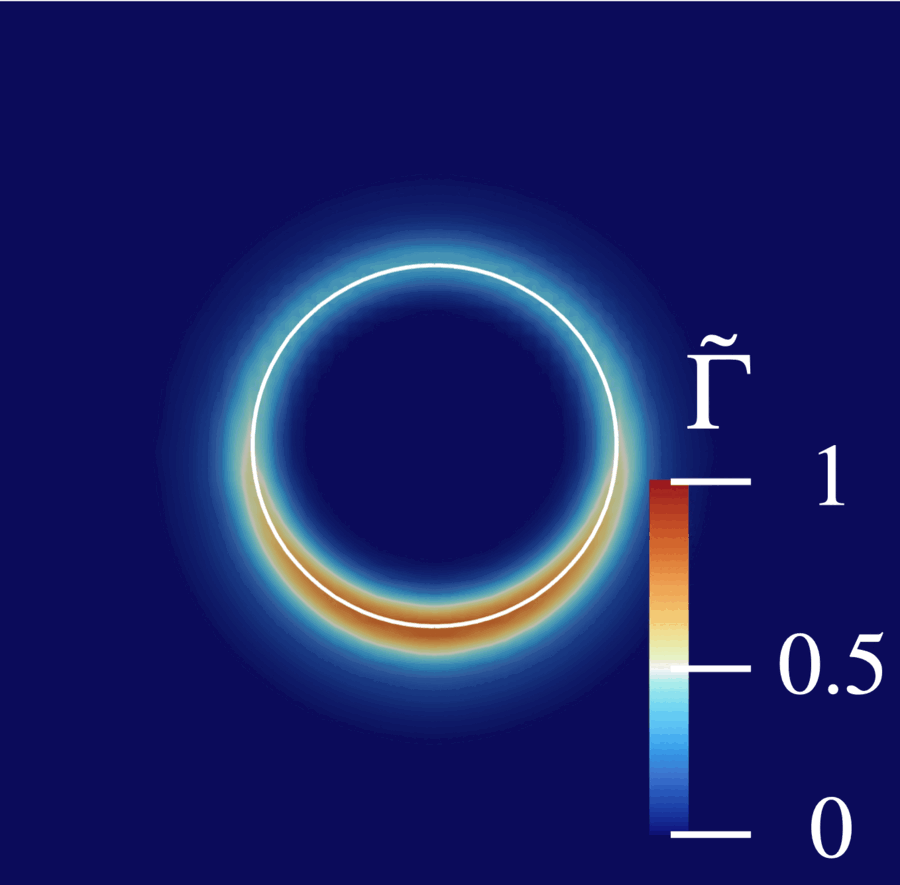}
    \caption{$\mathrm{Da}=0.5,\,\mathrm{Ma}=1$}
    \label{fig:sub3}
  \end{subfigure}
  \hfill
  \begin{subfigure}[t]{0.3\textwidth}
    \centering
    \includegraphics[width=\textwidth]{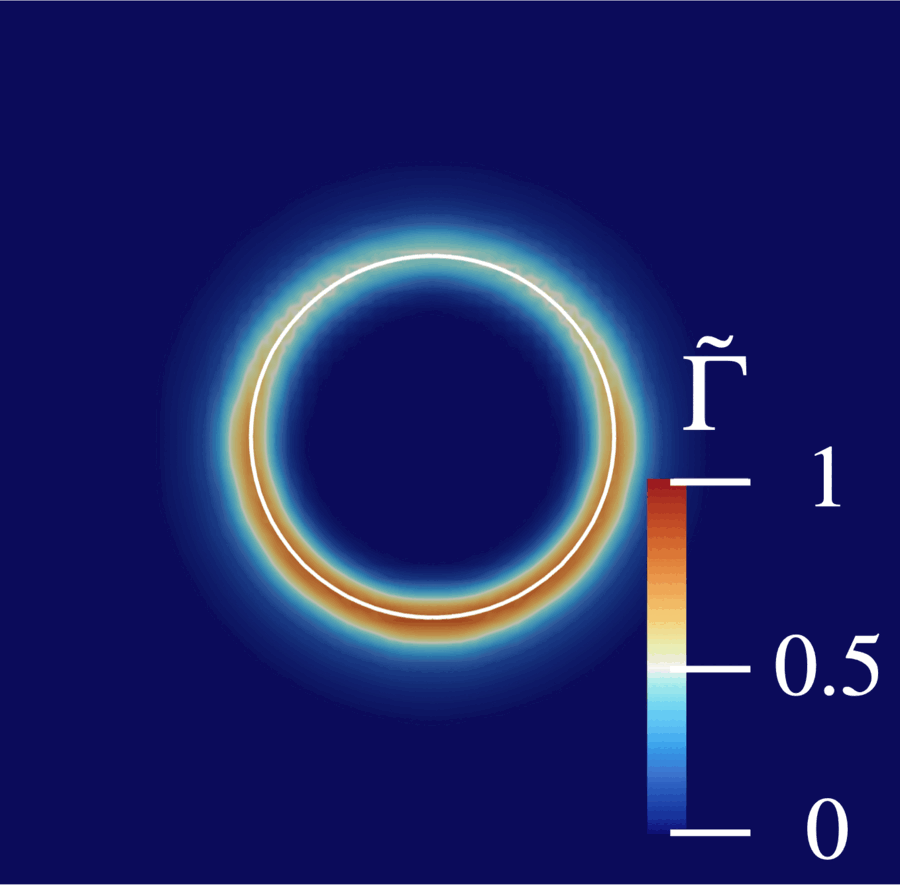}
    \caption{$\mathrm{Da}=0.5,\,\mathrm{Ma}=10$}
    \label{fig:sub3}
  \end{subfigure}

  \vspace{0.5cm} 

  \begin{subfigure}[t]{0.3\textwidth}
    \centering
    \includegraphics[width=\textwidth]{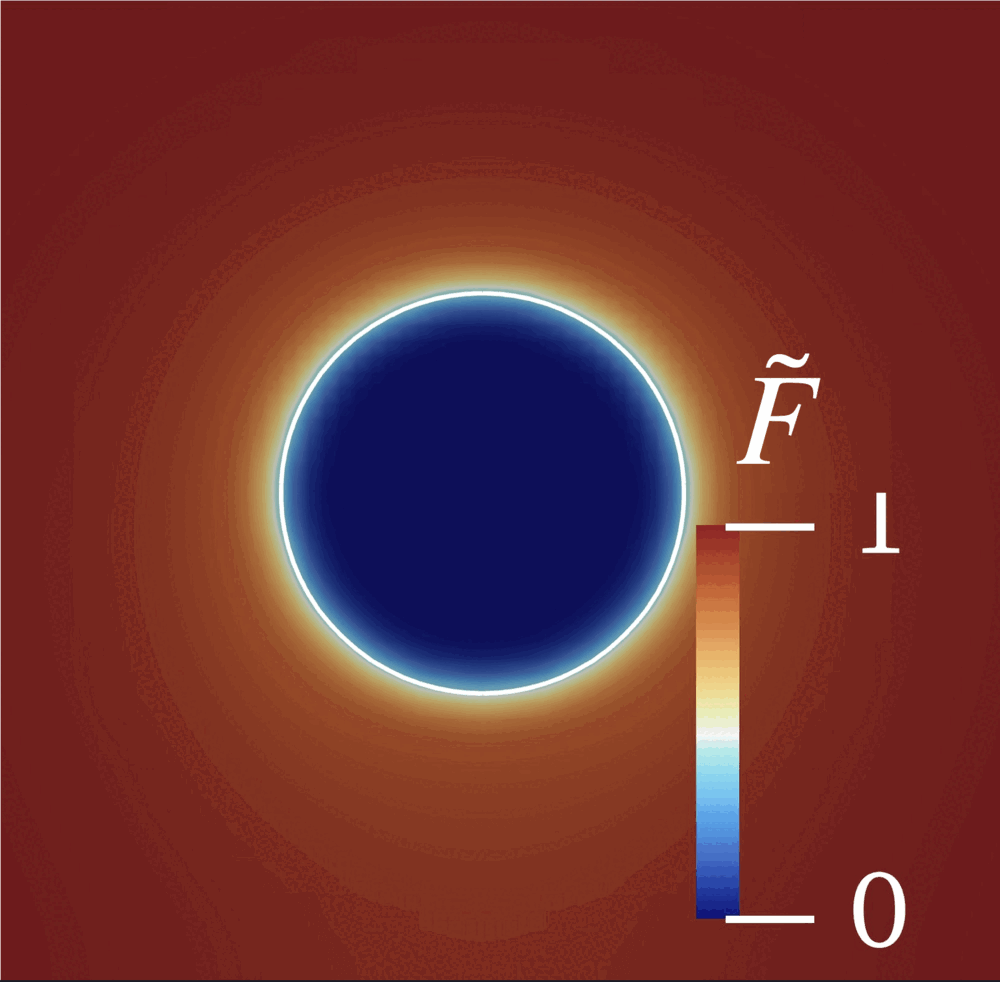}
    \caption{$\mathrm{Da}=0.5,\,\mathrm{Ma}=0.2$}
    \label{fig:sub5}
  \end{subfigure}
  \hfill
  \begin{subfigure}[t]{0.3\textwidth}
    \centering
    \includegraphics[width=\textwidth]{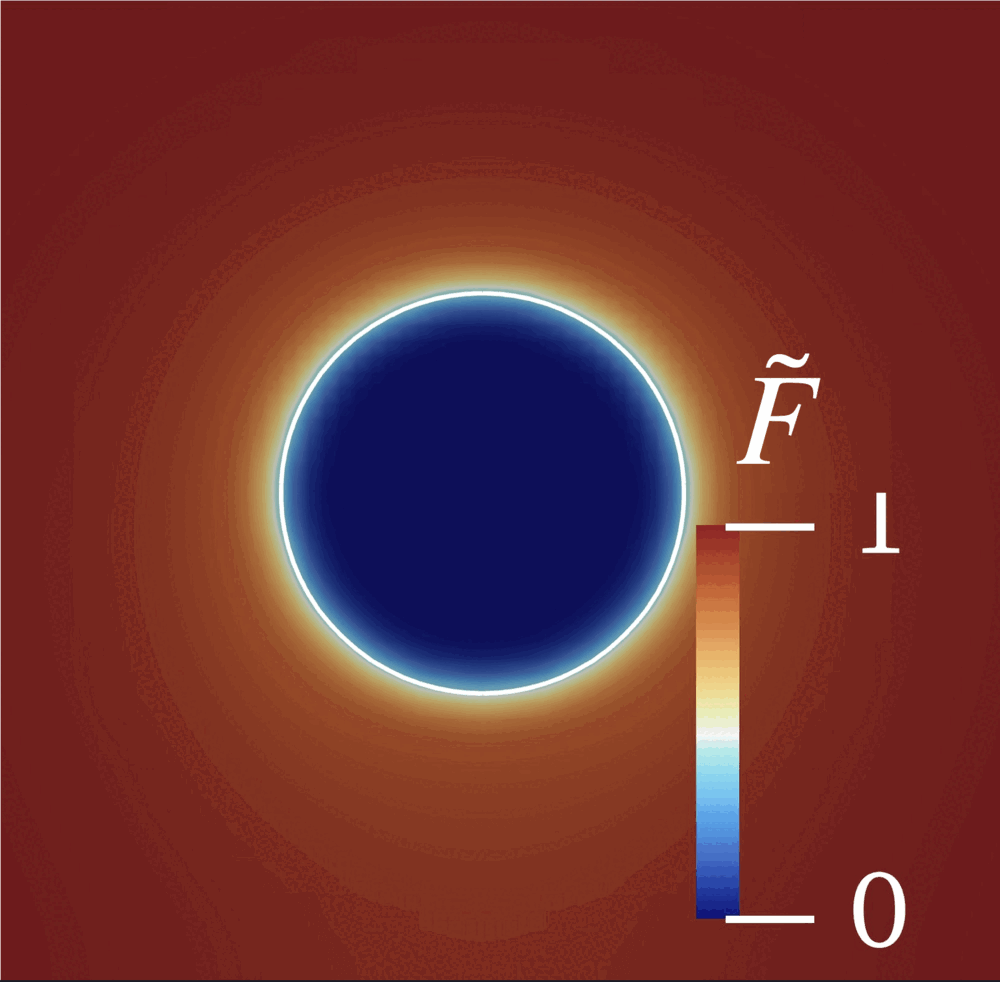}
    \caption{$\mathrm{Da}=0.5,\,\mathrm{Ma}=1$}
    \label{fig:sub6}
  \end{subfigure}
  \hfill
  \begin{subfigure}[t]{0.3\textwidth}
    \centering
    \includegraphics[width=\textwidth]{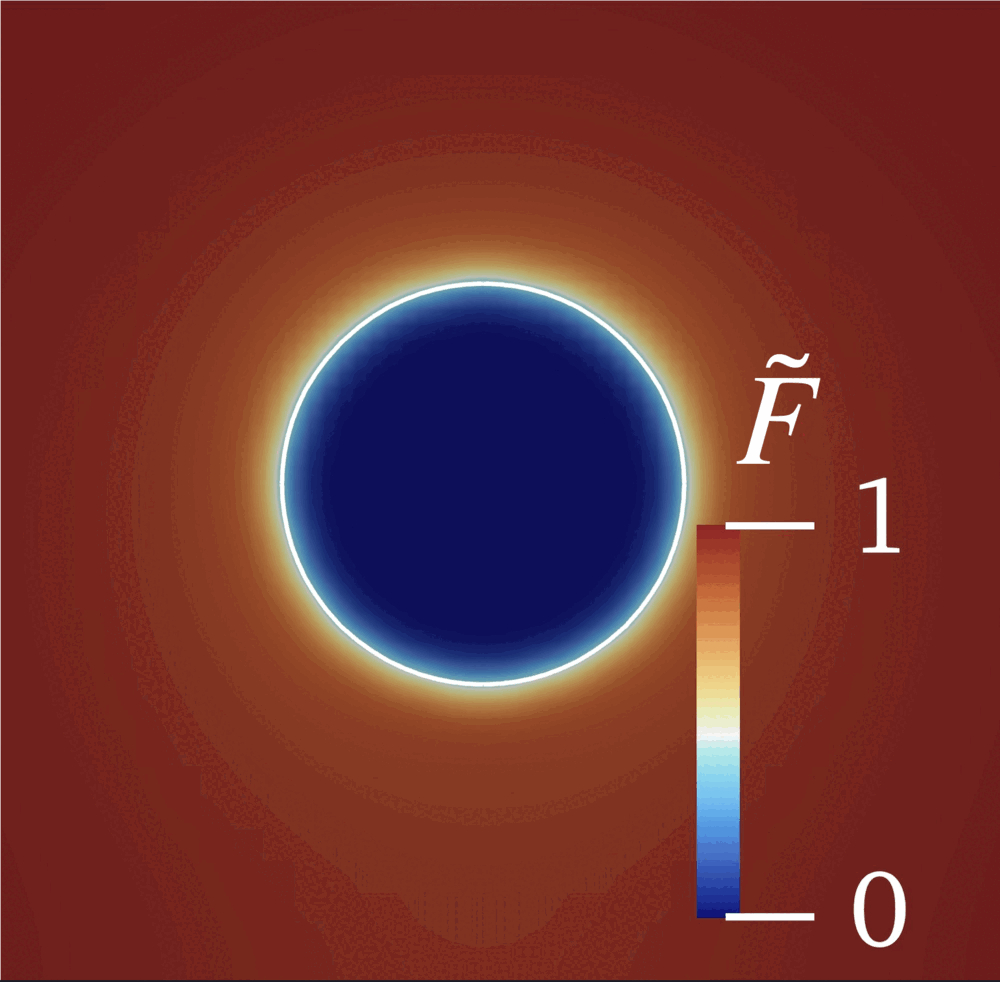}
    \caption{$\mathrm{Da}=0.5,\,\mathrm{Ma}=10$}
    \label{fig:sub3}
  \end{subfigure}

  \vspace{0.5cm} 

  \begin{subfigure}[t]{0.3\textwidth}
    \centering
    \includegraphics[width=\textwidth]{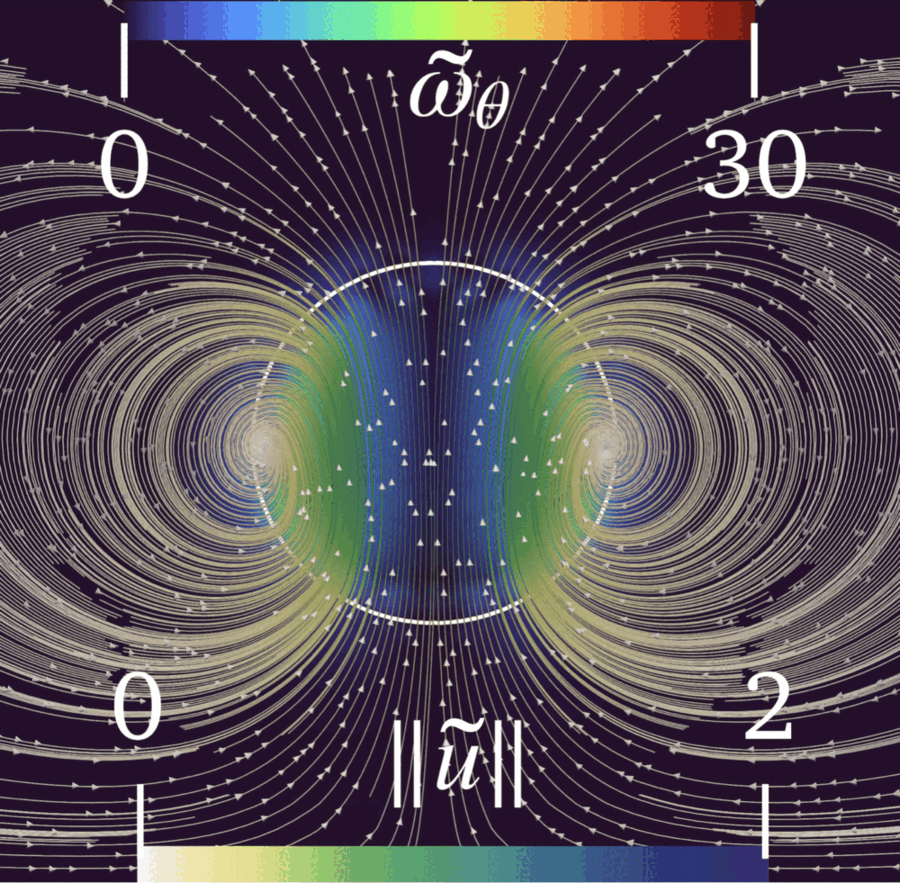}
    \caption{$\mathrm{Da}=0.5,\,\mathrm{Ma}=0.2$}
    \label{fig:sub5}
  \end{subfigure}
  \hfill
  \begin{subfigure}[t]{0.3\textwidth}
    \centering
    \includegraphics[width=\textwidth]{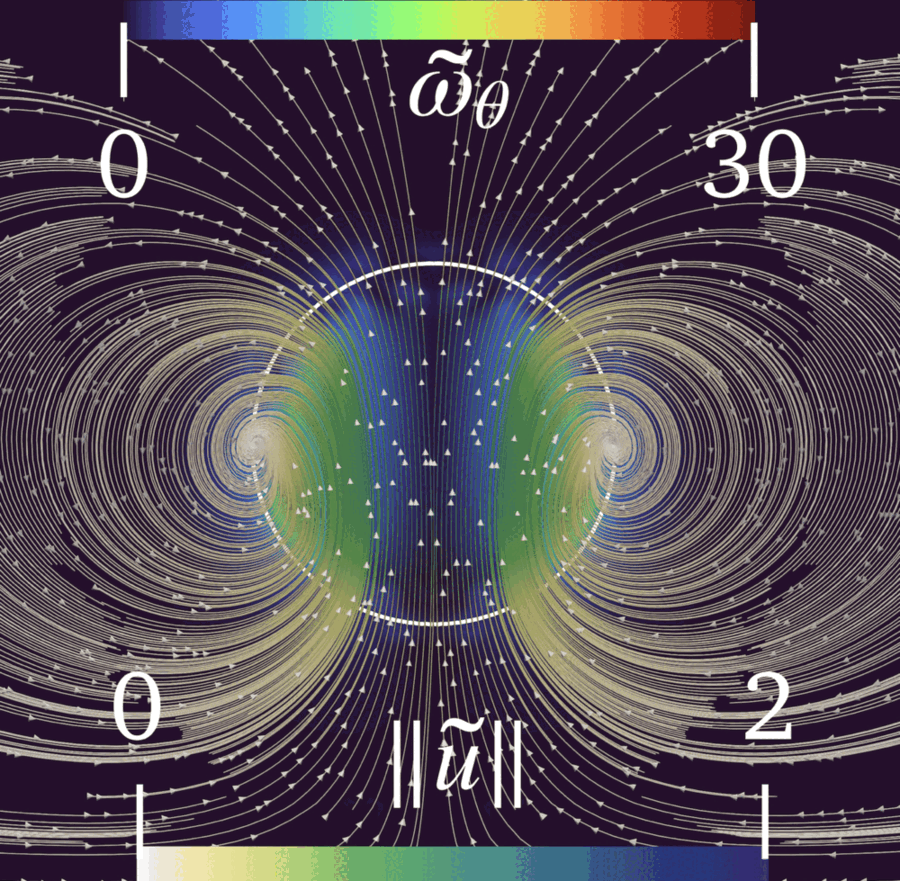}
    \caption{$\mathrm{Da}=0.5,\,\mathrm{Ma}=1$}
    \label{fig:sub6}
  \end{subfigure}
  \hfill
  \begin{subfigure}[t]{0.3\textwidth}
    \centering
    \includegraphics[width=\textwidth]{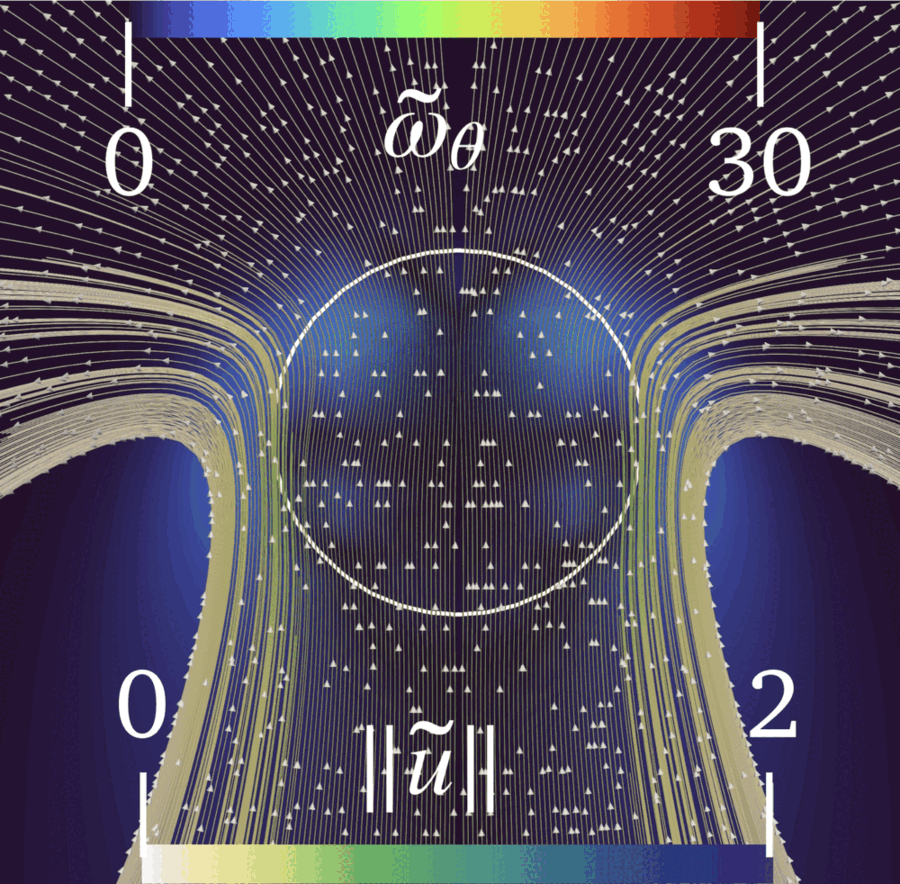}
    \caption{$\mathrm{Da}=0.5,\,\mathrm{Ma}=10$}
    \label{fig:sub3}
  \end{subfigure}

\caption{
Snapshots of the interfacial surfactant concentration \(\tilde{\Gamma}\) (top row), bulk concentration \(\tilde{F}\) (middle row), and vorticity field $\tilde{\omega}_\theta$ with streamlines (bottom row) for increasing Marangoni numbers \(\mathrm{Ma} = 0.2\), \(1\), and \(10\) at fixed adsorption number \(\mathrm{Da} = 0.5\). 
In the pure adsorption regime, increasing \(\mathrm{Ma}\) enhances the accumulation of surfactant at the interface, leading to stronger depletion in the bulk and the formation of a more pronounced stagnation cap. Higher \(\mathrm{Ma}\) values also suppress internal recirculation inside the bubble, as evidenced by the reduced vorticity magnitude and weakened streamline motion.
}
  \label{fig:adsorption_fields}
\end{figure}

Additional insight into the coupled transport and flow dynamics is provided in Fig.~\ref{fig:adsorption_fields}, which displays snapshots of the interfacial concentration, bulk concentration, and vorticity field for increasing Marangoni numbers at fixed adsorption rate. In the pure adsorption regime, surfactant is continuously transferred from the bulk to the interface, producing a depletion of the bulk concentration near the interface and the formation of a pronounced stagnant cap. As the Marangoni number increases, the angular extent of the stagnant cap increases, further suppressing the internal recirculation within the bubble and modifying the surrounding flow structure.

These results demonstrate that the proposed model captures the limiting behavior of soluble surfactants during bubble rise. By independently varying the desorption Biot number and the adsorption Damköhler number, the solver consistently recovers the classical asymptotic regimes: the clean interface limit for large $\mathrm{Bi}$ and the surface-incompressible interface limit for large $\mathrm{Da}$. The associated modifications of the flow structure and bubble dynamics highlight the central role of interfacial kinetics in regulating Marangoni stresses in rising bubbles. Having validated these mechanisms in the axisymmetric configuration, we now extend the analysis to fully three-dimensional simulations to further assess the capabilities of the method.

\section{Three-dimensional rising bubble with surfactant}

To finish, we now consider the classical problem of a single bubble rising under gravity in a three-dimensional liquid column, in the presence of either insoluble or soluble surfactants. This configuration provides a rich benchmark to explore the coupling between hydrodynamics, interfacial tension gradients, and surfactant transport mechanisms. In particular, we focus on the trajectory of the bubble and the development of interfacial instabilities that may arise during the rise process.

A bubble of initial diameter $d_b$ is placed at rest at the bottom of a computational domain of size $[-L/2,\,L/2] \times [-L/2,\,L/2] \times [0,\,L]$, with $L=20 d_b$. Periodic boundary conditions are applied in the vertical direction (z-axis).  The bubble rises because of buoyancy in a quiescent liquid.

The flow is governed by the incompressible Navier–Stokes equations coupled with the surfactant transport equations. For the insoluble case, only the interfacial concentration $\Gamma(\mathbf{x},t)$ is evolved. For the soluble case, both the bulk concentration $F(\mathbf{x},t)$ and the interfacial concentration $\Gamma(\mathbf{x},t)$ are computed using the fully coupled model described in Section~2.

Surface tension is modeled as a nonlinear function of the concentration of interfacial surfactant \(\Gamma(\mathbf{x}, t)\), using the equation of state proposed by \citep{farsoiya2024coupled}:
\begin{equation}
    \sigma(\mathbf{x}, t) = \sigma_0 \left(1 - \tanh\left[\mathrm{E}\frac{\Gamma(\mathbf{x}, t)}{\Gamma_\infty} \right] \right),
\end{equation}
where \(\sigma_0\) denotes the surface tension of a clean interface, \(\Gamma_\infty\) is a interfacial surfactant concentration at saturation, and \(\mathrm{E}\) corresponds to the dimensionless Gibbs elasticity number.

All equations are solved in a non-dimensional form, using $d_b$ as the characteristic length and $\sqrt{d_b/g}$ as the characteristic time. The governing dimensionless parameters include:
\begin{gather}
    \mathrm{Ma} =\frac{\mathcal{R}T\Gamma_\infty}{\mu_l\sqrt{gd_b}},\quad 
    \rho_r = \frac{\rho_l}{\rho_g},\quad 
    \mu_r = \frac{\mu_l}{\mu_g},\quad 
    \mathrm{Da} = \frac{r_a d_b F_{0}}{\sqrt{g d_b}},\quad 
    \mathrm{Bo} = \frac{\rho_l g d_b^2}{\sigma_0}, \nonumber\\
    \mathrm{Pe}_f = \frac{d_b \sqrt{g d_b}}{D_f},\quad 
    \mathrm{Pe}_F = \frac{d_b \sqrt{g d_b}}{D_F},\quad 
    \mathrm{Bi} = \frac{r_d d_b}{\sqrt{g d_b}},\quad 
    \mathrm{Ga} = \frac{\rho_l d_b \sqrt{g d_b}}{\mu_l}.
\end{gather}

For the simulations presented in this work, we consider the following non-dimensional parameters: Marangoni number \(\mathrm{Ma} = 1\), density ratio \(\rho_r = 1000\), viscosity ratio \(\mu_r = 100\), Bond number \(\mathrm{Bo} = 10\), interfacial Péclet number \(\mathrm{Pe}_f = 100\), bulk Péclet number \(\mathrm{Pe}_b = 10\), Biot number \(\mathrm{Bi} = 0.5\), Damköhler number \(\mathrm{Da} = 0\) and Galilei number \(\mathrm{Ga} = 100.25\).

\begin{figure}[h!]
    \centering
    \begin{overpic}[width=1\linewidth]{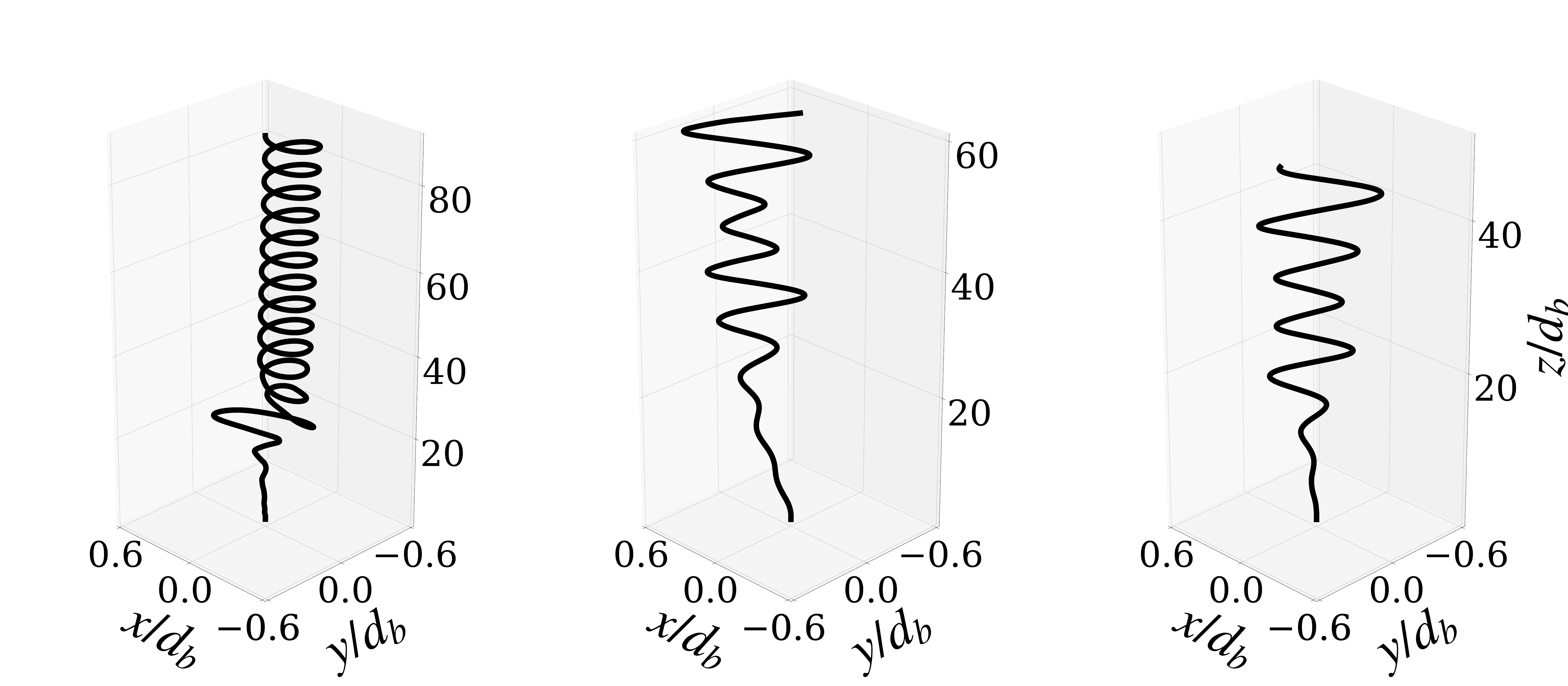}
        \put(12,-3){\scriptsize (a) $\tilde{t}_f=110$}
        \put(46,-3){\scriptsize (b) $\tilde{t}_f=71$}
        \put(80,-3){\scriptsize (c) $\tilde{t}_f=53$}
    \end{overpic}
    \captionsetup{skip=20pt}
    \caption{Vertical trajectories of the bubble centroid for three configurations: (a) clean interface, (b) insoluble surfactant, and (c) soluble surfactant.}
    \label{fig:3Dtrajectory}
\end{figure}

We track the center of mass of the bubble $(x_b(t),\,y_b(t),\,z_b(t))$ over time and analyze the trajectory and stability of its motion. Fig.~\ref{fig:3Dtrajectory} shows the trajectory of the bubble centroid for three configurations: clean interface, insoluble surfactant, and soluble surfactant. In the clean case Fig.~\ref{fig:3Dtrajectory}(a), the bubble rises along a slightly non-vertical path, with a gentle helicoidal motion. With insoluble surfactants Fig.~\ref{fig:3Dtrajectory}(b), surface tension gradients induce lateral Marangoni forces that lead to stronger deviations from verticality, producing asymmetric paths. In the soluble case Fig.~\ref{fig:3Dtrajectory}(c), the intermediate behavior reflects the balance between desorption kinetics and Marangoni effects.

\begin{figure}[H]
    \centering

    \begin{subfigure}{0.3\textwidth}
        \centering
        \includegraphics[width=\linewidth]{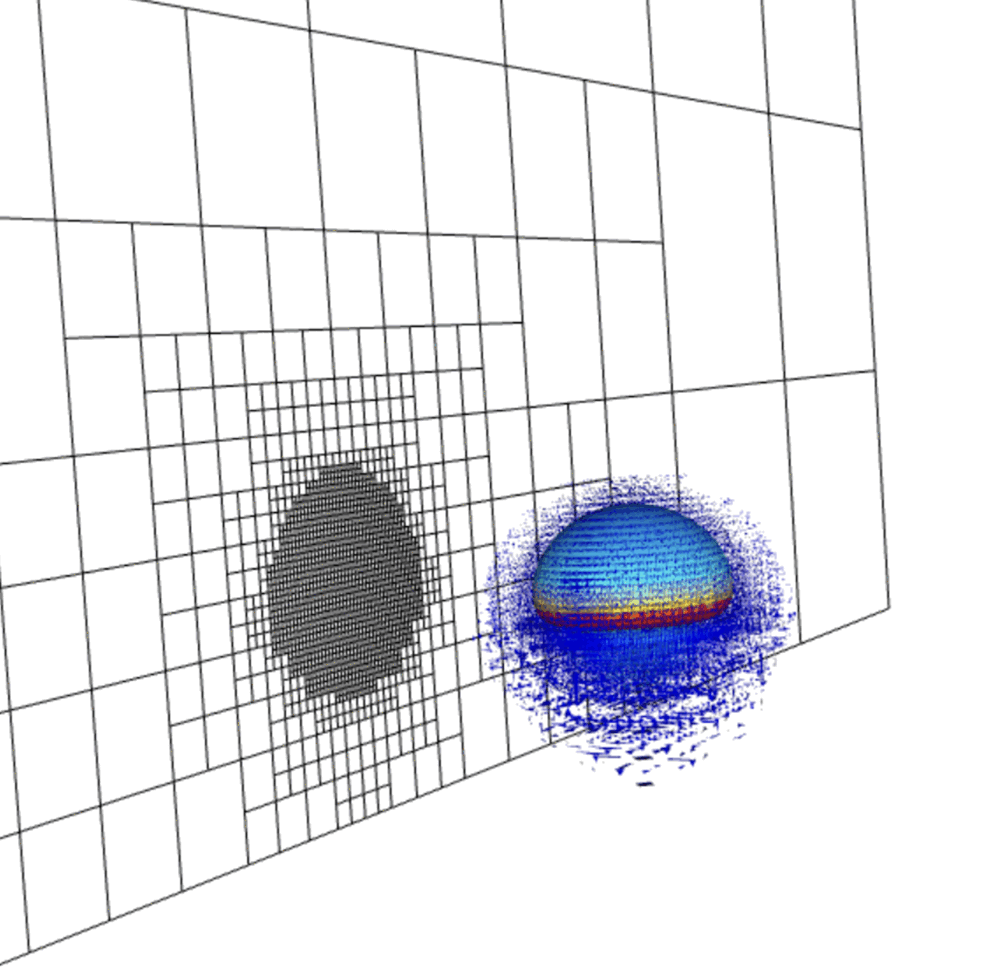}
        \caption{$\tilde{t}=1$}
    \end{subfigure}
    \begin{subfigure}{0.3\textwidth}
        \centering
        \includegraphics[width=\linewidth]{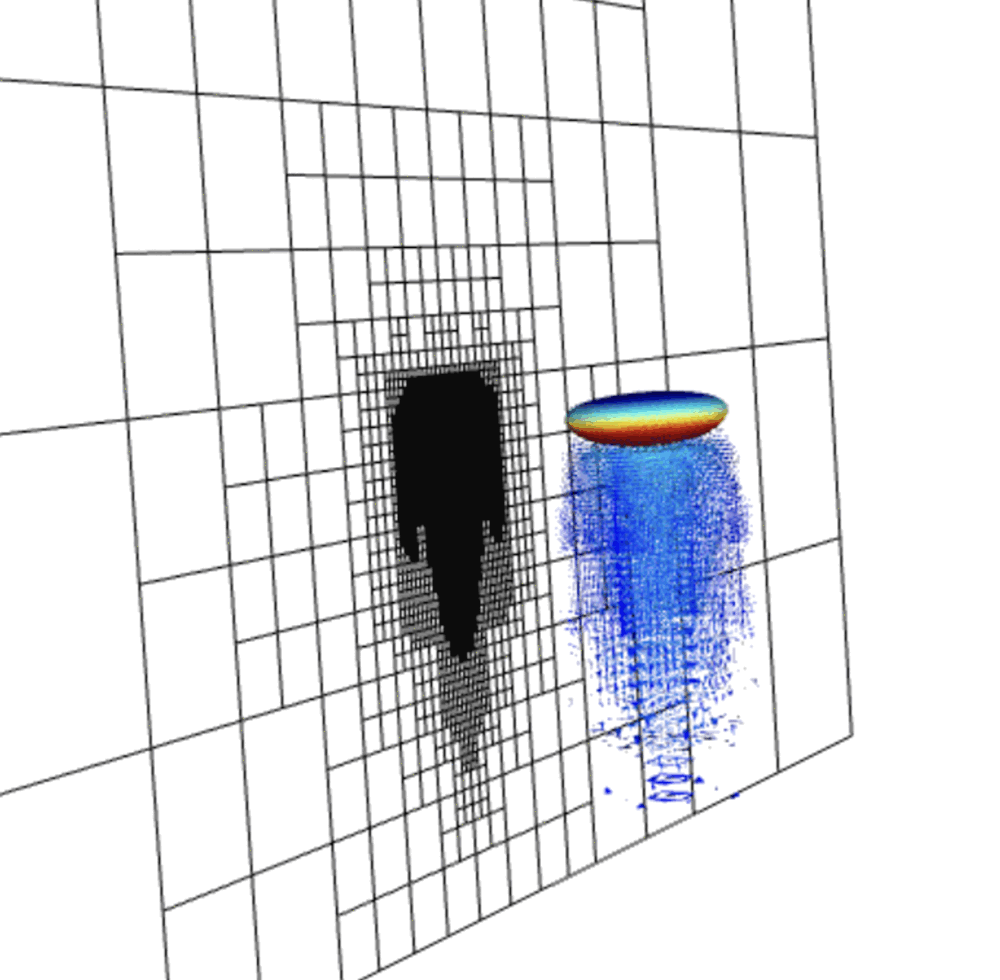}
        \caption{$\tilde{t}=10$}
    \end{subfigure}
    \begin{subfigure}{0.3\textwidth}
        \centering
        \includegraphics[width=\linewidth]{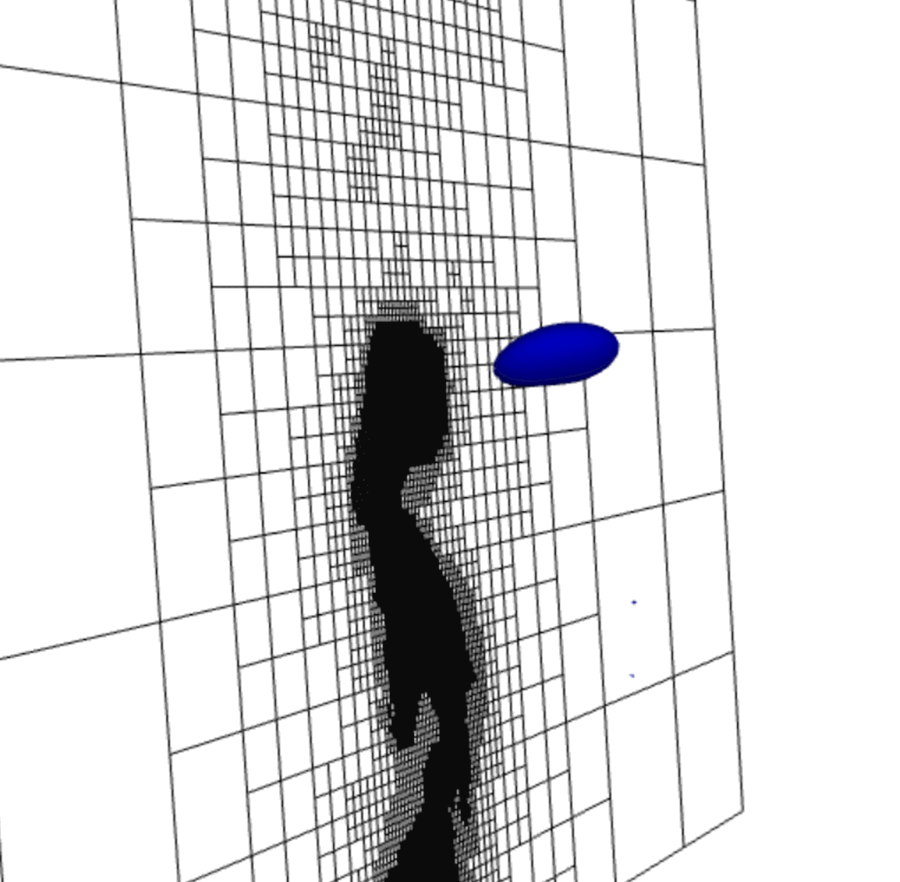}
        \caption{$\tilde{t}=53$}
    \end{subfigure}

    \begin{subfigure}{0.3\textwidth}
        \centering
        \includegraphics[width=\linewidth]{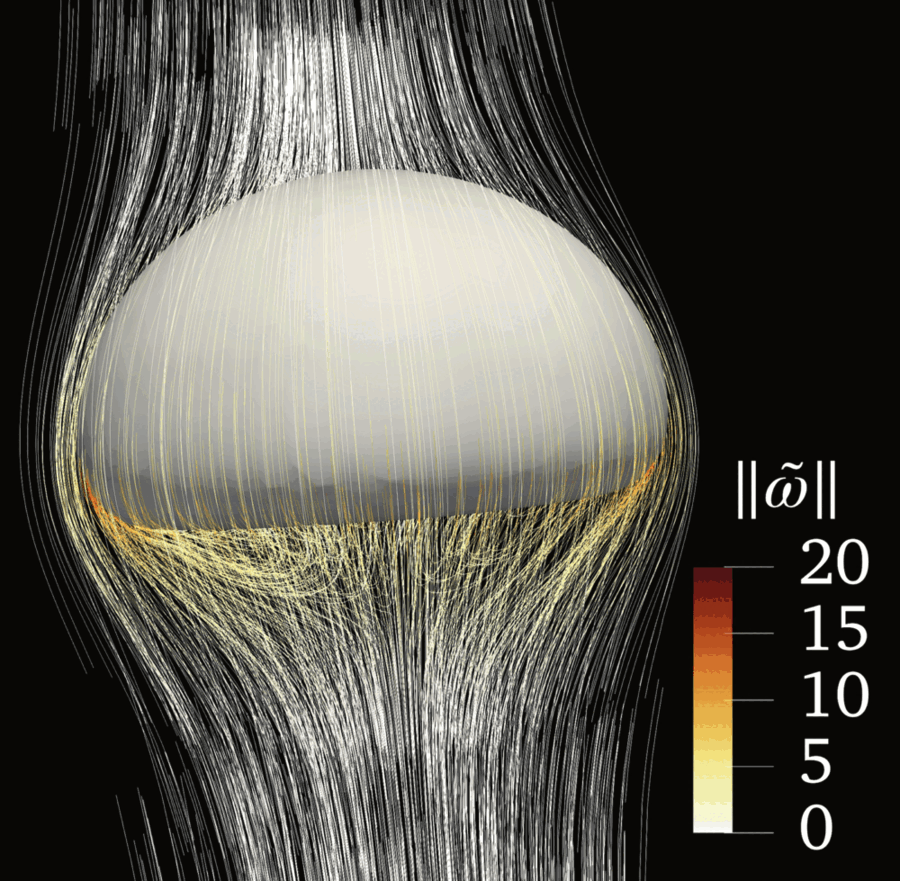}
        \caption{$\tilde{t}=1$}
    \end{subfigure}
    \begin{subfigure}{0.3\textwidth}
        \centering
        \includegraphics[width=\linewidth]{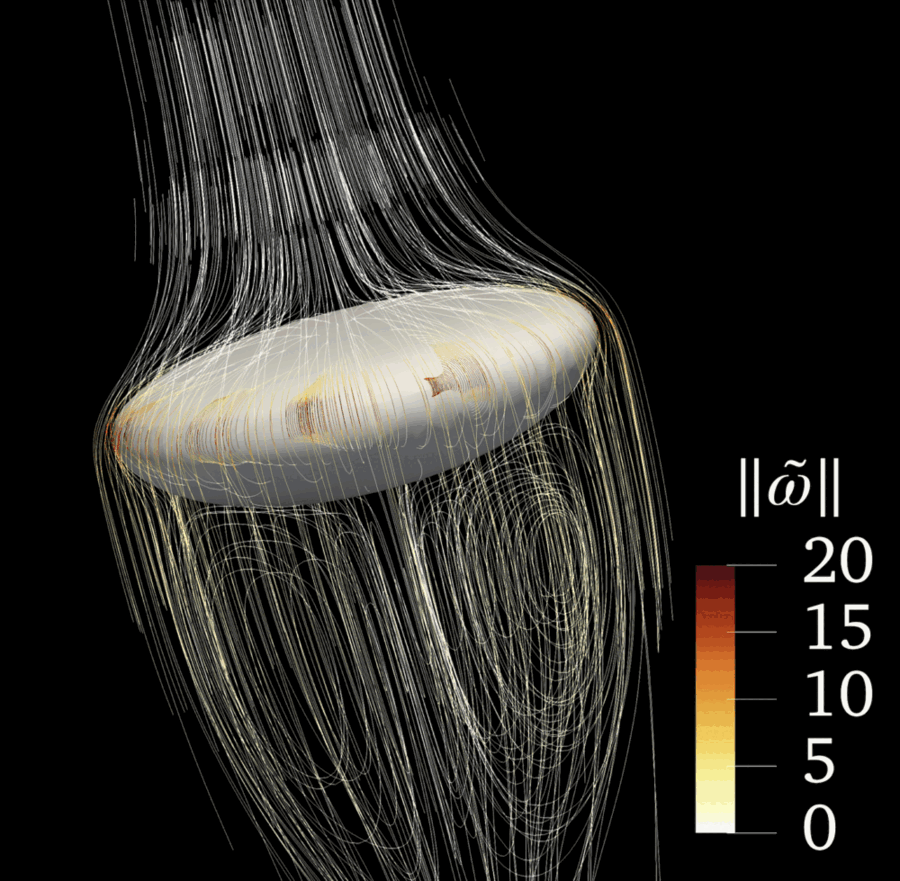}
        \caption{$\tilde{t}=10$}
    \end{subfigure}
    \begin{subfigure}{0.3\textwidth}
        \centering
        \includegraphics[width=\linewidth]{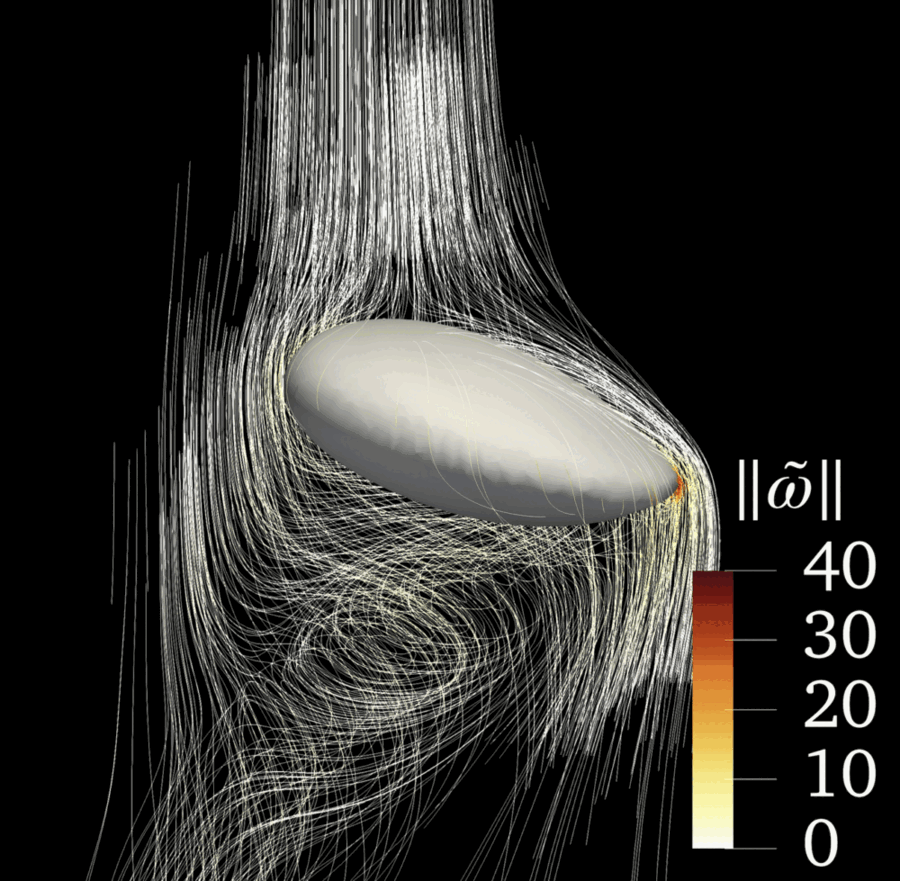}
        \caption{$\tilde{t}=53$}
    \end{subfigure}

    \caption{
    Time evolution of a rising bubble in the pure desorption regime (\(\mathrm{Da} = 0\)). Top row: interfacial surfactant concentration and adaptive mesh at three time instants \(\tilde{t} = 1,\ 10,\ 53\). Initially, surfactants are confined to the interface. As time progresses, desorption transfers surfactants to the bulk, forming a trailing plume. The adaptive mesh dynamically refines around the interface and regions of strong gradients. Bottom row: streamlines and dimensionless vorticy magnitude $\|\tilde{\boldsymbol{\omega}}\|$ defined in the bubble frame. At early times, Marangoni stresses significantly alter the flow. As the interface is depleted, the flow recovers a fore–aft symmetry characteristic of clean-interface dynamics.
    }
    \label{fig:3Dsoluble}
\end{figure}

Fig.~\ref{fig:3Dsoluble} illustrates the temporal evolution of the interfacial concentration, adaptive mesh, and flow field in the pure desorption regime (\(\mathrm{Da} = 0\)). Initially, surfactants are localized at the interface. Over time, desorption transfers them to the bulk, forming a characteristic trailing plume. The adaptive mesh automatically refines around the interface and regions of steep gradients. The bottom row shows streamlines and the flow vorticity magnitude $\|\tilde{\boldsymbol{\omega}}\|$ defined in the bubble frame of reference at successive times. At early times, the flow is strongly altered by interfacial Marangoni stresses. At late times, once the interface is depleted, the flow structure becomes symmetric and resembles the clean case.

This three-dimensional rising bubble example demonstrates the ability of the proposed numerical framework to simulate fully coupled surfactant-laden interfacial flows in complex geometries. The model captures the interplay between interfacial stresses, bulk transport, and kinetic exchanges, while handling deforming geometries and unsteady flow patterns. The results confirm that both the clean and insoluble limits are consistently recovered, and that the method can robustly simulate soluble surfactant dynamics in a fully three-dimensional setting. These capabilities make the method a promising tool for studying realistic multiphase systems involving surfactant-laden interfaces.

\section{Conclusion}

We have developed a robust and versatile numerical framework for the simulation of interfacial flows laden with surfactants, capable of treating both insoluble and soluble surfactant transport mechanisms, including adsorption and desorption kinetics. The model couples the incompressible Navier–Stokes equations with bulk and interfacial advection–diffusion equations, and incorporates Marangoni stresses via a nonlinear equation of state. It is designed to operate seamlessly in planar, axisymmetric and fully three-dimensional geometries.

A comprehensive set of canonical test cases were performed to assess the accuracy, consistency, and predictive capability of the approach. In the insoluble limit, we recovered classical results such as the reduction of the terminal rise velocity due to Marangoni stresses, and verified the accuracy of the interfacial transport solver. To evaluate the role of interfacial kinetics, we investigated pure desorption and pure adsorption regimes, systematically varying the Biot and Damk\"ohler numbers. These tests confirmed that the model recovers the correct asymptotic limits corresponding to clean and insoluble dynamics, thus validating the implementation of kinetic source terms.

We further explored the transition between these limits under soluble surfactant conditions. The coupled effects of bulk transport, interfacial accumulation, and hydrodynamic feedback were captured across a range of Marangoni numbers. To mitigate numerical artifacts in the velocity signal associated with sharp interfacial gradients, a spectral filtering strategy was employed, enabling reliable extraction of the terminal velocity without excessive computational cost.

The framework was finally applied to three-dimensional simulations of rising bubbles with soluble surfactants. These simulations captured intricate features such as shape deformation, lateral trajectory deviations, surfactant redistribution, and the formation of bulk plumes due to desorption. The results demonstrate the ability of the method to resolve complex interfacial dynamics in deforming geometries and to capture the interplay between surfactant transport and flow instabilities.

In summary, the proposed method provides a unified and flexible tool for the simulation of multiphase flows with surfactants, effectively bridging the gap between classical insoluble models and more realistic soluble formulations. Its capability to handle nonlinear coupling, evolving interfaces, and fully three-dimensional configurations makes it a promising open-source platform for future investigations of surfactant-driven phenomena, including droplet coalescence, film rupture, interfacial rheology, and chemically active flows in open or confined environments.

While the present framework captures a broad range of interfacial dynamics, its applicability is currently limited to moderate Péclet numbers due to numerical diffusion inherent in the discretization. Consequently, accurately resolving sharp interfacial gradients and accessing the asymptotic regime of large Péclet numbers remain challenging. Ongoing work aims to overcome this limitation through the development of a new formulation designed to minimize numerical diffusion and enable accurate simulations in the high Péclet or convection-dominated regime.


\section*{CRediT authorship contribution statement}
\textbf{Ilies Haouche:} Conceptualization, Methodology, Investigation, Software, Visualization, Formal analysis, Data curation, Writing – original draft. 
\textbf{Benjamin Reichert:} Supervision, Validation, Visualization, Formal analysis, Writing – review \& editing. 
\textbf{Micha\"el Baudoin:} Supervision, Validation, Visualization, Formal analysis, Funding acquisition, Writing – review \& editing. 
\textbf{Palas Kumar Farsoiya:} Supervision, Conceptualization, Methodology, Funding acquisition, Writing – review \& editing.

\section*{Declaration of competing interests}
The authors declare that they have no known competing financial interests or personal relationships that could have appeared to influence the work reported in this paper.

\section*{Funding}
This work was supported by the French National Research Agency (ANR) [grant ANR-23-CE30-0026 (ACOUSURF)]; the International Speaker Program grant (University of Lille, France); the Faculty Initiation Grant [FIG No.~101039] provided by the Indian Institute of Technology (IIT) Roorkee, Uttarakhand, India; and the Anusandhan National Research Foundation (ANRF) [grant ANRF/ARG/2025/003821/ENS].

\section*{Acknowledgments}
The authors would like to thank Guillaume Lagubeau and Alexis Duchesne for insightful discussions and valuable comments on the manuscript.

\section*{Data availability}
The developed code and scripts for the test cases used in the present study are openly available at:
\begin{center}
\url{https://basilisk.fr/sandbox/haouche/soluble-surfactants/}
\end{center}

\bibliographystyle{elsarticle-num} 
\bibliography{biblio}
\pagebreak

\appendix

\section{3D Finite-volume discretization}

We consider the three-dimensional extension of the finite-volume formulation introduced in Section~3.2. The governing transport equation reads
\begin{gather}
    \int_K \frac{\partial \xi}{\partial t} \, dV
    = \int_K \boldsymbol{\nabla} \cdot \left( \boldsymbol{\mathcal{D}} \boldsymbol{\nabla} \xi + \mathbf{B}\, \xi \right) dV
    + \int_K q \, dV,
\end{gather}
where \(\boldsymbol{\mathcal{D}} = D\mathbf{I}\), such that \(\boldsymbol{\mathcal{D}}\boldsymbol{\nabla}\xi = D\boldsymbol{\nabla}\xi\).

In three dimensions, the finite-volume method consists in integrating the equation over each control volume \(K\), and approximating the solution by its cell-average value:
\begin{equation}
    \xi_{i,j,k}(t) \approx \frac{1}{V} \int_K \xi(\mathbf{x}, t)\, dV,
\end{equation}
with \(V = \Delta x \Delta y \Delta z\).

Applying the divergence theorem yields
\begin{gather}
    \frac{d}{dt} \left( \xi_{i,j,k} V \right)
    =
    \oint_{\partial K}
    \left( D \boldsymbol{\nabla} \xi + \mathbf{B}\, \xi \right)\cdot \mathbf{n}\, dS
    + q_{i,j,k} V.
\end{gather}

The fluxes are evaluated as the sum over the six faces of the control volume:

\begin{gather}
    \sum_{\text{faces}} (D \boldsymbol{\nabla}\xi)\cdot \mathbf{n} S
    =
    \Delta y \Delta z \left[
    \left( D \frac{\partial \xi}{\partial x} \right)_{i+\frac{1}{2},j,k}
    -
    \left( D \frac{\partial \xi}{\partial x} \right)_{i-\frac{1}{2},j,k}
    \right]
    \nonumber\\
    + \Delta x \Delta z \left[
    \left( D \frac{\partial \xi}{\partial y} \right)_{i,j+\frac{1}{2},k}
    -
    \left( D \frac{\partial \xi}{\partial y} \right)_{i,j-\frac{1}{2},k}
    \right]
    \nonumber\\
    + \Delta x \Delta y \left[
    \left( D \frac{\partial \xi}{\partial z} \right)_{i,j,k+\frac{1}{2}}
    -
    \left( D \frac{\partial \xi}{\partial z} \right)_{i,j,k-\frac{1}{2}}
    \right],
\end{gather}

\begin{gather}
    \sum_{\text{faces}} (\mathbf{B}\xi)\cdot \mathbf{n} S
    =
    \Delta y \Delta z \left[
    (B_x \xi)_{i+\frac{1}{2},j,k}
    -
    (B_x \xi)_{i-\frac{1}{2},j,k}
    \right]
    \nonumber\\
    + \Delta x \Delta z \left[
    (B_y \xi)_{i,j+\frac{1}{2},k}
    -
    (B_y \xi)_{i,j-\frac{1}{2},k}
    \right]
    \nonumber\\
    + \Delta x \Delta y \left[
    (B_z \xi)_{i,j,k+\frac{1}{2}}
    -
    (B_z \xi)_{i,j,k-\frac{1}{2}}
    \right].
\end{gather}

Time integration is performed using an implicit Euler scheme:
\begin{gather}
    \frac{\xi_{i,j,k}^{n+1} - \xi_{i,j,k}^n}{\Delta t}
    =
    \frac{1}{\Delta x} \left[
    \left( D \frac{\partial \xi^{n+1}}{\partial x} \right)_{i+\frac{1}{2},j,k}
    -
    \left( D \frac{\partial \xi^{n+1}}{\partial x} \right)_{i-\frac{1}{2},j,k}
    \right]
    \nonumber\\
    + \frac{1}{\Delta y} \left[
    \left( D \frac{\partial \xi^{n+1}}{\partial y} \right)_{i,j+\frac{1}{2},k}
    -
    \left( D \frac{\partial \xi^{n+1}}{\partial y} \right)_{i,j-\frac{1}{2},k}
    \right]
    \nonumber\\
    + \frac{1}{\Delta z} \left[
    \left( D \frac{\partial \xi^{n+1}}{\partial z} \right)_{i,j,k+\frac{1}{2}}
    -
    \left( D \frac{\partial \xi^{n+1}}{\partial z} \right)_{i,j,k-\frac{1}{2}}
    \right]
    \nonumber\\
    + \frac{1}{\Delta x} \left[
    (B_x \xi^{n+1})_{i+\frac{1}{2},j,k}
    -
    (B_x \xi^{n+1})_{i-\frac{1}{2},j,k}
    \right]
    \nonumber\\
    + \frac{1}{\Delta y} \left[
    (B_y \xi^{n+1})_{i,j+\frac{1}{2},k}
    -
    (B_y \xi^{n+1})_{i,j-\frac{1}{2},k}
    \right]
    \nonumber\\
    + \frac{1}{\Delta z} \left[
    (B_z \xi^{n+1})_{i,j,k+\frac{1}{2}}
    -
    (B_z \xi^{n+1})_{i,j,k-\frac{1}{2}}
    \right]
    + q_{i,j,k}^{n+1}.
\end{gather}

As in the two-dimensional case, diffusive fluxes are approximated using centered differences, ensuring second-order spatial accuracy. The anti-diffusive contribution associated with \(\mathbf{B}\) is treated as a convective flux, with \(\mathbf{B}\) acting as an effective velocity field. The corresponding fluxes are discretized using a first-order upwind scheme based on the sign of the face-centered components of \(\mathbf{B}\).

This formulation preserves local conservation and extends naturally to adaptive octree meshes, enabling consistent simulations of surfactant transport in fully three-dimensional configurations.

\section{Finite-volume discretization in axisymmetric geometry}

We consider the transport equation
\begin{equation}
    \frac{\partial \xi}{\partial t}
    =
    \boldsymbol{\nabla}\cdot
    \left(
    D \boldsymbol{\nabla}\xi
    + \mathbf{B}\,\xi
    \right)
    + q.
\end{equation}

In axisymmetric coordinates \((r,z)\), the governing equation becomes
\begin{equation}
    \frac{\partial \xi}{\partial t}
    =
    \frac{1}{r}\frac{\partial}{\partial r}
    \left[
    r\left(D\frac{\partial \xi}{\partial r} + B_r \xi\right)
    \right]
    +
    \frac{\partial}{\partial z}
    \left(
    D\frac{\partial \xi}{\partial z} + B_z \xi
    \right)
    + q,
\end{equation}
where \((B_r,B_z)\) are the components of \(\mathbf{B}\).

The finite-volume formulation is obtained by integrating over a control volume \(K\), yielding
\begin{gather}
    \frac{d}{dt}\big(\xi_{i,j}V_{i,j}\big)
    =
    2\pi \Delta z
    \left[
    r_{i+\frac12}\left(D\frac{\partial \xi}{\partial r} + B_r \xi\right)_{i+\frac12,j}
    -
    r_{i-\frac12}\left(D\frac{\partial \xi}{\partial r} + B_r \xi\right)_{i-\frac12,j}
    \right]
    \nonumber\\
    +
    \pi\left(r_{i+\frac12}^2-r_{i-\frac12}^2\right)
    \left[
    \left(D\frac{\partial \xi}{\partial z} + B_z \xi\right)_{i,j+\frac12}
    -
    \left(D\frac{\partial \xi}{\partial z} + B_z \xi\right)_{i,j-\frac12}
    \right]
    +
    q_{i,j}V_{i,j},
\end{gather}
with the exact cell volume
\begin{equation}
    V_{i,j}
    =
    \pi\left(r_{i+\frac12}^2-r_{i-\frac12}^2\right)\Delta z.
\end{equation}

Time integration is performed using an implicit Euler scheme:
\begin{gather}
    \frac{\xi_{i,j}^{n+1}-\xi_{i,j}^{n}}{\Delta t}
    =
    \frac{2}{r_{i+\frac12}^2-r_{i-\frac12}^2}
    \left[
    r_{i+\frac12}\left(D\frac{\partial \xi^{n+1}}{\partial r} + B_r \xi^{n+1}\right)_{i+\frac12,j}
    -
    r_{i-\frac12}\left(D\frac{\partial \xi^{n+1}}{\partial r} + B_r \xi^{n+1}\right)_{i-\frac12,j}
    \right]
    \nonumber\\
    +
    \frac{1}{\Delta z}
    \left[
    \left(D\frac{\partial \xi^{n+1}}{\partial z} + B_z \xi^{n+1}\right)_{i,j+\frac12}
    -
    \left(D\frac{\partial \xi^{n+1}}{\partial z} + B_z \xi^{n+1}\right)_{i,j-\frac12}
    \right]
    +
    q_{i,j}^{n+1}.
\end{gather}

The diffusive fluxes are approximated using centered differences:
\begin{gather}
    \left(\frac{\partial \xi}{\partial r}\right)_{i+\frac12,j}
    \approx
    \frac{\xi_{i+1,j}^{n+1}-\xi_{i,j}^{n+1}}{\Delta r},
    \qquad
    \left(\frac{\partial \xi}{\partial z}\right)_{i,j+\frac12}
    \approx
    \frac{\xi_{i,j+1}^{n+1}-\xi_{i,j}^{n+1}}{\Delta z}.
\end{gather}

The anti-diffusive contribution is evaluated in conservative form through the face-centered fluxes
\begin{gather}
    (B_r \xi)_{i+\frac12,j}^{n+1},
    \qquad
    (B_z \xi)_{i,j+\frac12}^{n+1}.
\end{gather}

As in the Cartesian formulation, the anti-diffusive term is treated as a convective flux, with \(\mathbf{B}\) acting as an effective velocity field. The corresponding fluxes are discretized using a first-order upwind scheme based on the sign of the face-centered components of \(\mathbf{B}\):
\begin{gather}
    \xi_{i+\frac12,j} =
    \begin{cases}
    \xi_{i,j} & \text{if } (B_r)_{i+\frac12,j} > 0, \\
    \xi_{i+1,j} & \text{otherwise},
    \end{cases}
\end{gather}
and similarly in the \(z\)-direction.

The face-centered values of \(\mathbf{B}\) are obtained by interpolation:
\begin{gather}
    (B_r)_{i+\frac12,j}
    \approx
    \frac{(B_r)_{i+1,j}+(B_r)_{i,j}}{2},
    \qquad
    (B_z)_{i,j+\frac12}
    \approx
    \frac{(B_z)_{i,j+1}+(B_z)_{i,j}}{2}.
\end{gather}

This formulation preserves local conservation and consistently accounts for the geometrical effects associated with axisymmetric coordinates.

\section{Coupled discretization with non-constant source term}

We consider the coupled transport equations for the interfacial and bulk surfactant concentrations, denoted here by \(f(\mathbf{x},t)\) and \(F(\mathbf{x},t)\), respectively. The source term is given by \(q(\mathbf{x},t)=j\delta_\phi\), where \(j\) represents the adsorption–desorption kinetics.

The governing equations read
\begin{equation}
    \frac{\partial f}{\partial t}
    =
    \boldsymbol{\nabla}\cdot \left(D_f \boldsymbol{\nabla} f + \mathbf{B}_f f\right)
    + q,
\end{equation}
with \(f = \Gamma \delta_\phi\) and \(\delta_\phi = \phi(1-\phi)/\epsilon\), and
\begin{equation}
    \frac{\partial F}{\partial t}
    =
    \boldsymbol{\nabla}\cdot \left(D_F \boldsymbol{\nabla} F + \mathbf{B}_F F\right)
    - q.
\end{equation}

The adsorption–desorption source term is defined as
\begin{equation}
    q = r_a \frac{F}{\phi}\left(f_{\infty} - f\right) - r_d f,
\end{equation}
where \(r_a\) and \(r_d\) denote the adsorption and desorption rates, respectively.

The anti-diffusive fluxes are defined by
\begin{equation}
    \mathbf{B}_f = -D_f \frac{2(0.5 - \phi)}{\epsilon}\mathbf{n},
    \qquad
    \mathbf{B}_F = -D_F \frac{1 - \phi}{\epsilon}\mathbf{n}.
\end{equation}

A fully implicit time discretization is employed, leading to the coupled nonlinear system
\begin{gather}
    \frac{f^{n+1} - f^n}{\Delta t}
    =
    \boldsymbol{\nabla}\cdot \left(D_f \boldsymbol{\nabla} f^{n+1} + \mathbf{B}_f f^{n+1}\right)
    + r_a \frac{F^{n+1}}{\phi^{n+1}}\left(f_{\infty} - f^{n+1}\right)
    - r_d f^{n+1},
\end{gather}
\begin{gather}
    \frac{F^{n+1} - F^n}{\Delta t}
    =
    \boldsymbol{\nabla}\cdot \left(D_F \boldsymbol{\nabla} F^{n+1} + \mathbf{B}_F F^{n+1}\right)
    - r_a \frac{F^{n+1}}{\phi^{n+1}}\left(f_{\infty} - f^{n+1}\right)
    + r_d f^{n+1}.
\end{gather}

These equations can be recast into a generalized elliptic form suitable for the Poisson solver.

For the interfacial concentration \(f(\mathbf{x},t)\), we obtain
\begin{gather}
    \boldsymbol{\nabla}\cdot \left(\alpha_f \boldsymbol{\nabla} f^{n+1} + \boldsymbol{\beta}_f f^{n+1}\right)
    + \lambda_f f^{n+1}
    =
    b_f,
\end{gather}
with
\begin{gather}
    \alpha_f = -D_f, \quad
    \boldsymbol{\beta}_f = -\mathbf{B}_f, \nonumber\\
    \lambda_f = \frac{1}{\Delta t} + r_d + r_a \frac{F^{n+1}}{\phi^{n+1}}, \nonumber\\
    b_f = \frac{f^n}{\Delta t} + r_a \frac{F^{n+1}}{\phi^{n+1}} f_{\infty}.
\end{gather}

Similarly, for the bulk concentration \(F(\mathbf{x},t)\), we obtain
\begin{gather}
    \boldsymbol{\nabla}\cdot \left(\alpha_F \boldsymbol{\nabla} F^{n+1} + \boldsymbol{\beta}_F F^{n+1}\right)
    + \lambda_F F^{n+1}
    =
    b_F,
\end{gather}
with
\begin{gather}
    \alpha_F = -D_F, \quad
    \boldsymbol{\beta}_F = -\mathbf{B}_F, \nonumber\\
    \lambda_F = \frac{1}{\Delta t} + r_a \frac{1}{\phi^{n+1}}\left(f_{\infty} - f^{n+1}\right), \nonumber\\
    b_F = \frac{F^n}{\Delta t} + r_d f^{n+1}.
\end{gather}

This formulation highlights that the coupling between the bulk and interfacial concentrations arises solely through the nonlinear source term, which is treated implicitly to ensure stability in stiff regimes.

\end{document}